\documentclass[12pt]{article}

\usepackage[colorlinks=true]{hyperref}
\usepackage{authblk}
\usepackage[toc,page]{appendix}

\usepackage{amsmath,amsfonts}
\usepackage{tensor}
\usepackage{slashed}
\usepackage{xcolor}
\usepackage{graphicx}
\usepackage{nicematrix}
\usepackage{tikz}
\usepackage{cancel}

\newcommand{\se}{\slashed{e}}

\numberwithin{equation}{section}

\newcommand{\ussvaldivia}{Facultad de Ingenier\'ia, Universidad San Sebasti\'an, General Lagos 1163, Valdivia, Chile}
\newcommand{\iowastate}{Department of Mathematics, Iowa State University, Ames IA 50011, USA}
\newcommand{\cecs}{Centro de Estudios Cient\'ificos, Universidad San Sebasti\'an, Avenida Arturo Prat 514, Valdivia, Chile}

\begin{document}

\title{Gravity and Matter from Soldered Chiral Superconnections}

\author[1,3]{P. D. Alvarez \thanks{E-mail \href{mailto:pedro.alvarez@uss.cl}{\nolinkurl{pedro.alvarez@uss.cl}}}}

\author[2]{J. T. Hartwig\thanks{E-mail \href{mailto:jth@iastate.edu}{\nolinkurl{jth@iastate.edu}}}}

\author[1,3]{A. Sharma \thanks{E-mail \href{mailto:ext.aditya.sharma@uss.cl}{\nolinkurl{ext.aditya.sharma@uss.cl}}}}

\affil[1]{\ussvaldivia}
\affil[2]{\iowastate}
\affil[3]{\cecs}

\date{}
\maketitle

\begin{abstract}
We present a reformulation of four-dimensional chiral gravity in terms of a chiral gauge connection and a Clifford-valued coframe, $\slashed{e}=e^{a}\Gamma_{a}$. The exterior algebra generated by $\slashed{e}$, together with the gauge curvature $F$, provides a natural and systematic construction of gauge-covariant differential forms.
We show that suitable combinations of the chiral actions reproduce Einstein--Cartan--Holst gravity with a cosmological term.
The original superconnection invariants yield only algebraic fermionic equations, so the fermionic one-forms do not propagate but instead generate a mass-like mixing between the two chiralities. Spinor propagation is supplied by a spinorial BF-type term, which under the matter ansatz reduces to the standard first-order Weyl kinetic action, including the torsion coupling.
We also review the symmetry and variational properties of the matter ansatz, including its Lorentz equivariance, its relation to residual supersymmetry and twistor spinors, and the connection between the reduced and unrestricted field equations.
\end{abstract}

\section{Introduction}

The formulation of gravity in terms of differential forms reveals a close connection between gravitational theories and gauge theories. In the Einstein--Cartan formulation, the fundamental gravitational variables are the Lorentz connection $\omega^{ab}$ and the coframe one-form $e^{a}$, treated as independent fields \cite{Cartan:1923zea,Sciama:1964jqa,Kibble:1961ba,Hehl:1976kj}. Their associated covariant field strengths are the Lorentz curvature and the torsion, respectively,
\begin{equation}
    R^{ab}
    =
    d\omega^{ab}
    +
    \omega^{a}{}_{c}\wedge\omega^{cb},
    \qquad
    T^{a}
    =
    De^{a}
    =
    de^{a}
    +
    \omega^{a}{}_{b}\wedge e^{b}.
\end{equation}
The Einstein--Cartan action can then be written as
\begin{equation}
    S_{\mathrm{EC}}
    =
    \frac{1}{4\kappa}
    \int
    \epsilon_{abcd}\,
    e^{a}\wedge e^{b}\wedge
    \left(
        R^{cd}
        -
        \frac{\Lambda}{6}
        e^{c}\wedge e^{d}
    \right).
\end{equation}
This formulation is gauge covariant under local Lorentz transformations and naturally accommodates spinorial matter, whose intrinsic angular momentum can source torsion.

Despite their appearance together in the action, the connection and the frame have conceptually different geometric roles. The spin connection is a genuine connection on the Lorentz principal bundle and therefore transforms inhomogeneously,
\begin{equation}
    \omega
    \longrightarrow
    g^{-1}\omega g+g^{-1}dg.
\end{equation}
By contrast, the coframe transforms homogeneously in the Lorentz vector representation,
\begin{equation}
    e^{a}
    \longrightarrow
    \Lambda^{a}{}_{b}\,e^{b}.
\end{equation}
Its purpose is to solder the internal Lorentz vector space to the tangent space of the spacetime manifold. For an invertible frame,
\begin{equation}
    e_x:T_xM\longrightarrow V_{3,1}
\end{equation}
is an isomorphism at every point $x\in M$. It is precisely this soldering property that distinguishes gravity from an ordinary Yang--Mills theory: the internal representation space is identified pointwise with the spacetime tangent space.

This distinction is especially transparent in Weyl's formulation of spinors in curved spacetime \cite{Weyl:1929fm}. A spinor $\psi$ transforms under a local Lorentz transformation in the spinorial representation, and its covariant derivative is defined using the spin connection,
\begin{equation}
    D\psi
    =
    d\psi
    +
    \frac{1}{2}\omega^{ab}\Sigma_{ab}\psi.
\end{equation}
The spin connection compares spinorial frames at neighboring spacetime points. The frame, on the other hand, relates Lorentz tensors and spinors to spacetime tensors. In particular, it enters the curved-space Dirac operator through
\begin{equation}
    \slashed{D}
    =
    \gamma^{a}e_{a}{}^{\mu}D_{\mu}.
\end{equation}
Equivalently, one may introduce the Clifford-valued soldering form
\begin{equation}
    \slashed{e}
    =
    e^{a}\gamma_{a}.
\end{equation}
The spin connection determines parallel transport, whereas $\slashed{e}$ converts spacetime differential forms into Clifford-algebra-valued objects. Weyl's construction therefore provides an early and particularly clear example in which the gauge connection and the soldering form perform complementary but sharply distinguished tasks\footnote{This local Lorentz invariance should be distinguished from the additional local phase invariance that led Weyl to the gauge description of electromagnetism.}.

A different, and remarkably successful, approach consists in enlarging the Lorentz algebra to the Poincar\'e or anti-de Sitter algebra. In these formulations the spin connection and the frame are assembled into a single spacetime connection,
\begin{equation}
    \mathcal{A}
    =
    \frac{1}{2}\omega^{ab}J_{ab}
    +
    \frac{1}{\ell}e^{a}P_{a}
    +\cdots,
\end{equation}
where $P_a$ denotes a translational or anti-de Sitter boost generator. The MacDowell--Mansouri construction showed that the Einstein--Cartan action with cosmological constant can be obtained from the curvature of an enlarged connection after selecting an appropriate Lorentz invariant from the full anti-de Sitter curvature \cite{MacDowell:1977jt,Townsend:1977xw}. Its supersymmetric extension similarly combines the spin connection, frame, and gravitino into a superconnection.

This gauge-theoretic viewpoint was developed further in the Chern--Simons supergravities constructed by Troncoso and Zanelli \cite{Troncoso:1997va,Troncoso:1998ng,Zanelli:2005sa}. In odd dimensions, the gravitational and matter fields can be organized as components of a connection for an appropriate supersymmetric extension of the anti-de Sitter algebra. The Chern--Simons action is then invariant under the full gauge supergroup, and the local supersymmetry algebra closes off shell without requiring the usual set of auxiliary fields.

These ideas also motivated the unconventional supersymmetry models of Alvarez, Valenzuela, Zanelli, and collaborators \cite{Alvarez:2011gd,Alvarez:2013tga,Alvarez:2020qmy}. In these constructions the fields are organized using a gauge superconnection, but the fermionic component is related to a spin-$\tfrac12$ field by a soldering condition of the form
\begin{equation}
    \Psi
    =
    \slashed{e}\,\chi.
\end{equation}
The Clifford-valued frame converts the spin-$\tfrac12$ field $\chi$ into a spinor-valued one-form $\Psi$. Consequently, models with gauge symmetries and rigid supersymmetry can be realized without introducing an independent spin-$\tfrac32$ gravitino. This formulation provides a direct relation between gauge superalgebras, spacetime geometry, and ordinary fermionic matter.

There is, however, a conceptual point that deserves further attention. When gravity is described by a Poincar\'e or anti-de Sitter superconnection, the frame is incorporated into the same enlarged connection as the Lorentz connection. This packaging is extremely useful and is natural from the perspective of Cartan geometry. Nevertheless, it can obscure the distinction between their respective roles: the spin connection defines parallel transport in the internal Lorentz bundle, whereas the frame establishes the soldering between that bundle and spacetime. In particular, the frame transforms as a component of the enlarged connection under translations or anti-de Sitter boosts, even though, from the Lorentz-bundle perspective, it is a tensorial one-form rather than a Lorentz connection.

The purpose of this work is to examine the same gauge-gravity strategy in a setting where the distinction between soldering forms and spin connection remain manifest. We consider theories whose underlying gauge symmetry is the complexified chiral Lorentz symmetry,
\begin{equation}
    \mathfrak{so}(3,1)_{\mathbb C}
    \simeq
    \mathfrak{sl}(2,\mathbb C)_{L}
    \oplus
    \mathfrak{sl}(2,\mathbb C)_{R},
\end{equation}
together with its orthosymplectic extension. The left- and right-handed spin connections are components of the gauge superconnection, while the vielbein is not associated with any generator of the gauge algebra. Instead, it is introduced independently and exclusively as a Clifford-valued soldering form,
\begin{equation}
    \slashed{e}
    =
    e^{a}
    \left(
        \Gamma^{LR}_{a}
        +
        \Gamma^{RL}_{a}
    \right),
\end{equation}
where the off-diagonal Clifford generators intertwine the left- and right-handed Lorentz representations, see sections \ref{sec:cliiford} and \ref{sec:actionpreliminaries}. Thus, the gauge connection and the frame retain the distinct roles already visible in the Einstein--Cartan--Weyl description.

Chiral formulations of four-dimensional gravity originate in the constrained two-form formulation of Plebański and in Ashtekar's self-dual variables \cite{Plebanski:1977zz,Ashtekar:1986yd,Jacobson:1988yy}. Their metric-free and pure-connection formulations were subsequently developed in Refs.~\cite{Capovilla:1989ac,Capovilla:1991qb,Krasnov:2011pp}. From a geometric perspective, a nondegenerate triple of chiral two-forms determines a conformal metric through the Urbantke construction \cite{Urbantke:1984eb}, while related gauge-theoretic characterizations of Einstein four-manifolds were developed in Ref.~\cite{Fine:2013qta}.

Once the coframe is regarded as a Clifford-valued one-form, its exterior algebra generates the covariant structures required by chiral gravity. In particular, the two-form $\slashed{e}\wedge\slashed{e}$ is block diagonal and decomposes into the self-dual and anti-self-dual Lorentz sectors. Its components are the Plebanski two-forms
\begin{equation}
    B^{i}
    =
    \frac{1}{2}\epsilon^{i}{}_{jk}\,
    e^{j}\wedge e^{k}
    +
    i\,e^{0}\wedge e^{i},
    \qquad
    \bar{B}^{\,i}
    =
    \frac{1}{2}\epsilon^{i}{}_{jk}\,
    e^{j}\wedge e^{k}
    -
    i\,e^{0}\wedge e^{i}.\label{PlebanskiTwoForm}
\end{equation}
Because these forms arise directly from the exterior square of the soldering form, their simplicity properties are built into the construction, see section \ref{sec:eterioralegbracoframe}
for more details.

Exterior products of $\se$ with the gauge curvature $F$ then produce the natural invariant four-forms. Thus, the exterior algebra of the Clifford-valued frame provides the cosmological, chiral gravitational, fermionic, and curvature-squared structures while preserving a clear separation between the gauge connection and the soldering form. Clifford-valued coframes and their exterior products have appeared in quadratic-spinor and Clifford-bundle formulations of gravity \cite{Tung:1999ic,Chou:2005ht,Lisi:2010td}. Moreover, spinor-valued one-forms have previously been used both in super-$SL(2,\mathbb C)$ gauge formulations and as alternative descriptions of propagating spin-$\tfrac12$ fields \cite{Tung:1999ic,Torres-Gomez:2012tka}. The specific construction developed here differs in combining an independent Clifford-valued coframe with the doubled chiral $\operatorname{OSp}(1|2,\mathbb C)$ connection, using the exterior algebra of the coframe to generate both chiral Plebański sectors and the invariant four-forms, and introducing a soldered spinorial BF-type pairing for the propagating Weyl sector.

The paper is organized as follows.
Section~\ref{sec:algebra} introduces the left- and right-handed orthosymplectic algebras, their matrix representations, and the corresponding gauge superconnections and curvatures.
Section~\ref{sec:cliiford} constructs the Clifford-valued soldering form and develops its exterior algebra, showing in particular how its square produces the self-dual and anti-self-dual Plebanski two-forms.
The covariance properties of the soldering form and its relation to the Hodge dual are also discussed.
Section~\ref{sec:actionpreliminaries} constructs and evaluates the three natural invariant four-forms \(\langle\slashed e^{\,4}\rangle\), \(\langle\slashed e^{\,2} \wedge F\rangle\), and \(\langle F\wedge F\rangle\), including their bosonic and fermionic components.
Section~\ref{sec:action} combines these invariants into the action, identifies the chiral weights that reproduce Einstein--Cartan--Holst gravity with a cosmological term, and derives the gravitational and fermionic field equations.
Section~\ref{sec:matter-ansatz-spinor-dynamics} introduces the matter ansatz for the fermionic one-forms and examines the resulting spin-\(\tfrac12\) dynamics. We show that the original superconnection invariants produce only algebraic fermionic equations and introduce a spinorial BF-type invariant that yields a first-order Weyl kinetic term, a mass-like left--right mixing, and the standard algebraic coupling to torsion.
Section~\ref{sec:matter-ansatz-equivariance} analyzes the kinematical, symmetry, and dynamical status of the matter ansatz. In particular, we establish its Lorentz equivariance, derive the twistor-spinor condition for residual supersymmetry, determine the additional condition for background invariance, and clarify why the ansatz generally defines a reduced theory rather than an automatic consistent truncation of the unrestricted vector-spinor system. The appendices collect our conventions, identities for the Plebanski two-forms, the explicit \(6\times6\) graded realization of the orthosymplectic algebra, the three-coframe Clifford identity, and the geometry of the matter-ansatz subspace.

\section{Orthosymplectic algebra}\label{sec:algebra}

We consider the direct sum $\mathfrak{g} = \mathfrak{osp}(1|2)_{L}\oplus\mathfrak{osp}(1|2)_{R}$ in a matrix representation adapted to the decomposition into left- and right-handed Weyl-spinor spaces. Each orthosymplectic factor acts on a three-dimensional graded vector space consisting of a two-dimensional Weyl-spinor block and a one-dimensional odd block. The reasons for this choice of representation is to allow for a lifting of Dirac gamma matrices, we will explain this in details in section \ref{liftingofse}. The complete representation therefore acts on the direct sum $\mathbb{C}^{2|1}_{L}\oplus\mathbb{C}^{2|1}_{R}$.

In this basis, the left- and right-handed Lorentz generators are represented by block-diagonal matrices,
\begin{equation}
J^L_{ab}
=
\left(
\begin{array}{c|c}
\begin{matrix}
(\Sigma_{ab})^{\alpha}{}_{\beta} & 0\\
0 & 0
\end{matrix}
&
0_{3\times 3}
\\ \hline
0_{3\times 3}
&
0_{3\times 3}
\end{array}
\right),
\qquad
J^R_{ab}
=
\left(
\begin{array}{c|c}
0_{3\times 3}
&
0_{3\times 3}
\\ \hline
0_{3\times 3}
&
\begin{matrix}
(\bar\Sigma_{ab})_{\dot\alpha}{}^{\dot\beta} & 0\\
0 & 0
\end{matrix}
\end{array}
\right).
\end{equation}

The generators $J^L_{ab}$ act nontrivially only on the left-handed Weyl-spinor block, whereas $J^R_{ab}$ act only on the right-handed block. The remaining one-dimensional directions are the odd components of the fundamental orthosymplectic representations. Since the two sets of generators occupy disjoint matrix blocks, they commute and define two independent chiral factors.

The corresponding orthosymplectic algebras are
\begin{align}
\mathfrak{osp}(1|2)_{L}:&& \mathfrak{osp}(1|2)_{R}:&&\\
[J^L_{ab},J^L_{cd}] &= -\eta_{ac}J^L_{bd}+\cdots, & [J^R_{ab}, J^R_{cd}] &= -\eta_{ac} J^R_{bd}+\cdots, \\
[J^L_{ab},Q_{\alpha}]
&=
Q_{\beta}(\Sigma_{ab})^{\beta}{}_{\alpha},
&
[J^R_{ab},\bar Q^{\dot\alpha}]
&=
\bar Q^{\dot\beta} (\bar\Sigma_{ab})_{\dot\beta}{}^{\dot\alpha}
,
\\
\{Q_{\alpha},Q_{\beta}\}
&=
-(\Sigma^{ab})_{\alpha\beta}J^L_{ab},
&
\{\bar Q^{\dot\alpha},\bar Q^{\dot\beta}\}
&=
-(\bar\Sigma^{ab})^{\dot\alpha\dot\beta} J^R_{ab}.
\end{align}

The even subalgebra of each orthosymplectic factor is the corresponding chiral Lorentz algebra. The commutators with $Q_{\alpha}$ and $\bar Q^{\dot\alpha}$ show that the odd generators transform as left- and right-handed Weyl spinors, respectively. Their anticommutators close on the appropriate chiral Lorentz generators. In particular, no generator associated with spacetime translations or anti-de Sitter boosts is introduced. The frame will consequently remain external to the gauge superconnection.

The brackets $\langle G \ G' \rangle$, $G, G' \in \mathfrak{g}$, denote the invariant bilinear form induced by the supertrace in the chosen representation. This bilinear form is
\begin{align}
\langle J^L_{ab}J^L_{cd}\rangle
&=
-\frac{1}{2}\eta_{ac}\eta_{bd}
+\frac{1}{2}\eta_{ad}\eta_{bc}
-\frac{i}{2}\epsilon_{abcd}, \label{JJtraceL}\\
\langle J^L_{ab}Q_{\alpha}\rangle
&=
0,
\\
\langle Q_{\alpha}Q_{\beta}\rangle
&=
-2\epsilon_{\alpha\beta},
\\
\langle J^R_{ab} J^R_{cd}\rangle
&=
-\frac{1}{2}\eta_{ac}\eta_{bd}
+\frac{1}{2}\eta_{ad}\eta_{bc}
+\frac{i}{2}\epsilon_{abcd},\label{JJtraceR}\\
\langle J^R_{ab}\bar Q^{\dot\alpha}\rangle
&=
0,
\\
\langle\bar Q^{\dot\alpha}\bar Q^{\dot\beta}\rangle
&=
+2\epsilon^{\dot\alpha\dot\beta}.
\end{align}
Because the left- and right-handed generators act on disjoint matrix blocks, mixed traces vanish:
\[
\langle G_{L}\, G_{R}\rangle=0.
\]
This orthogonality is responsible for the separation of the two chiral sectors in all the invariant forms constructed below.

The two Lorentz-invariant tensors appearing in these expressions are the contraction constructed from $\eta_{ab}$ and the Levi-Civita tensor $\epsilon_{abcd}$. On a chiral spinor representation these structures are related by internal duality. Defining the internal Hodge dual of a Lorentz bivector by
\begin{equation}\label{internalHodgedual}
    (\star X)_{ab}
    :=
    \frac{1}{2}\epsilon_{ab}{}^{cd}X_{cd},
\end{equation}
one has $\star^{2}=-1$ in Lorentzian signature. The left- and right-handed Lorentz generators therefore belong to the two complex eigenspaces of the internal Hodge operator:
\begin{equation}
    \frac{1}{2}\epsilon_{ab}{}^{cd}
    (\Sigma_{cd})_{\alpha\beta}
    =
    -i(\Sigma_{ab})_{\alpha\beta},
    \qquad
    \frac{1}{2}\epsilon_{ab}{}^{cd}
    (\bar\Sigma_{cd})_{\dot\alpha\dot\beta}
    =
    +i(\bar\Sigma_{ab})_{\dot\alpha\dot\beta}. \label{starSigmas}
\end{equation}
Equivalently, $\star\Sigma_{ab}=-i\Sigma_{ab}$ and $\star\bar\Sigma_{ab}=+i\bar\Sigma_{ab}$. Thus, in the conventions used here, $\Sigma_{ab}$ and $\bar\Sigma_{ab}$ span the $-i$ and $+i$ eigenspaces of the internal Hodge operator, respectively. The opposite signs of the Levi-Civita terms in the invariant traces encode precisely these opposite duality properties. Each chiral bilinear is therefore complex, as is natural in a Lorentzian formulation based on self-dual and anti-self-dual variables.

\subsection{Gauge fields and curvatures}\label{sec:gaugefields}

We introduce independent gauge superconnections for the two orthosymplectic factors. They are Lie-superalgebra-valued one-forms of the form
\begin{equation}
A_{L}
=
\frac{1}{2}\omega_{L}^{ab}J_{ab}^{L}
+ \Psi_{L}^{\alpha} Q_{\alpha},
\end{equation}
and
\begin{equation}
A_{R}
=
\frac{1}{2}\omega_{R}^{ab}J_{ab}^{R}
+ \Psi_{R \, \dot\alpha} \bar Q^{\dot\alpha}.
\end{equation}

Here $J_{ab}^{L}$ and $Q_{\alpha}$ are infinitesimal generators of $\mathfrak{osp}(1|2)_{L}$, while $J_{ab}^{R}$ and $\bar Q^{\dot\alpha}$ are generators of $\mathfrak{osp}(1|2)_{R}$. The fields $\omega_{L}^{ab}$ and $\omega_{R}^{ab}$ are the left-handed self-dual and right-handed anti-self-dual parts of the Lorentz connection. They are genuine gauge potentials and transform inhomogeneously under their respective chiral gauge transformations.

The fields $\Psi_{L}^{\alpha}$ and $\Psi_{R\, \dot\alpha}$ are Weyl-spinor-valued one-forms. Their Grassmann parity combines with their differential-form degree according to the standard graded multiplication rules for superconnection-valued differential forms.

Depending on the theory under consideration, we shall work with either a single chiral sector or with both chiral sectors simultaneously. A purely left-handed model is obtained by setting $A_{R}=0$, while a purely right-handed model is obtained by setting $A_{L}=0$. Alternatively, retaining both $A_{L}$ and $A_{R}$ gives a left--right model. A left--right symmetric theory is obtained when the two sectors are assigned the same couplings and are related by the appropriate exchange or complex-conjugation conditions. This notation therefore allows the chiral and left--right symmetric theories to be treated within the same algebraic framework.

The curvature two-form associated with the complete gauge connection is
\begin{equation}
F
=
dA+A\wedge A
=
d(A_{L}+A_{R})
+
(A_{L}+A_{R})\wedge(A_{L}+A_{R}).
\end{equation}

Since the full gauge algebra is a direct sum and the left and right matrix blocks are mutually orthogonal, the mixed commutators vanish. The curvature consequently decomposes as $F=F_{L}+F_{R}$, where each chiral curvature depends only on its corresponding gauge superconnection.

The curvature can be expanded on its components as follows,
\begin{equation}
F_{L} = \frac{1}{2}\mathcal{F}_{L}^{ab}J^{L}_{ab} + \rho_{L}^{\alpha} Q_{\alpha},
\qquad
F_{R} = \frac{1}{2}\mathcal{F}_{R}^{ab}J^{R}_{ab} + \rho_{R\dot{\alpha}} \bar{Q}^{\dot{\alpha}}.
\end{equation}
The curvature of a superconnection contains both even and odd components. The coefficient of $J_{ab}$ defines the generalized chiral Lorentz curvature, while the coefficient of $Q_{\alpha}$ is the Lorentz-covariant derivative of the fermionic one-form. The fermion bilinear in the even component arises from the anticommutator of two odd generators. Explicitly,
\begin{align}
\mathcal{F}_{L}^{ab} &= R_{L}^{ab} - (\Sigma^{ab})_{\alpha \beta} \Psi_{L}^{\alpha}\wedge\Psi_{L}^{\beta},\label{calFabL}\\
\mathcal{F}_{R}^{ab} &= R_{R}^{ab} - (\Sigma^{ab})^{\dot{\alpha} \dot{\beta}} \Psi_{L}^{\dot{\alpha}}\wedge\Psi_{L \, \dot{\beta}}.\label{calFabR}
\end{align}
The second term is a spinor bilinear transforming in the same chiral bivector representation as $R_{L}^{ab}$, where $R_{L}^{ab}$ is the curvature of the left-handed spin connection
\begin{align}
R_{L}^{ab} &= d\omega_{L}^{ab} + \omega_{L}^{a}{}_{c}\wedge\omega_{L}^{cb}, \\
R_{R}^{ab} &= d\omega_{R}^{ab} + \omega_{R}^{a}{}_{c}\wedge\omega_{R}^{cb}.
\end{align}

The Lorentz-covariant derivative is
\begin{align}
\rho_{L}^{\alpha}
:=
D_{L}\Psi_{L}^{\alpha}
&=
d\Psi_{L}^{\alpha}
+
\frac{1}{2}\,
\omega_{L}^{ab}\wedge
\left(\Sigma_{ab}\right)^{\alpha}{}_{\beta}
\Psi_{L}^{\beta}, \label{rhoalpha-definition}
\\
\rho_{R\dot{\alpha}}
:=
D_{R}\Psi_{R\dot{\alpha}}
&=
d\Psi_{R\dot{\alpha}}
+
\frac{1}{2}\,
\omega_{R}^{ab}\wedge
\left(\bar{\Sigma}_{ab}\right)_{\dot{\alpha}}{}^{\dot{\beta}}
\Psi_{R\dot{\beta}}. \label{rhoalphadot-definition}
\end{align}

This derivative acts simultaneously on the differential-form and spinorial structures of $\Psi_{L}^{\alpha}$. Its covariance follows from the transformation of $\Psi_{L}^{\alpha}$ in the Weyl-spinor representation and the inhomogeneous transformation of $\omega_{L}^{ab}$. The right-handed curvature is obtained analogously by replacing undotted indices and $\Sigma_{ab}$ with dotted indices and $\bar\Sigma_{ab}$.

The Lorentz-valued components of the curvatures inherit the chirality properties of the corresponding left- and right-handed Lorentz connections. The Hodge dual in this statement acts on the internal Lorentz indices, and not on the spacetime differential-form indices. With the conventions adopted here, the left-handed curvature is anti-self-dual, whereas the right-handed curvature is self-dual:
\begin{align}
\left(\star\mathcal{F}_{L}\right)^{ab}
&:=
\frac{1}{2}\epsilon^{ab}{}_{cd}\,
\mathcal{F}_{L}^{cd}
=
-i\,\mathcal{F}_{L}^{ab},
\\
\left(\star\mathcal{F}_{R}\right)^{ab}
&:=
\frac{1}{2}\epsilon^{ab}{}_{cd}\,
\mathcal{F}_{R}^{cd}
=
+i\,\mathcal{F}_{R}^{ab}.
\end{align}
Indeed, the purely bosonic curvatures satisfy
\begin{align}
\frac{1}{2}\epsilon^{ab}{}_{cd}\,R_{L}^{cd}
&=
-i\,R_{L}^{ab},
&
\frac{1}{2}\epsilon^{ab}{}_{cd}\,R_{R}^{cd}
&=
+i\,R_{R}^{ab},
\end{align}
while the chiral spinor matrices obey \eqref{starSigmas}. Consequently, the fermionic bilinears appearing in \eqref{calFabL} and \eqref{calFabR} have the same duality properties as their respective bosonic curvatures. Therefore, the complete supercovariant curvature components remain in the irreducible chiral Lorentz sectors,
\begin{equation}
\mathcal{F}_{L}^{ab}\in(1,0),
\qquad
\mathcal{F}_{R}^{ab}\in(0,1).
\end{equation}

\section{Clifford-valued soldering form}\label{sec:cliiford}

The introduction of the Clifford-valued soldering form is already motivated by the kinetic terms of ordinary spin-$\tfrac12$ fields. In a chiral basis, a Dirac spinor may be assembled from two independent Weyl spinors as
\begin{equation}
    \Psi_D
    =
    \begin{pmatrix}
        \xi^\alpha\\[1mm]
        \eta^\dagger_{\dot\alpha}
    \end{pmatrix},
    \qquad
    \overline{\Psi}_D
    =
    \Psi_D^\dagger \mathcal{A}_D
    =
    \begin{pmatrix}
        \eta_\alpha
        &
        \xi^{\dagger\, \dot\alpha}
    \end{pmatrix}.
\end{equation}
where $\mathcal{A}_D$ is the hermitizing matrix. In the representation adopted here, $\mathcal{A}_D=\gamma^0$.
The Dirac matrices in the same basis are off diagonal,
\begin{equation}
    \gamma^a
    =
    \begin{pmatrix}
        0
        &
        (\sigma^a)^{\alpha\dot\beta}
        \\[1mm]
        (\bar\sigma^a)_{\dot\alpha\beta}
        &
        0
    \end{pmatrix}.
\end{equation}
Consequently, the Dirac kinetic term decomposes into two independent Weyl kinetic terms:
\begin{equation}
    i\overline{\Psi}_D\gamma^\mu D_\mu\Psi_D
    =
    i\eta_\alpha
    (\sigma^\mu)^{\alpha\dot\beta}
    D_\mu\eta^{\dagger\dot\beta}
    +
    i\xi^{\dagger\, \dot\alpha}
    (\bar\sigma^\mu)_{\dot\alpha\beta}
    D_\mu\xi^\beta.
\end{equation}
The essential algebraic ingredients are therefore the two intertwiners $(\sigma^a)^{\alpha\dot\beta}$ and $(\bar\sigma^a)_{\dot\alpha\beta}$. In Minkowski spacetime one normally writes $\sigma^\mu$ and $\bar\sigma^\mu$, tacitly identifying the coordinate index $\mu$ with the Lorentz index $a$. No such canonical identification exists on a generic curved spacetime. It must instead be supplied by the frame:
\begin{equation}
    (\sigma^\mu)^{\alpha\dot\beta}
    =
    e_a{}^\mu
    (\sigma^a)^{\alpha\dot\beta},
    \qquad
    (\bar\sigma^\mu)_{\dot\alpha\beta}
    =
    e_a{}^\mu
    (\bar\sigma^a)_{\dot\alpha\beta}.
\end{equation}
The inverse frame converts the spacetime index carried by the covariant derivative into an internal Lorentz index, while the Pauli matrices convert that Lorentz index into a pair of Weyl-spinor indices. The curved-space Weyl kinetic terms are therefore
\begin{equation}
    S_{\mathrm{Weyl}}
    =
    i\int d^4x\,e
    \left[
        \eta_\alpha
        (\sigma^a)^{\alpha\dot\beta}
        e_a{}^\mu
        D_\mu\eta^\dagger_{\dot\beta}
        +
        \xi^{\dagger\,\dot\alpha}
        (\bar\sigma^a)_{\dot\alpha\beta}
        e_a{}^\mu
        D_\mu\xi^\beta
    \right],
\end{equation}
where $e=\det(e^a{}_\mu)$. This expression clearly separates the two geometric operations: the spin connection contained in $D_\mu$ defines parallel transport in the Weyl-spinor bundles, whereas the frame relates the spacetime tangent bundle to the internal Lorentz representations.

The same action can be written entirely in terms of differential forms. Introducing
\begin{equation}
    \ast e^a
    =
    \frac{1}{3!}
    \epsilon^a{}_{bcd}\,
    e^b\wedge e^c\wedge e^d,
\end{equation}
the Weyl kinetic terms become
\begin{equation}
    S_{\mathrm{Weyl}}
    =
    i\int
    \left[
        \eta_\alpha
        (\sigma_a)^{\alpha\dot\beta}
        D\eta^\dagger_{\dot\beta}
        \wedge\ast e^a
        +
        \xi^{\dagger\,\dot\alpha}
        (\bar\sigma_a)_{\dot\alpha\beta}
        D\xi^\beta
        \wedge\ast e^a
    \right].\label{weylkineticterm}
\end{equation}
The frame appearing in this expression is not another gauge connection. It is the soldering field required to convert the spacetime one-form $D\xi$ or $D\eta^\dagger$ into a Lorentz-invariant top form. The Weyl action therefore provides a direct physical motivation for introducing the frame independently of the chiral spin connections.

Equivalently, one may combine the frame and the Pauli intertwiners into the bispinor-valued one-forms
\begin{equation}
    e^{\alpha\dot\alpha}
    =
    e^a(\sigma_a)^{\alpha\dot\alpha},
    \qquad
    \bar e_{\dot\alpha\alpha}
    =
    e^a(\bar\sigma_a)_{\dot\alpha\alpha}. \label{solderingoneforms}
\end{equation}
These objects are the spinorial realization of the soldering form. They map the two Weyl-spinor bundles into one another:
\begin{equation}
    e_{\alpha}{}^{\dot\alpha}
    :
    S_R\longrightarrow
    \Omega^1(M,S_L),
    \qquad
    \bar e^{\dot\alpha}{}_{\alpha}
    :
    S_L\longrightarrow
    \Omega^1(M,S_R).
\end{equation}
This explains why a spinor-valued one-form can be constructed from a Weyl-spinor zero-form without introducing an independent vector-spinor. Identifying $\xi^\alpha=\psi^\alpha$ and $\eta^{\dagger\, \dot\alpha}=\chi^{\dot\alpha}$, the corresponding soldered one-forms are
\begin{equation}
    \Psi_L^\alpha
    =
    e^a(\sigma_a)^{\alpha\dot\alpha}
    \chi_{\dot\alpha},
    \qquad
    \Psi_{R\, \dot\alpha}
    =
    e^a(\bar\sigma_a)_{\dot\alpha\alpha}
    \psi^\alpha.
\end{equation}
The opposite-chirality spinor index carried by the frame is contracted with the original Weyl field, leaving a one-form with a single free spinor index. In representation-theoretic language, the soldering form supplies the additional $(\frac{1}{2},\frac{1}{2})$ structure required to relate the two inequivalent Weyl representations. There is no nonzero Lorentz-equivariant map from a single Weyl-spinor space directly to the vector representation without this additional opposite-chirality structure, more details on Section \ref{sec:covariance}.

The two soldering maps are naturally assembled into the Clifford-valued coframe
\begin{equation}
    \slashed e
    =
    e^a\gamma_a
    =
    \begin{pmatrix}
        0
        &
        e^a(\sigma_a)^{\alpha \dot\alpha}
        \\[2mm]
        e^a(\bar\sigma_a)_{\dot\alpha \alpha}
        &
        0
    \end{pmatrix}. \label{solderedgammaa}
\end{equation}
Thus, $\slashed e$ is not introduced merely as a convenient algebraic device. Its off-diagonal blocks are precisely the soldering structures already required to formulate the kinetic terms of left- and right-handed Weyl fermions in curved spacetime. The exterior algebra generated by $\slashed e$ then extends this same construction to higher-degree covariant forms, including the chiral two-forms and invariant four-forms needed for the gravitational action.

\subsection{Exterior algebra of the Clifford-valued soldering form}
\label{sec:eterioralegbracoframe}

Before introducing the explicit embedding into the orthosymplectic matrix representation, it is useful to describe the exterior algebra generated by the Clifford-valued frame. Let $S_L$ and $S_R$ denote the left- and right-handed Weyl-spinor bundles, transforming in the $(\frac{1}{2},0)$ and $(0,\frac{1}{2})$ representations of the complexified Lorentz group, respectively. Their direct sum
\begin{equation}
    S_D
    =
    S_L\oplus S_R \label{DiracSpaceDirectSum}
\end{equation}
is the Dirac-spinor bundle. The complexified Lorentz-vector representation satisfies
\begin{equation}
    V_{\mathbb C}
    \simeq
    S_L\otimes S_R
    \simeq
    \left(\tfrac{1}{2},\tfrac{1}{2}\right),
\end{equation}
where the invariant antisymmetric tensors $\epsilon_{\alpha\beta}$ and $\epsilon_{\dot\alpha\dot\beta}$ have been used to identify each fundamental spinor representation with its dual.

The Pauli matrices provide the corresponding Lorentz-equivariant intertwiners,
\begin{equation}
    \sigma:
    V_{\mathbb C}
    \longrightarrow
    \operatorname{Hom}(S_R,S_L),
    \qquad
    \bar\sigma:
    V_{\mathbb C}
    \longrightarrow
    \operatorname{Hom}(S_L,S_R). \label{intertwinermap}
\end{equation}
wich are maps defined by
\begin{equation}
    v^a
    \longmapsto
    v^a\sigma_a\,
    \qquad
        v^a
    \longmapsto
    v^a\bar{\sigma}_a.
\end{equation}
The notation $\operatorname{Hom}(S_R,S_L)$ denotes the vector space of all complex-linear maps from the right-handed Weyl-spinor space $S_R$ to the left-handed Weyl-spinor space $S_L$:
\begin{equation}
    \operatorname{Hom}(S_R,S_L)
    =
    \left\{
        X:S_R\longrightarrow S_L
        \,\middle|\,
        X\ \text{is complex linear}
    \right\}.
\end{equation}
Thus, $(\sigma_a)^{\alpha \dot\alpha}$ and $(\bar\sigma_a)_{\dot\alpha\alpha}$ convert a Lorentz-vector index into a pair of Weyl-spinor indices. They should not be interpreted as Lorentz-equivariant maps from a single Weyl-spinor representation to the vector representation. Such a map would require an additional spinor of the opposite chirality. Their natural role is instead to identify a vector with a bispinor or, equivalently, with a homomorphism between the two Weyl-spinor spaces.

Since $S_R$ and $S_L$ are both two-dimensional complex vector spaces, an element of $\operatorname{Hom}(S_R,S_L)$ is represented by a $2\times2$ complex matrix. In spinor notation, such an element is written as $X^{\alpha \dot\beta}
    \in
    \operatorname{Hom}(S_R,S_L).$
Its action on a right-handed Weyl spinor $\chi_{\dot\beta}\in S_R$ is
\begin{equation}
    X^{\alpha\dot\beta}:
    \chi_{\dot\beta}
    \longmapsto
    (X\chi)^\alpha
    =
    X^{\alpha \dot\beta}\chi_{\dot\beta}.
\end{equation}
The resulting object has an undotted index and therefore belongs to $S_L$.

Equivalently, the bundle of homomorphisms can be written as
\begin{equation}
    \operatorname{Hom}(S_R,S_L)
    \simeq
    S_L\otimes S_R^*.
\end{equation}
This expression reflects the fact that a linear map has one output index in $S_L$ and one dual input index in $S_R^*$. The invariant tensor $\epsilon_{\dot\alpha\dot\beta}$ can be used to identify $S_R^*$ with $S_R$ and to raise or lower the dotted index.

Therefore, the two soldering mapsm  $e^a\sigma_a$ and $e^a\bar\sigma_a$, the two Weyl-spinor bundles into one another, while simultaneously supplying the spacetime one-form index.

The coframe $e^a\in\Omega^1(M,V)$ therefore defines the spinor-valued one-forms $e^{\alpha \dot{\alpha}}$ and $e_{\dot{\alpha} \alpha}$, see \eqref{solderingoneforms}. The first is a one-form valued in $\operatorname{Hom}(S_R,S_L)$, while the second is valued in $\operatorname{Hom}(S_L,S_R)$.
Accordingly,
\begin{equation}
    \slashed e
    \in
    \Omega^1\!\left(M,\operatorname{End}(S_D)\right).
\end{equation}
The object $\slashed e$ is not itself a section of the Dirac-spinor bundle. Rather, it is an endomorphism-valued differential form acting on sections of that bundle. The algebra generated by its exterior powers is therefore a subalgebra of the algebra of Clifford-valued differential forms,
\begin{equation}
    \Omega^\bullet\!\left(M,\operatorname{End}(S_D)\right),
\end{equation}
where multiplication combines the exterior product of differential forms with matrix composition in $\operatorname{End}(S_D)$.

The exterior powers of the Clifford-valued coframe are defined by
\begin{equation}
    \slashed e^{\,p}
    :=
    \underbrace{
    \slashed e\wedge\cdots\wedge\slashed e
    }_{p\ {\rm factors}}.
\end{equation}
Because the form coefficients are completely antisymmetrized by the exterior product, only the antisymmetric Clifford products contribute:
\begin{equation}
    \slashed e^{\,p}
    =
    e^{a_1}\wedge\cdots\wedge e^{a_p}
    \gamma_{a_1\cdots a_p}.
\end{equation}
where
\begin{equation}
    \gamma_{a_1\cdots a_p}
    :=
    \gamma_{[a_1}\cdots\gamma_{a_p]}.
\end{equation}
For a nondegenerate frame, the collection of these antisymmetrized products realizes the usual vector-space identification, under which
\begin{equation}
    e^{a_1}\wedge\cdots\wedge e^{a_p}
    \longleftrightarrow
    \gamma^{a_1 \cdots a_p},
\end{equation}
between the exterior algebra and the complexified Clifford algebra, and therefore
\begin{equation}
    \Lambda^\bullet V_{\mathbb C}^{*}
    \simeq
    \operatorname{Cl}(V_{\mathbb C},\eta)
\end{equation}
as vector spaces. This identification does not, however, equate the exterior product with the full Clifford product, since the latter also contains contractions induced by the metric. In exterior powers of $\slashed e$, these symmetric metric contributions vanish automatically because of the antisymmetry of the wedge product.

The Clifford-valued frame is off diagonal in the chiral basis. Consequently, its odd exterior powers interchange chirality, whereas its even exterior powers preserve chirality.
This alternation follows immediately from the block-matrix product: a single off-diagonal matrix changes chirality, two such matrices return to the original chiral subspace, and the pattern repeats for higher powers.

The exterior square is explicitly
\begin{equation}
    \slashed e\wedge\slashed e
    =
    \begin{pmatrix}
        e^a\wedge e^b\,
        \sigma_a\bar\sigma_b
        &
        0
        \\[2mm]
        0
        &
        e^a\wedge e^b\,
        \bar\sigma_a\sigma_b
    \end{pmatrix}.
\end{equation}
Using the Clifford relations, the products $\sigma_a\bar\sigma_b$ and $\bar\sigma_a\sigma_b$ decompose into a symmetric part proportional to $\eta_{ab}$ and an antisymmetric part proportional to the chiral Lorentz generators, which leads to
\begin{equation}
    \slashed e\wedge\slashed e
    =
    -2e^a\wedge e^b
    \begin{pmatrix}
        \Sigma_{ab} & 0
        \\[2mm]
        0 & \bar\Sigma_{ab}
    \end{pmatrix},\label{sese}
\end{equation}
where symmetric contribution vanishes upon contraction with $e^a\wedge e^b$. The minus sign on \eqref{sese} comes from raising and lowering spinor indices with the conventions explained in the Appendix \ref{app:preliminaries}.

The matrices $\Sigma_{ab}$ and $\bar\Sigma_{ab}$ are precisely the bivector soldering forms that convert an antisymmetric Lorentz pair $[ab]$ into a symmetric pair of undotted or dotted spinor indices:
\begin{equation}
    \Sigma_{ab}:
    \Lambda^2 V_{\mathbb C}
    \longrightarrow
    \operatorname{Sym}^2(S_L),
    \qquad
    \bar\Sigma_{ab}:
    \Lambda^2 V_{\mathbb C}
    \longrightarrow
    \operatorname{Sym}^2(S_R).
\end{equation}
They project a general complex Lorentz bivector onto its two irreducible
chiral components,
\begin{equation}
    \Lambda^2 V_{\mathbb C}
    \simeq
    (1,0)\oplus(0,1)
    \simeq
    \operatorname{Sym}^2(S_L)
    \oplus
    \operatorname{Sym}^2(S_R).
\end{equation}

Accordingly, the two chiral two-forms obtained from the frame are
\begin{equation}
    \mathcal B_{\alpha\beta}^{L}
    :=
    -e^a\wedge e^b
    (\Sigma_{ab})_{\alpha\beta},
    \qquad
    \mathcal B_{\dot\alpha\dot\beta}^{R}
    :=
    -e^a\wedge e^b
    (\bar\Sigma_{ab})_{\dot\alpha\dot\beta}.
\end{equation}
They are symmetric in their spinor indices,
\begin{equation}
    \mathcal B_{\alpha\beta}^{L}
    =
    \mathcal B_{\beta\alpha}^{L},
    \qquad
    \mathcal B_{\dot\alpha\dot\beta}^{R}
    =
    \mathcal B_{\dot\beta\dot\alpha}^{R},
\end{equation}
and constitute the correct soldered representatives of the self-dual and anti-self-dual two-form sectors. These are the spinorial forms of the Plebański self-dual and anti-self-dual two-forms \cite{Plebanski:1977zz,Capovilla:1991qb}.

Indeed, using the Lorentzian duality conventions adopted here, see \eqref{starSigmas}, one finds that the spacetime Hodge dual
\begin{equation}
    \ast\mathcal B_{\alpha\beta}^{L}
    =
    -i\mathcal B_{\alpha\beta}^{L},
    \qquad
    \ast\mathcal B_{\dot\alpha\dot\beta}^{R}
    =
    +i\mathcal B_{\dot\alpha\dot\beta}^{R}.
\end{equation}
is compatible with the internal hodge dual \eqref{internalHodgedual}. Thus, the undotted soldered two-form belongs to the $-i$ chiral eigenspace, while the dotted soldered two-form belongs to the $+i$ chiral eigenspace. Consequently,
\begin{equation}
    \slashed e\wedge\slashed e
    =
    2
    \begin{pmatrix}
        \mathcal B^{L} & 0\\[2mm]
        0 & \mathcal B^{R}
    \end{pmatrix},
\end{equation}
where the spinor indices on the two diagonal blocks are understood. The exterior square of the Clifford-valued soldering form therefore produces directly the properly soldered self-dual and anti-self-dual two-forms required in the two chiral gravitational sectors.

Because the chiral two-forms are constructed from a single frame, they automatically satisfy the Plebański simplicity relations. In an orthonormal chiral basis $B^i$, see \ref{PlebanskiTwoForm}, these relations take the form
\begin{equation}\label{simplicityrelation}
    B^i\wedge B^j
    =
    \frac{1}{3}\delta^{ij}
    B^k\wedge B_k.
\end{equation}

Thus, the exterior square of $\slashed e$ produces simple chiral two-forms without introducing an independent $B$ field or imposing simplicity through a separate Lagrange multiplier. Conversely, a nondegenerate triple of two-forms satisfying the simplicity relations determines the conformal metric, as in the Urbantke construction \cite{Urbantke:1984eb}.

The exterior cube is again off diagonal:
\begin{equation}
    \slashed e^{\,3}
    =
    \begin{pmatrix}
        0
        &
        e^a\wedge e^b\wedge e^c\,
        \sigma_a\bar\sigma_b\sigma_c
        \\[2mm]
        e^a\wedge e^b\wedge e^c\,
        \bar\sigma_a\sigma_b\bar\sigma_c
        &
        0
    \end{pmatrix}.
\end{equation}
In four dimensions, three-forms are Hodge dual to one-forms. Accordingly,
\begin{equation}
    \Lambda^3 V_{\mathbb C}
    \simeq
    \Lambda^1 V_{\mathbb C}
    \simeq
    \left(\frac{1}{2},\frac{1}{2}\right).
\end{equation}
The exterior cube therefore transforms again as a Lorentz vector and interchanges the two chiral spinor spaces, just as $\slashed e$ itself does.

The fourth exterior power is block diagonal and proportional to the spacetime volume form:
\begin{equation}
    \slashed e^{\,4}
    =
    e^a\wedge e^b\wedge e^c\wedge e^d\,
    \gamma_{[a}\gamma_b\gamma_c\gamma_{d]}.
\end{equation}
Since
\begin{equation}
    \Lambda^4 V_{\mathbb C}
    \simeq
    (0,0),
\end{equation}
the top-degree element is a Lorentz singlet. In the Clifford representation it is proportional to the chirality matrix and acts with opposite signs on $S_L$ and $S_R$. After applying the appropriate invariant trace, it gives the volume or cosmological contribution to the action. All higher exterior powers vanish identically in four dimensions.

The complete Lorentz-representation content of the exterior algebra is therefore
\begin{equation}
\begin{array}{c|c|c}
p
&
\Lambda^p V_{\mathbb C}
&
\text{chiral block structure}
\\ \hline
0
&
(0,0)
&
\text{diagonal}
\\[1mm]
1
&
\left(\frac{1}{2},\frac{1}{2}\right)
&
\text{off diagonal}
\\[1mm]
2
&
(1,0)\oplus(0,1)
&
\text{diagonal}
\\[1mm]
3
&
\left(\frac{1}{2},\frac{1}{2}\right)
&
\text{off diagonal}
\\[1mm]
4
&
(0,0)
&
\text{diagonal}.
\end{array}
\end{equation}
The dimensions of the successive exterior powers are $1$, $4$, $6$, $4$, and $1$, giving the expected total dimension
\begin{equation}
    \dim_{\mathbb C}\Lambda^\bullet V_{\mathbb C}
    =
    1+4+6+4+1
    =
    16.
\end{equation}
This is also the complex dimension of the four-dimensional Dirac Clifford algebra.

\subsection{Hodge dual operator}

The exterior algebra is closed under the spacetime Hodge operator. In four dimensions, the Hodge map exchanges degrees zero and four, exchanges degrees one and three, and preserves degree two:
\begin{equation}
    \ast:
    \Lambda^p V_{\mathbb C}^{*}
    \longrightarrow
    \Lambda^{4-p}V_{\mathbb C}^{*}.
\end{equation}
At degree two, it separates the $(1,0)$ and $(0,1)$ chiral eigenspaces. In the Clifford realization, this operation can equivalently be expressed, up to convention-dependent signs and factors of $i$, through multiplication by the chirality or Clifford-volume element. The ordinary differential-form Hodge operator will nevertheless be retained in what follows in order to keep the geometric meaning of the construction transparent.

\subsection{Covariance}
\label{sec:covariance}

As emphasized by Weyl, a fundamental property of the coframe is that it transforms in the Lorentz vector representation $(\tfrac12,\tfrac12)$. Infinitesimally, we adopt the convention
\begin{equation}
    \delta_{\lambda} e^c
    =
    -\frac{1}{2}\lambda^{ab}
    \bigl(J^{(1)}_{ab}\bigr)^c{}_{d}\,e^d ,
    \qquad
    \lambda^{ab}=-\lambda^{ba},
\end{equation}
where the Lorentz generators in the vector representation are
\begin{equation}
    \bigl(J^{(1)}_{ab}\bigr)^c{}_{d}
    =
    \delta_a^c\eta_{bd}
    -
    \delta_b^c\eta_{ad}.
\end{equation}
Consequently,
\begin{equation}
    \bigl(J^{(1)}_{ab}\bigr)^c{}_{d}e^d
    =
    \delta_a^c e_b-\delta_b^c e_a,
\end{equation}
and hence
\begin{equation}
    \delta_{\lambda}e^c
    =
    -\lambda^c{}_{d}e^d.
\end{equation}

At the level of chiral spinor representations, the vector representation is obtained from the left- and right-handed Weyl representations. Schematically,
\begin{equation}
    J^{(1)}_{ab}
    =
    \Sigma_{ab}\otimes\mathbf 1
    -
    \mathbf 1\otimes\overline{\Sigma}_{ab}
    \in
    \left(\frac12,\frac12\right).
\end{equation}
Here the second factor is understood to act in the dual representation. Depending on whether the dotted index is written upstairs or downstairs, this dual action may equivalently appear as a sum after raising or lowering the index with the spinor symplectic form.

On the Dirac space $S_D$, defined in \eqref{DiracSpaceDirectSum}, in the chiral basis, define
\begin{equation}
    \gamma_a
    =
    \begin{pmatrix}
        0 & \sigma_a\\
        \overline{\sigma}_a & 0
    \end{pmatrix},
    \qquad
    J_{ab}
    =
    \Sigma_{ab}\oplus\overline{\Sigma}_{ab}
    =
    \begin{pmatrix}
        \Sigma_{ab} & 0\\
        0 & \overline{\Sigma}_{ab}
    \end{pmatrix},
\end{equation}

The compatibility between the vector and Dirac representations is encoded in the intertwining identity
\begin{equation}
    [J_{ab},\gamma_d]
    =
    \bigl(J^{(1)}_{ab}\bigr)^c{}_{d}\gamma_c.
\end{equation}
Introducing the Dirac-representation Lorentz parameter
\begin{equation}
    \lambda
    :=
    \frac{1}{2}\lambda^{ab}J_{ab},
\end{equation}
we therefore find
\begin{align}
    \delta_{\lambda}\slashed e
    &=
    (\delta_{\lambda}e^c)\gamma_c
    \nonumber\\
    &=
    -\frac{1}{2}\lambda^{ab}e^d
    \bigl(J^{(1)}_{ab}\bigr)^c{}_{d}\gamma_c
    \nonumber\\
    &=
    -\frac{1}{2}\lambda^{ab}e^d
    [J_{ab},\gamma_d]
    \nonumber\\
    &=
    [\slashed e,\lambda]. \label{lorentzcompatibility}
\end{align}
We will refer to this relation to gauge covariance-compatibility identity.

\emph{Thus the Lorentz transformation of the coframe and the adjoint transformation of the Clifford-valued coframe are precisely equivalent.}

With the finite transformation written as
\begin{equation}
    S=\exp(\lambda)
    =
    \operatorname{diag}(S_L,S_R),
\end{equation}
the Clifford-valued frame transforms homogeneously:
\begin{equation}
    \slashed e
    \longrightarrow
    S^{-1}\slashed e S.
\end{equation}
The coframe is therefore a covariant tensorial one-form, rather than a gauge connection.

Since $S$ is a zero-form, conjugation is compatible with the wedge
product. Every exterior power therefore transforms covariantly:
\begin{equation}
    \slashed e^{\wedge p}
    \longrightarrow
    S^{-1}\slashed e^{\wedge p}S,
\end{equation}
or, infinitesimally,
\begin{equation}
    \delta_{\lambda}
    \bigl(\slashed e^{\wedge p}\bigr)
    =
    [\slashed e^{\wedge p},\lambda]. \label{lorentzcompatibility2}
\end{equation}
This guarantees the covariance of forms of arbitrary degree in the
exterior algebra generated by $\slashed e$.

The even-degree forms make this covariance particularly transparent.
Since $\slashed e$ is off diagonal in the chiral decomposition,
\begin{equation}
    \slashed e^{\wedge 2k}
    \in
    \operatorname{End}(S_L)
    \oplus
    \operatorname{End}(S_R),
\end{equation}
whereas
\begin{equation}
    \slashed e^{\wedge(2k+1)}
    \in
    \operatorname{Hom}(S_R,S_L)
    \oplus
    \operatorname{Hom}(S_L,S_R).
\end{equation}
Thus an even exterior power has the block-diagonal form
\begin{equation}
    \slashed e^{\wedge 2k}
    =
    \begin{pmatrix}
        E_L^{(2k)} & 0\\
        0 & E_R^{(2k)}
    \end{pmatrix},
\end{equation}
and its covariance can be checked independently in the two chiral sectors:
\begin{equation}
    E_L^{(2k)}
    \longrightarrow
    S_L^{-1} E_L^{(2k)}S_L,
    \qquad
    E_R^{(2k)}
    \longrightarrow
    S_R^{-1} E_R^{(2k)}S_R.
\end{equation}

As a simple illustration of this covariance, the covariant exterior derivative of the Clifford-valued coframe is precisely the Clifford-algebra representative of the torsion:
\begin{equation}
    D_{\omega}\slashed e
    =
    d\slashed e+[\boldsymbol{\omega},\slashed e]_{\mathrm{gr}}
    =
    (D_{\omega}e^a)\gamma_a
    =
    T^a\gamma_a,
\end{equation}
where
\begin{equation}
    \boldsymbol{\omega}
    =
    \frac{1}{2}\omega^{ab}J_{ab}
\end{equation}
and the bracket is the graded commutator of differential forms.

In bispinor notation of \eqref{solderingoneforms} the equivalent relation becomes
\begin{equation}
    D e^{\alpha\dot\alpha}
    =
    d e^{\alpha\dot\alpha}
    +
    (\omega_L)^\alpha{}_{\beta}
    \wedge e^{\beta\dot\alpha}
    +
    (\omega_R)^{\dot\alpha}{}_{\dot\beta}
    \wedge e^{\alpha\dot\beta} =: T^{\alpha\dot\alpha}.\label{Talphaalphadot}
\end{equation}
This displays explicitly the complementary roles of the two chiral spin connections: the left connection acts on the undotted index, the right connection acts on the dotted index, and the coframe intertwines the two chiral representations.

All these covariance statements concern the bosonic chiral Lorentz subgroup. A transformation law under the odd orthosymplectic generators would require an extension of the field content and additional transformation rules.

The exact compatibility identity \eqref{lorentzcompatibility} (or more generally \eqref{lorentzcompatibility2}) is special to Lorentz transformations. In contrast, one should not expect the supersymmetry transformations to satisfy an analogous intertwining identity within the vector sector alone, but instead a compatibility condition may emerge in an enlarge setting. Indeed, the graded commutator of a supercharge with a Clifford generator is fermionic and, in general, does not belong to the linear span of the matrices \(\gamma_a\):
\begin{equation}
    [Q_\alpha,\gamma_a]_{\mathrm{gr}}
    \notin \operatorname{span}\{\gamma_b\}.
\end{equation}
Consequently, there is no transformation acting only on the vector index of \(e^a\) that can reproduce the adjoint supersymmetry transformation of \(\slashed e\). The precise form of this obstruction, and the enlarged representation required to describe it, can be seen explicitly in Appendix~\ref{app:osp-six-dimensional-realization}.

This obstruction, however, does not preclude supersymmetry from being realized as a symmetry of a particular background. Such a symmetry is contingent rather than structural: it arises only when the background fields satisfy the conditions required for their supersymmetry variations to vanish, typically through the existence of an appropriate Killing or twistor spinor \cite{deMedeiros:2012sb,Festuccia:2011ws}. This is analogous to the translational invariance of Minkowski spacetime. Although translations are not isometries of a generic solution of General Relativity, they emerge as symmetries of the Minkowski background because of its particular geometry. Likewise, a bosonic background may preserve a subset of the supersymmetry transformations even though no supersymmetric intertwining identity exists within the vector sector at the level of the general representation. See section \ref{sec:matter-ansatz-equivariance} for more details.

\section{Invariant terms in the orthosymplectic algebra}\label{sec:actionpreliminaries}

The even part of the exterior algebra can be paired directly with the block-diagonal gauge curvature. In particular, $\slashed e^{\,2}$ contains the chiral Plebański two-forms, while $F=F_L+F_R$ contains the corresponding chiral curvatures. Their invariant product produces the gravitational terms. Similarly, $\slashed e^{\,4}$ produces the volume contribution, and $F^2$ gives the curvature-squared Chern--Weil structure. The natural four-forms are consequently
\begin{equation}
    \left\langle\slashed e^{\,4}\right\rangle,
    \qquad
    \left\langle\slashed e^{\,2} \wedge F\right\rangle,
    \qquad
    \left\langle F\wedge F \right\rangle. \label{invforms}
\end{equation}

The invariant forms \eqref{invforms} may be restricted to either chiral sector or evaluated on the full left–right connection introduced in Sec. \ref{sec:gaugefields}.
The same Clifford-valued exterior algebra therefore provides a common geometric framework for the left-chiral, right-chiral, and left--right symmetric constructions.

\subsection{Lifting of $\slashed e$}\label{liftingofse}

The gauge superconnections introduced above contain the chiral spin connections and the fermionic one-forms, but they do not contain the frame. We introduce the latter independently as a Clifford-valued soldering form,
\begin{equation}
\slashed e = e^a\Gamma_{a}^{LR} + e^a \Gamma_{a}^{RL},
\end{equation}
where
\begin{equation}
\Gamma_{a}^{LR}
=
\left(
\begin{array}{c|c}
0_{3\times 3}
&
\begin{matrix}
(\sigma_{a})^{\alpha\dot\alpha} & 0\\
0 & 0
\end{matrix}
\\ \hline
0_{3\times 3}
&
0_{3\times 3}
\end{array}
\right),
\qquad
\Gamma_{a}^{RL}
=
\left(
\begin{array}{c|c}
0_{3\times 3}
&
0_{3\times 3}
\\ \hline
\begin{matrix}
(\bar\sigma_{a})_{\dot\alpha\alpha} & 0\\
0 & 0
\end{matrix}
&
0_{3\times 3}
\end{array}
\right).
\end{equation}

The block off-diagonal matrices $\Gamma_{a}^{LR}$ and $\Gamma_{a}^{RL}$ can be seen as intertwiners between the two chiral Weyl-spinor spaces. The first maps the right-handed spinor block into the left-handed block, while the second implements the reverse map. Their nonzero components are the Pauli matrices $\sigma_{a}$ and $\bar\sigma_{a}$. They act trivially on the one-dimensional odd directions of the two orthosymplectic representations.

The Clifford matrices satisfy
\begin{equation}
\Gamma_{a}^{LR}\Gamma_{b}^{LR}
=
\Gamma_{a}^{RL}\Gamma_{b}^{RL}
=
0,
\end{equation}
together with
\begin{align}
\Gamma_{a}^{LR}\Gamma_{b}^{RL} &= -\eta_{ab}\operatorname{diag}(1,1,0;0,0,0) - 2J_{ab}^{L},\\
\Gamma_{a}^{RL}\Gamma_{b}^{LR} &= -\eta_{ab}\operatorname{diag}(0,0,0;1,1,0) - 2J_{ab}^{R}.
\end{align}

These relations are the block-matrix realization of the Clifford algebra. Products of two intertwiners with the same orientation vanish because the image of the first map does not belong to the domain of the second. Alternating products return to the original chiral subspace and decompose into an identity contribution and a chiral Lorentz generator.

In full generality,
\begin{align}\label{sese-coframe}
\slashed e\wedge\slashed e
&=
-2 e^{a}\wedge e^{b} (J_{ab}^{L} + J_{ab}^{R}) \nonumber \\
&= - 4e^{0}\wedge e^{i} (J_{0i}^{L} + J_{0i}^{R}) - 2e^{i}\wedge e^{j} (J_{ij}^{L}+J_{ij}^{R}).
\end{align}
Thus, $\slashed e\wedge\slashed e$ is a bivector-valued two-form belonging entirely to the block-diagonal chiral Lorentz sector. Equivalently, in terms of \eqref{PlebanskiTwoForm},
\begin{equation}
e^{0}\wedge e^{i} = -\frac{i}{2}(B^{i}-\bar{B}^{i}),
\qquad 
e^{i}\wedge e^{j} = \frac{1}{2}\epsilon^{ij}{}_{k} (B^{k}+\bar{B}^{k}).
\end{equation}
The forms $B^{i}$ and $\bar{B}^{i}$ are the self-dual and anti-self-dual projections of $e^{a}\wedge e^{b}$, according to the duality conventions adopted here. Because they are constructed from a single frame, their simplicity properties are automatic and do not need to be imposed by independent Lagrange multipliers.

Substituting these expressions into the exterior square of the frame gives
\begin{equation}
\slashed e\wedge\slashed e = 2i(B^{i}-\bar{B}^{i}) (J_{0i}^{L}+J_{0i}^{R}) - \epsilon^{ij}{}_{k} (B^{k}+\bar{B}^{k}) (J_{ij}^{L}+J_{ij}^{R}).
\end{equation}

This equation makes explicit the relation between the Clifford-valued exterior algebra and the chiral decomposition of gravity. Although the fundamental frame is an off-diagonal intertwiner between the two Weyl-spinor representations, its exterior square produces precisely the block-diagonal anti-self-dual and self-dual two-forms required in the Plebański formulation.

The internal duality properties of the chiral Lorentz generators imply
\begin{align}
\epsilon^{ij}{}_{k}J_{ij}^{L}
&=
-2iJ_{0k}^{L},
&
\epsilon^{ij}{}_{k}J_{ij}^{R}
&=
+2iJ_{0k}^{R}.
\end{align}
Consequently, the exterior square separates into its two chiral
components:
\begin{equation}\label{sesePlebanski}
\se\wedge\se = 4iB^{i}J_{0i}^{L} - 4i\bar{B}^{i}J_{0i}^{R}.
\end{equation}
Thus, the anti-self-dual Plebanski form \(B^{i}\) belongs to the
left-handed Lorentz sector, while the self-dual form \(\bar{B}^{i}\)
belongs to the right-handed sector.

\subsection{ Computation of the $\langle \se^4 \rangle$ term}

Using \eqref{sese} and the fact that the two chiral blocks have vanishing mixed products, $J_{ab}^{L}J_{cd}^{R}=0$, one obtains
\begin{equation}
\se^{4}
=
-16B^{i}\wedge B^{j}J_{0i}^{L}J_{0j}^{L}
-
16\bar{B}^{i}\wedge\bar{B}^{j}
J_{0i}^{R}J_{0j}^{R}.
\end{equation}
The invariant tensors satisfy
\begin{equation}
\left\langle J_{0i}^{L}J_{0j}^{L}\right\rangle
=
\left\langle J_{0i}^{R}J_{0j}^{R}\right\rangle
=
\frac{1}{2}\delta_{ij},
\end{equation}
and therefore
\begin{align}
\left\langle\se^{4}\right\rangle_{L}
&=
-8B^{i}\wedge B_{i},
\\
\left\langle\se^{4}\right\rangle_{R}
&=
-8\bar{B}^{i}\wedge\bar{B}_{i}.
\end{align}

The Plebanski two-forms obey the simplicity relations
\begin{align}
B^{i}\wedge B^{j}
&=
2i\delta^{ij}\,\mathrm{vol}_{e},
\\
\bar{B}^{i}\wedge\bar{B}^{j}
&=
-2i\delta^{ij}\,\mathrm{vol}_{e},
\\
B^{i}\wedge\bar{B}^{j}
&=
0,
\end{align}
where
\begin{equation}
\mathrm{vol}_{e}
=
e^{0}\wedge e^{1}\wedge e^{2}\wedge e^{3}
=
\frac{1}{4!}\epsilon_{abcd} \ e^{a}\wedge e^{b}\wedge e^{c}\wedge e^{d}.
\end{equation}
Consequently,
\begin{align}
\left\langle\se^{4}\right\rangle_{L}
&=
-48i\,\mathrm{vol}_{e}
=
-2i\epsilon_{abcd} \ e^{a}\wedge e^{b}\wedge e^{c}\wedge e^{d},
\\
\left\langle\se^{4}\right\rangle_{R}
&=
+48i\,\mathrm{vol}_{e}
=
+2i\epsilon_{abcd}\ e^{a}\wedge e^{b}\wedge e^{c}\wedge e^{d}.
\end{align}
Thus, for the invariant on the full direct sum,
\begin{equation}
\left\langle\se^{4}\right\rangle  = -8\left( B^{i}\wedge B_{i} + \bar{B}^{i}\wedge\bar{B}_{i} \right) = 0.
\end{equation}
This result may seem odd, but let us recall that the coframe is real, the two chiral Plebanski forms are complex conjugates, $\bar{B}^{i}=\left(B^{i}\right)^{*}$. Indeed,
writing
\begin{equation}
B^{i}=S^{i}+iE^{i},
\qquad
\bar{B}^{i}=S^{i}-iE^{i},
\end{equation}
with
\begin{equation}
S^{i}
=
\frac{1}{2}\epsilon^{i}{}_{jk}\,
e^{j}\wedge e^{k},
\qquad
E^{i}=e^{0}\wedge e^{i},
\end{equation}
one has
\begin{equation}
S^{i}\wedge S^{j}=0,
\qquad
E^{i}\wedge E^{j}=0,
\qquad
S^{i}\wedge E^{j}
=
\delta^{ij}\,\mathrm{vol}_{e}.
\end{equation}

Thus \(B^{i}\wedge B_{i}\) is purely imaginary and
\begin{equation}
\left(B^{i}\wedge B_{i}\right)^{*}
=
\bar{B}^{i}\wedge\bar{B}_{i}
=
-B^{i}\wedge B_{i}.
\end{equation}
whereas their difference gives the volume form,
\begin{equation}
\mathrm{vol}_{e}
=
\frac{1}{12i}
\left(
B^{i}\wedge B_{i}
-
\bar{B}^{i}\wedge\bar{B}_{i}
\right).
\end{equation}

\subsection{Computation of the $\langle \se^2 \wedge F \rangle$ term}

Using \eqref{sesePlebanski} and noticing that the fermionic components do not contribute, since $\left\langle J_{ab}^{L}Q_{\alpha}\right\rangle
=
\left\langle J_{ab}^{R}\bar{Q}^{\dot{\alpha}}\right\rangle
=
0.$
The invariant tensors, together with the anti-self-duality and self-duality conditions,
\begin{align}\label{selfdualconditionsF}
\frac{1}{2}\epsilon_{abcd}\mathcal{F}_{L}^{cd}
&=
-i\mathcal{F}_{L\,ab},
&
\frac{1}{2}\epsilon_{abcd}\mathcal{F}_{R}^{cd}
&=
+i\mathcal{F}_{R\,ab},
\end{align}
give
\begin{align}
\left\langle
J_{0i}^{L}\,
\frac{1}{2}\mathcal{F}_{L}^{ab}J_{ab}^{L}
\right\rangle
&=
-\mathcal{F}_{L\,0i},
\\
\left\langle
J_{0i}^{R}\,
\frac{1}{2}\mathcal{F}_{R}^{ab}J_{ab}^{R}
\right\rangle
&=
-\mathcal{F}_{R\,0i}.
\end{align}
Therefore,
\begin{equation}
\left\langle\se^{2} \wedge F\right\rangle
=
4i\left(
\bar{B}^{i}\wedge\mathcal{F}_{R\,0i}
-
B^{i}\wedge\mathcal{F}_{L\,0i}
\right).
\end{equation}

Equivalently, starting directly from the first line in \eqref{sese-coframe} one obtains
\begin{align}
\left\langle\se^{2} \wedge F\right\rangle
={}&
e^{a}\wedge e^{b}\wedge\mathcal{F}_{L\,ab}
+
\frac{i}{2}\epsilon_{abcd}\,
e^{a}\wedge e^{b}\wedge\mathcal{F}_{L}^{cd}
\nonumber\\
&+
e^{a}\wedge e^{b}\wedge\mathcal{F}_{R\,ab}
-
\frac{i}{2}\epsilon_{abcd}\,
e^{a}\wedge e^{b}\wedge\mathcal{F}_{R}^{cd}.
\end{align}
Using the chiral duality properties of the curvature, this reduces to
\begin{equation}
\left\langle\se^{2} \wedge F\right\rangle
=
2e^{a}\wedge e^{b}\wedge
\left(
\mathcal{F}_{L\,ab}
+
\mathcal{F}_{R\,ab}
\right).
\end{equation}
Hence,
\begin{equation}
4i\left(
\bar{B}^{i}\wedge\mathcal{F}_{R\,0i}
-
B^{i}\wedge\mathcal{F}_{L\,0i}
\right)
=
2e^{a}\wedge e^{b}\wedge
\left(
\mathcal{F}_{L\,ab}
+
\mathcal{F}_{R\,ab}
\right).
\end{equation}

Introducing the independent chiral components of the curvature,
\begin{equation}
\mathcal{F}_{i}:=\mathcal{F}_{L\,0i},
\qquad
\bar{\mathcal{F}}_{i}:=\mathcal{F}_{R\,0i},
\end{equation}
the anti-self-duality and self-duality conditions determine the
remaining Lorentz components,
\begin{align}
\mathcal{F}_{L\,ij}
&=
i\,\epsilon_{ij}{}^{k}\mathcal{F}_{k},
\\
\mathcal{F}_{R\,ij}
&=
-i\,\epsilon_{ij}{}^{k}\bar{\mathcal{F}}_{k}.
\end{align}
The invariant can then be written entirely in terms of the chiral Plebanski two-forms and the corresponding chiral curvature
components:
\begin{equation}
\left\langle\se^{2} \wedge F \right\rangle
=
4i\left(
\bar{B}^{i}\wedge\bar{\mathcal{F}}_{i}
-
B^{i}\wedge\mathcal{F}_{i}
\right).
\end{equation}
For a real Lorentz connection the two sectors are related by complex conjugation, $\bar{B}^{i}=(B^{i})^{*}$, $\bar{\mathcal{F}}_{i}=(\mathcal{F}_{i})^{*}$, so that the complete invariant is real.

\subsection{Computation of the $\langle F\wedge F \rangle$ term}

Since the two chiral algebras form a direct sum, all mixed
left--right contractions vanish. Furthermore,
\(\langle J_{ab}Q_{\alpha}\rangle=0\), and therefore the invariant
quadratic form becomes
\begin{align}
\left\langle F\wedge F\right\rangle
={}&
-\frac{1}{4}
\mathcal{F}_{L}^{ab}\wedge\mathcal{F}_{L\,ab}
-\frac{i}{8}\epsilon_{abcd}\,
\mathcal{F}_{L}^{ab}\wedge\mathcal{F}_{L}^{cd}
-2\epsilon_{\alpha\beta}\,
\rho_{L}^{\alpha}\wedge\rho_{L}^{\beta}
\nonumber\\
&-
\frac{1}{4}
\mathcal{F}_{R}^{ab}\wedge\mathcal{F}_{R\,ab}
+\frac{i}{8}\epsilon_{abcd}\,
\mathcal{F}_{R}^{ab}\wedge\mathcal{F}_{R}^{cd}
+2\epsilon^{\dot{\alpha}\dot{\beta}}\,
\rho_{R\dot{\alpha}}\wedge\rho_{R\dot{\beta}}.
\end{align}
where $\rho_{L}^{\alpha}$ and $\rho_{R\dot{\alpha}}$ were defined in \eqref{rhoalpha-definition} and \eqref{rhoalphadot-definition}.
Using the anti-self-duality and self-duality conditions \eqref{selfdualconditionsF}, this reduces to
\begin{align}
\left\langle F\wedge F\right\rangle
={}&
-\frac{1}{2}
\mathcal{F}_{L}^{ab}\wedge\mathcal{F}_{L\,ab}
-\frac{1}{2}
\mathcal{F}_{R}^{ab}\wedge\mathcal{F}_{R\,ab}
\nonumber\\
&-
2\epsilon_{\alpha\beta}\,
\rho_{L}^{\alpha}\wedge\rho_{L}^{\beta}
+
2\epsilon^{\dot{\alpha}\dot{\beta}}\,
\rho_{R\dot{\alpha}}\wedge\rho_{R\dot{\beta}}.
\end{align}
In terms of the independent chiral curvature components it follows that
\begin{align}
\mathcal{F}_{L}^{ab}\wedge\mathcal{F}_{L\,ab}
&=
-4\mathcal{F}^{i}\wedge\mathcal{F}_{i},
&
\mathcal{F}_{R}^{ab}\wedge\mathcal{F}_{R\,ab}
&=
-4\bar{\mathcal{F}}^{\,i}\wedge\bar{\mathcal{F}}_{i}.
\end{align}
Consequently,
\begin{equation}
\left\langle F\wedge F\right\rangle
=
2\mathcal{F}^{i}\wedge\mathcal{F}_{i}
+
2\bar{\mathcal{F}}^{\,i}\wedge\bar{\mathcal{F}}_{i}
-
2\epsilon_{\alpha\beta}\rho_{L}^{\alpha}\wedge\rho_{L}^{\beta}
+
2\epsilon^{\dot{\alpha}\dot{\beta}}
\rho_{R\dot{\alpha}}\wedge\rho_{R\dot{\beta}}.
\end{equation}

\section{The action}\label{sec:action}

We introduce three independent coefficients and consider the action
\begin{equation}
S
=
\int_{M}
\left[
\lambda_{4}\left\langle\se^{4}\right\rangle
+
\lambda_{2}\left\langle\se^{2}F\right\rangle
+
\lambda_{0}\left\langle F \, F \right\rangle
\right].
\end{equation}
The coefficients \(\lambda_{4}\), \(\lambda_{2}\), and \(\lambda_{0}\) multiply, respectively, the cosmological or Plebanski \(B^{2}\) term, the \(B F\) term, and the quadratic Chern--Weil term.

For the purely left-handed model, \(A_{R}=0\), restriction to the left-handed invariant gives
\begin{equation}
S_{L}
=
\int_{M}
\left[
-8\lambda_{4}B^{i}\wedge B_{i}
-4i\lambda_{2}B^{i}\wedge\mathcal{F}_{i}
+2\lambda_{0}\mathcal{F}^{i}\wedge\mathcal{F}_{i}
-2\lambda_{0}\epsilon_{\alpha\beta}
\rho_{L}^{\alpha}\wedge\rho_{L}^{\beta}
\right].
\end{equation}
The first term is $-8\lambda_{4}B^{i}\wedge B_{i} = -48i\lambda_{4}\,\mathrm{vol}_{e}$.

For comparison with chiral Plebanski gravity, we consider the
bosonic truncation and choose the normalization
\begin{equation}
S_{\mathrm{Pl},L}
=
-\frac{1}{8\pi G}
\int_M
\left[
B^i\wedge\mathcal{F}_i
-\frac{i\Lambda}{12}
B^i\wedge B_i
\right].
\end{equation}
Comparison with the action above gives
\begin{equation}
\lambda_2=-\frac{i}{32\pi G},
\qquad
\lambda_4=-\frac{i\Lambda}{768\pi G},
\qquad
\lambda_0=0,
\end{equation}
and, in particular,
\begin{equation}
\lambda_4=\frac{\Lambda}{24}\lambda_2.
\end{equation}
Since in the present construction the Plebanski two-forms are constructed directly from the coframe, their simplicity conditions are already solved. Thus this action corresponds to the reduced chiral Plebanski, or self-dual Einstein, formulation. The coefficient \(\lambda_0\) may alternatively be kept arbitrary, since the \(\langle F \, F\rangle_L\) contribution is a Chern--Weil term and does not modify the local bulk field equations.

For the purely right-handed model, \(A_{L}=0\), one obtains
\begin{equation}
S_{R}
=
\int_{M}
\left[
-8\lambda_{4}\bar{B}^{i}\wedge\bar{B}_{i}
+4i\lambda_{2}\bar{B}^{i}\wedge\bar{\mathcal{F}}_{i}
+2\lambda_{0}\bar{\mathcal{F}}^{\,i}
\wedge\bar{\mathcal{F}}_{i}
+2\lambda_{0}\epsilon^{\dot{\alpha}\dot{\beta}}
\rho_{R\dot{\alpha}}\wedge\rho_{R\dot{\beta}}
\right].
\end{equation}
The first term is now $-8\lambda_{4}\bar{B}^{i}\wedge\bar{B}_{i} = +48i\lambda_{4}\,\mathrm{vol}_{e}$.

For equal coefficients in the two chiral sectors, the \(\lambda_{4}\) contribution cancels identically and the full left--right symmetric action is
\begin{align}
S_{LR}
=
\int_{M}\Big[
&
4i\lambda_{2}
\left(
\bar{B}^{i}\wedge\bar{\mathcal{F}}_{i}
-
B^{i}\wedge\mathcal{F}_{i}
\right)
\nonumber\\
&+
2\lambda_{0}
\left(
\mathcal{F}^{i}\wedge\mathcal{F}_{i}
+
\bar{\mathcal{F}}^{\,i}\wedge\bar{\mathcal{F}}_{i}
\right)
\nonumber\\
&-
2\lambda_{0}\epsilon_{\alpha\beta}
\rho_{L}^{\alpha}\wedge\rho_{L}^{\beta}
+
2\lambda_{0}\epsilon^{\dot{\alpha}\dot{\beta}}
\rho_{R\dot{\alpha}}\wedge\rho_{R\dot{\beta}}
\Big].
\end{align}
Although this action is symmetric under the bare exchange \(L\leftrightarrow R\), it is not parity invariant, since parity also reverses spatial orientation and therefore selects the opposite-weight chiral combination as the parity-even one. Indeed,
\begin{equation} \label{HolstContraction}
\left\langle\se^{2} F\right\rangle_{L}
+
\left\langle\se^{2} F\right\rangle_{R}
=
2e^{a}\wedge e^{b}\wedge R_{ab},
\end{equation}
which is the Holst-type contraction, while
\begin{equation} \label{equalweightCC}
\left\langle\se^{4}\right\rangle_{L}
+
\left\langle\se^{4}\right\rangle_{R}
=
0.
\end{equation}
The Einstein--Hilbert and cosmological terms are instead selected by the difference of the two chiral invariant forms. With the conventions adopted here,
\begin{align}
\left\langle\se^{2} F\right\rangle_{L}
-
\left\langle\se^{2} F\right\rangle_{R}
&=
i\epsilon_{abcd}\,
e^{a}\wedge e^{b}\wedge R^{cd},\label{EHContraction}
\\
\left\langle\se^{4}\right\rangle_{L}
-
\left\langle\se^{4}\right\rangle_{R}
&=
-4i\epsilon_{abcd}\,
e^{a}\wedge e^{b}\wedge e^{c}\wedge e^{d}.\label{oppositeweightCC}
\end{align}
Consequently, the Einstein--Hilbert action with cosmological constant,
\begin{equation}
S_{\mathrm{EH}+\Lambda}
=
\frac{1}{32\pi G}
\int
\epsilon_{abcd}
\left(
e^{a}\wedge e^{b}\wedge R^{cd}
-
\frac{\Lambda}{6}
e^{a}\wedge e^{b}\wedge e^{c}\wedge e^{d}
\right),
\end{equation}
is reproduced by
\begin{equation}
S_{\mathrm{EH}+\Lambda}
=
-\frac{i}{32\pi G}
\int
\left[
\left(
\left\langle\se^{2} F\right\rangle_{L}
-
\left\langle\se^{2} F\right\rangle_{R}
\right)
+
\frac{\Lambda}{24}
\left(
\left\langle\se^{4}\right\rangle_{L}
-
\left\langle\se^{4}\right\rangle_{R}
\right)
\right].
\end{equation}
Equivalently, the required coefficients satisfy
\begin{align}
\lambda_{2}^{L}
&=
-\frac{i}{32\pi G},
&
\lambda_{2}^{R}
&=
+\frac{i}{32\pi G},
\\
\lambda_{4}^{L}
&=
-\frac{i\Lambda}{768\pi G},
&
\lambda_{4}^{R}
&=
+\frac{i\Lambda}{768\pi G},
\end{align}
or, in arbitrary units,
\begin{equation}
\lambda_{2}^{R}=-\lambda_{2}^{L},
\qquad
\lambda_{4}^{R}=-\lambda_{4}^{L},
\qquad
\frac{\lambda_{4}^{L}}{\lambda_{2}^{L}}
=
\frac{\Lambda}{24}.
\end{equation}

In summary, a bare interchange of the two chiral sectors differs from a parity transformation. The Einstein--Hilbert action is not obtained from the equal-weight combination of the left- and right-handed invariant forms. Rather, with the conventions adopted here, \eqref{HolstContraction} is the Holst-type contraction, whereas the \eqref{EHContraction} is the Einstein--Cartan term. Similarly, the equal-weight sum of the two chiral contributions vanishes \eqref{equalweightCC}, while the opposite-weight sum of the two chiral contributions provides a nonvanishing cosmological constant term \eqref{oppositeweightCC}. More generally, the two independent combinations may be organized as
\begin{equation}
I_{L}+I_{R} \sim e^a e^b R_{ab},
\qquad
i\left(I_{L}-I_{R}\right) \sim i e^a e^b \star(R_{ab}),
\end{equation}
corresponding respectively to the Holst and Einstein sectors. The Einstein--Cartan--Holst action therefore assigns the chiral coefficients
\begin{equation}
\lambda_{2}^{L}
=
\frac{1}{32\pi G}
\left(\frac{1}{\gamma}-i\right),
\qquad
\lambda_{2}^{R}
=
\frac{1}{32\pi G}
\left(\frac{1}{\gamma}+i\right),
\end{equation}
where the pure Einstein limit is obtained for
\(\gamma^{-1}=0\).

Thus, although the Einstein--Hilbert plus cosmological-constant action is antisymmetric under the bare algebraic exchange $L \leftrightarrow R$, it is parity invariant. Parity exchanges the two chiral Lorentz representations and simultaneously reverses spatial orientation. The relative sign between the two chiral invariant forms is therefore precisely what reconstructs the Levi--Civita contraction of the parity-even Einstein action. In Dirac language, this difference is equivalent to the insertion of the chirality operator,
\begin{equation}
\operatorname{Tr}_{L}-\operatorname{Tr}_{R}
\sim
\operatorname{Tr}(\gamma_{5},\cdots),
\end{equation}
which generates the Levi--Civita tensor. This observation closely parallels the MacDowell--Mansouri formulation of gravity \cite{MacDowell:1977jt}, where the Einstein--Hilbert and cosmological terms arise from a curvature-squared action after selecting the Lorentz invariant defined by the Levi--Civita tensor, or equivalently, in a Dirac-matrix representation, by an insertion of $\gamma_{5}$. In that construction one schematically has
\begin{equation}
S_{\rm MM}
\sim
\int
\operatorname{Tr}(
\gamma_{5},
\mathcal{F}\wedge\mathcal{F}
),
\end{equation}
and, after decomposing the enlarged (A)dS curvature into its Lorentz-curvature and coframe components, the cross term produces the Einstein--Hilbert action while the purely coframe contribution produces the cosmological term, together with a topological curvature-squared term.

The similarity is particularly transparent in the chiral basis: the insertion of $\gamma_{5}$ assigns opposite signs to the two Weyl blocks and therefore implements precisely the difference between the left- and right-handed invariant pairings. There is, however, an important geometric distinction. In the MacDowell--Mansouri construction the coframe appears as part of an enlarged (A)dS connection, whereas in the present formulation it remains an independent Clifford-valued soldering form. Consequently, the same $\gamma_{5}$-type chiral weighting emerges here directly from the decomposition
\begin{equation}
\se^{2}\in(1,0)\oplus(0,1),
\end{equation}
without embedding the coframe into the gauge connection.

The relation obtained above between the relative weights of the left- and right-handed sectors and the Einstein--Cartan--Holst action is closely related to the chiral decompositions used in the study of the Barbero--Immirzi parameter. In particular, Mercuri showed that the first-order action can be reorganized in terms of dynamically independent self-dual and anti-self-dual sectors carrying different weights determined by the Immirzi parameter \cite{Mercuri:2006um}. Related constructions by Chou, Tung, and Yu, and by Chagoya and Sabido, combine the self-dual and anti-self-dual curvature sectors within first-order gauge-theoretic formulations and relate their relative weights to the Immirzi ambiguity \cite{Chou:2005ht,Chagoya:2016zhy}. In the present formulation this structure follows directly from the representation content of the Clifford-valued soldering form: since $\se^{2}$ takes values in $(1,0)\oplus(0,1)$, the equal-weight combination of the two chiral invariant forms implements the identity pairing on Lorentz bivectors and produces the Holst contraction, whereas the opposite-weight combination implements the internal duality operator, equivalently the insertion of $\gamma_{5}$ in the Dirac representation, and reconstructs the Levi--Civita contraction of the Einstein--Hilbert term.

A closely related supergravity perspective was developed by Eder and Sahlmann in their Holst--MacDowell--Mansouri formulation of $\mathcal N=1,2$ AdS supergravity \cite{Eder:2021rgt}. There the Barbero--Immirzi dependence is incorporated through a deformed invariant pairing on superalgebra-valued forms, while in the chiral limit the corresponding operator becomes a genuine chiral projector and exposes an $\operatorname{OSp}(\mathcal N|2,\mathbb C)$ gauge symmetry. The present construction shares this organization into chiral orthosymplectic sectors, but differs geometrically in that the coframe is not incorporated into an enlarged Cartan superconnection. Instead, it remains an independent Clifford-valued soldering form whose exterior square generates the two chiral bivector sectors. The Einstein and Holst contractions are therefore obtained here by the two possible relative weightings of the same left--right chiral pairing, while the respective roles of the gauge connection and the soldering form remain manifestly distinct.

\subsection{Gravitational field equations}
\label{sec:gravitational-field-equations}

For completeness, we summarize the standard gravitational field equations of the Einstein--Cartan--Holst sector in the conventions used above \cite{Holst:1995pc,Immirzi:1996di}. Restricting to the bosonic sector and omitting the Chern--Weil term, which does not affect the local bulk equations, the action may be written as
\begin{equation}
S_{\mathrm{ECH}}
=
\frac{1}{32\pi G}
\int_M
\left[
\epsilon_{abcd}\,e^a\wedge e^b\wedge R^{cd}
+\frac{2}{\gamma}\,e^a\wedge e^b\wedge R_{ab}
-\frac{\Lambda}{6}\epsilon_{abcd}\,
e^a\wedge e^b\wedge e^c\wedge e^d
\right].
\label{eq:ECHaction}
\end{equation}
This is precisely the combination obtained from the chiral coefficients above. It is useful to introduce the linear map on Lorentz bivectors
\begin{equation}
P_{ab}{}^{cd}
:=
\frac{1}{2}\epsilon_{ab}{}^{cd}
+\frac{1}{\gamma}\delta^c_{[a}\delta^d_{b]}.
\label{eq:HolstPoperator}
\end{equation}
Variation with respect to the independent Lorentz connection, using $\delta R^{ab}=D\delta\omega^{ab}$, gives
\begin{equation}
D\left(P_{ab}{}^{cd}\,e^a\wedge e^b\right)=0.
\label{eq:connection-ECH}
\end{equation}
Equivalently, in terms of the torsion,
\begin{equation}
\epsilon_{abcd}\,e^c\wedge T^d
+\frac{2}{\gamma}\,e_{[a}\wedge T_{b]}
=0.
\label{eq:torsion-ECH}
\end{equation}
For Lorentzian signature, $\star^2=-1$ on internal bivectors, and therefore $P$ is invertible for $\gamma^2\neq -1$. For a nondegenerate coframe, Eq.~\eqref{eq:connection-ECH} then implies
\begin{equation}
T^a=De^a=0,
\end{equation}
so that the independent spin connection reduces to the Levi--Civita spin connection. At the special values $\gamma=\pm i$, one of the two chiral coefficients vanishes and $P$ becomes a chiral projector; the invertibility argument is then replaced by the usual self-dual or anti-self-dual formulation.

Variation of Eq.~\eqref{eq:ECHaction} with respect to the coframe gives
\begin{equation}
\epsilon_{abcd}\,e^b\wedge R^{cd}
+\frac{2}{\gamma}\,e^b\wedge R_{ab}
-\frac{\Lambda}{3}\epsilon_{abcd}\,
e^b\wedge e^c\wedge e^d
=0.
\label{eq:coframe-ECH}
\end{equation}
Once the connection equation has imposed vanishing torsion, the first Bianchi identity gives
\begin{equation}
R^a{}_{b}\wedge e^b=0,
\end{equation}
and hence the term proportional to $\gamma^{-1}$ in Eq.~\eqref{eq:coframe-ECH} vanishes. The remaining equation is
\begin{equation}
\epsilon_{abcd}\,e^b\wedge
\left(
R^{cd}-\frac{\Lambda}{3}e^c\wedge e^d
\right)=0,
\end{equation}
which is equivalent, for an invertible coframe, to the vacuum Einstein equations with cosmological constant. Thus the Immirzi parameter does not modify the local classical equations of pure torsionless gravity. In the same torsionless sector the Holst density itself vanishes,
\begin{equation}
e^a\wedge e^b\wedge R_{ab}=0.
\end{equation}

The latter statement should not be confused with the Holst density being a topological invariant off shell. The Nieh--Yan identity reads
\begin{equation}
d\left(e^a\wedge T_a\right)
=
T^a\wedge T_a
-
e^a\wedge e^b\wedge R_{ab},
\label{eq:NiehYanIdentity}
\end{equation}
so the Holst density differs from the exact Nieh--Yan form by a torsion-squared term. Consequently, when spinorial matter sources torsion, the connection equation is modified and the Immirzi parameter can enter the resulting effective fermionic interactions for standard minimal couplings \cite{Perez:2005pm}. For suitable nonminimal fermion couplings the Holst contribution can instead combine with the fermionic sector into the Nieh--Yan invariant, removing this dependence from the classical effective dynamics \cite{Mercuri:2006um}. Since the fermionic one-forms in the present construction arise from the orthosymplectic gauge connection rather than from a standard minimally coupled Dirac action, their effect on the torsion equation must be determined from the corresponding field equations rather than inferred directly from the usual Einstein--Cartan--Holst matter coupling.

The special cases $\gamma=\pm i$ can be analyzed particularly transparently in the present left--right formulation. With the Lorentzian duality conventions of Eq.~\eqref{starSigmas}. Since $P=\star+\gamma^{-1}$, one obtains
\begin{equation}
P\big|_{\gamma=+i}
=
\star-i
=
-2i\Pi_L,
\qquad
P\big|_{\gamma=-i}
=
\star+i
=
+2i\Pi_R,
\label{eq:HolstP-chiral-limits}
\end{equation}
where $\Pi_{L,R}$ are chiral projectors or \eqref{eq:chiral-projectors-Immirzi}. Thus $\gamma=+i$ retains the left $\star=-i$ sector, whereas $\gamma=-i$ retains the right $\star=+i$ sector. Equivalently, using the coefficients obtained above,
\begin{equation}
\gamma=+i:
\quad
\lambda_2^L=-\frac{i}{16\pi G},
\quad
\lambda_2^R=0,
\qquad
\gamma=-i:
\quad
\lambda_2^L=0,
\quad
\lambda_2^R=+\frac{i}{16\pi G}.
\label{eq:chiral-Immirzi-coefficients}
\end{equation}

Accordingly, the connection equation no longer determines both chiralities at once. In the chiral basis it reduces to
\begin{equation}
D_L B^i=0
\qquad (\gamma=+i),
\qquad
D_R\bar B^i=0
\qquad (\gamma=-i).
\label{eq:chiral-compatibility-Immirzi}
\end{equation}
Because $B^i$ and $\bar B^i$ are constructed directly from the same nondegenerate coframe, see \eqref{PlebanskiTwoForm}, their simplicity relations, \eqref{simplicityrelation}, are already solved (since they are identities in the current setup). The corresponding compatibility equation therefore fixes the active chiral connection uniquely as the appropriate projection of the Levi--Civita connection,
\begin{equation}
\omega_L=\Pi_L\omega_{\mathrm{LC}}(e)
\qquad (\gamma=+i),
\qquad
\omega_R=\Pi_R\omega_{\mathrm{LC}}(e)
\qquad (\gamma=-i).
\label{eq:chiral-Levi-Civita-connection}
\end{equation}
This is similar to the usual self-dual or anti-self-dual first-order formulation.

\subsection{Gravitational field equations with the fermionic one-form}
\label{sec:gravitational-field-equations-fermionic}

We now retain the odd components of the orthosymplectic connection and vary the fermionic one-forms as independent fields. No matter ansatz is imposed: the coframe $e^{a}$, the Lorentz connection, and the vector-spinors $\Psi_{L}^{\alpha}$ and $\Psi_{R\dot\alpha}$ are treated as independent variables. Define
\begin{align}
\Theta_{L}^{ab}
&:=
(\Sigma^{ab})_{\alpha\beta}\,
\Psi_{L}^{\alpha}\wedge\Psi_{L}^{\beta},
&
\Theta_{R}^{ab}
&:=
(\bar\Sigma^{ab})^{\dot\alpha\dot\beta}\,
\Psi_{R\dot\alpha}\wedge\Psi_{R\dot\beta},
\end{align}
so that
\begin{equation}
\mathcal F_{L}^{ab}=R_{L}^{ab}-\Theta_{L}^{ab},
\qquad
\mathcal F_{R}^{ab}=R_{R}^{ab}-\Theta_{R}^{ab}.
\end{equation}
It is convenient to combine the even curvatures as
\begin{equation}
\mathcal F^{ab}
:=
\mathcal F_{L}^{ab}+\mathcal F_{R}^{ab}
=
R^{ab}-\Theta^{ab},
\qquad
\Theta^{ab}:=\Theta_{L}^{ab}+\Theta_{R}^{ab}.
\label{eq:supercovariant-even-curvature}
\end{equation}
Thus the local Einstein--Cartan--Holst equations are obtained from the preceding bosonic equations by replacing $R^{ab}$ with $\mathcal F^{ab}$ wherever the even supercurvature enters.

The $\lambda_0$ sector requires a separate comment. The curvature-square and $\rho\wedge\rho$ contributions are not independent bulk terms but the component expansion of the single Chern--Weil invariant $\langle F\wedge F\rangle$. Indeed,
\begin{equation}
\delta\int_M\langle F\wedge F\rangle
=
2\int_M d\langle\delta A\wedge F\rangle,
\end{equation}
where the bulk contribution vanishes by the super-Bianchi identity $DF=0$. Hence, with the usual boundary conditions, the complete $\lambda_0$ sector produces no local field equation. In particular, the derivative-containing $\rho\wedge\rho$ term cannot by itself be interpreted as a kinetic term for $\Psi$.

Variation with respect to the coframe gives the modified gravitational equation
\begin{equation}
\epsilon_{abcd}\,e^{b}\wedge\mathcal F^{cd}
+\frac{2}{\gamma}\,e^{b}\wedge\mathcal F_{ab}
-\frac{\Lambda}{3}\epsilon_{abcd}\,
e^{b}\wedge e^{c}\wedge e^{d}
=0,
\label{eq:coframe-ECH-Psi}
\end{equation}
or, separating the ordinary Lorentz curvature,
\begin{align}
&\epsilon_{abcd}\,e^{b}\wedge R^{cd}
+\frac{2}{\gamma}\,e^{b}\wedge R_{ab}
-\frac{\Lambda}{3}\epsilon_{abcd}\,
e^{b}\wedge e^{c}\wedge e^{d}
\nonumber\\
&\hspace{2cm}=
\epsilon_{abcd}\,e^{b}\wedge\Theta^{cd}
+\frac{2}{\gamma}\,e^{b}\wedge\Theta_{ab}.
\label{eq:coframe-ECH-Psi-source}
\end{align}
Thus the unrestricted vector-spinor contributes algebraically to the coframe equation through the bilinear $\Theta^{ab}$.

The Lorentz-connection equation is unchanged. Since $\Theta^{ab}$ contains no Lorentz connection when $\Psi$ is treated as an independent one-form, and the $\lambda_0$ sector is Chern--Weil, variation with respect to $\omega^{ab}$ gives precisely Eq.~\eqref{eq:connection-ECH}, or equivalently Eq.~\eqref{eq:torsion-ECH}. Therefore, for a nondegenerate coframe and generic Lorentzian $\gamma$, the present fermionic one-form does not source torsion and one again obtains $T^a=0$. At $\gamma=\pm i$ the connection equation instead reduces to the corresponding chiral compatibility equation in Eq.~\eqref{eq:chiral-compatibility-Immirzi}.

The strongest restriction follows from varying the fermionic one-forms themselves. Since the $\lambda_4$ term is independent of $\Psi$ and the $\lambda_0$ term is topological, the only local fermionic equation comes from $\langle\se^2F\rangle$. Using
\begin{equation}
\delta\Theta_L^{ab}
=
2(\Sigma^{ab})_{\alpha\beta}\,
\delta\Psi_L^{\alpha}\wedge\Psi_L^{\beta},
\end{equation}
with the analogous right-handed expression, one obtains
\begin{align}
\lambda_2^L\,
e^a\wedge e^b\wedge
(\Sigma_{ab})_{\alpha\beta}\Psi_L^{\beta}
&=0,
\label{eq:Psi-left-general}
\\
\lambda_2^R\,
e^a\wedge e^b\wedge
(\bar\Sigma_{ab})_{\dot\alpha\dot\beta}\Psi_R^{\dot\beta}
&=0.
\label{eq:Psi-right-general}
\end{align}
In these equations we used the (anti-)self-duality of the chiral Lorentz generators, so that the $\epsilon_{abcd}$ contribution coming from \eqref{JJtraceL} and \eqref{JJtraceL} reduce to the metric contraction and is absorbed into the overall normalization. For the Einstein--Cartan--Holst weights these become
\begin{align}
\left(\frac{1}{\gamma}-i\right)
e^a\wedge e^b\wedge
(\Sigma_{ab})_{\alpha\beta}\Psi_L^{\beta}
&=0,
\label{eq:Psi-left-ECH}
\\
\left(\frac{1}{\gamma}+i\right)
e^a\wedge e^b\wedge
(\bar\Sigma_{ab})_{\dot\alpha\dot\beta}\Psi_R^{\dot\beta}
&=0.
\label{eq:Psi-right-ECH}
\end{align}
These are algebraic three-form equations rather than Rarita--Schwinger-type propagation equations. For an invertible coframe, write $\Psi=e^c\Psi_c$ and Hodge-dualize the three-form equation. Using the (anti-)self-duality of the corresponding Lorentz generators gives, up to a nonzero overall factor,
\begin{equation}
\gamma^{dc}\Psi_c=0,
\qquad
\gamma^{dc}:=\frac{1}{2}[\gamma^d,\gamma^c],
\label{eq:vector-spinor-algebraic-kernel}
\end{equation}
and using the Clifford algebra,
\begin{equation}
\gamma^{dc}\Psi_c
=
\frac{1}{2}[\gamma^d,\gamma^c]\Psi_c
=
\gamma^d\gamma^c\Psi_c
-
\eta^{dc}\Psi_c.
\end{equation}
Defining $\chi:=\gamma^c\Psi_c$, 
\begin{equation}
\gamma^{dc}\Psi_c
=
\gamma^d\chi-\Psi_d,
\end{equation}
Therefore,
\begin{equation}
\gamma^{dc}\Psi_c=0
\quad\Longrightarrow\quad
\Psi_d=\gamma_d\chi.
\end{equation}
Taking the gamma trace then gives $\chi=4\chi$, hence $\chi=0$ and therefore $\Psi_d=0$. Consequently, whenever the corresponding chiral coefficient is nonzero, that vector-spinor vanishes. For generic $\gamma$ both coefficients are nonzero and
\begin{equation}
\Psi_L=0,
\qquad
\Psi_R=0.
\label{eq:Psi-trivial-solution}
\end{equation}

The exceptional values are consistent with the chiral reduction described above. At $\gamma=+i$, $\lambda_2^R=0$ while $\lambda_2^L\neq0$, so the local bulk equations impose $\Psi_L=0$ whereas $\Psi_R$ receives no equation from either the vanished $\lambda_2^R$ term or the topological $\lambda_0$ sector. At $\gamma=-i$ the roles are reversed: $\Psi_R=0$ while $\Psi_L$ is absent from the local bulk equations. Thus the chiral Immirzi values do not generate fermionic propagation; they simply remove one of the two algebraic constraints together with the corresponding inactive chiral sector.

The outcome is therefore restrictive. With only $\langle\se^4\rangle$, $\langle\se^2F\rangle$, and $\langle F F\rangle$, an unrestricted fermionic one-form has no propagating bulk equation. For generic $\gamma$ its algebraic field equation trivializes both chiral vector-spinors, after which $\Theta^{ab}=0$ and the gravitational equations reduce to the pure Einstein--Cartan--Holst equations above. A nontrivial propagating vector-spinor sector requires an additional non-topological invariant containing $D\Psi$.

\section{Matter ansatz and spinor dynamics}
\label{sec:matter-ansatz-spinor-dynamics}

The preceding analysis shows that the original geometric invariants do not provide propagating bulk dynamics for an independent fermionic one-form: its field equation is algebraic. Here we take a different route and focus on spin-$\tfrac12$ degrees of freedom carried by spinor-valued zero-forms. The coframe provides the soldering map needed to embed such fields into the spinorial differential forms appearing naturally in the orthosymplectic construction. In particular, a Weyl spinor zero-form can be converted into a spinor-valued one-form by multiplication with $\se$, while higher powers of $\se$ generate the corresponding higher-degree spinorial forms. This suggests restricting the fermionic sector to a matter-ansatz subspace and supplementing the original geometric action by a non-topological invariant involving $D\Psi$. The purpose of this section is to show that this extension produces ordinary first-order Weyl dynamics while retaining a direct geometric relation with the fermionic sector of the superconnection.

\subsection{Weyl spinor dynamics}
\label{sec:matter-ansatz-spinor-dynamics-intro}

To construct the required kinetic invariant, we use the soldering form not only to identify the fermionic one-form with a Weyl spinor zero-form, but also to construct a companion spinorial two-form from the same Dirac spinor. Since $\se$ reverses Weyl chirality whereas $\se\wedge\se$ preserves it, the left-handed one-form and two-form necessarily involve opposite Weyl components. Their invariant pairing,
$\langle\boldsymbol{\Psi}^{(2)}_L\wedge D_L\boldsymbol{\Psi}_L\rangle$,
is then a four-form of the appropriate degree for an action. After imposing the matter ansaetze, the three coframes supplied by $\Psi_L^{(2)}$ and $\Psi_L$ combine through the Clifford identities into the Hodge-dual coframe appearing in the standard Weyl action. Consequently, this spinorial BF-type invariant reduces to the usual first-order Weyl kinetic term, together with the corresponding torsion coupling. The two-form $\Psi_L^{(2)}$ should therefore not be interpreted, after imposing the ansatz, as a new independent propagating field; rather, it provides the geometric structure required to express the spin-$\tfrac12$ kinetic term in the same chiral language as the original fermionic one-form.

[--[We impose the soldering ansatz
\begin{equation}
\Psi_{L}^{\alpha}
=
e^{a}(\sigma_{a})^{\alpha\dot\alpha}\chi_{\dot\alpha},
\qquad
\Psi_{R\, \dot\alpha}
=
e^{a}(\bar\sigma_{a})_{\dot\alpha\alpha}\psi^{\alpha}.
\end{equation}
The fields $\chi_{\dot\alpha}$ and $\psi_{\alpha}$ are spinor-valued zero-forms. Contraction with the frame and the Pauli intertwiners converts them into spinor-valued one-forms. Thus, $\Psi_{L}^{\alpha}$ and $\Psi_{R}^{\dot\alpha}$ are not independent Rarita--Schwinger fields: their spacetime one-form index is supplied entirely by the frame, while their independent fermionic content is carried by the spin-$\tfrac12$ fields $\chi$ and $\psi$.]--]

Let
\begin{equation}
 \psi=\begin{pmatrix}\psi_L^\alpha\\ \psi_{R\dot\alpha}\end{pmatrix}
 \in\Omega^0(M,S_L\oplus S_R)
\end{equation}
be a Dirac spinor zero-form. Since $\se$ reverses Weyl chirality, whereas $\se\wedge\se$ preserves it, the consistent left-handed ansaetze are
\begin{align}
 \Psi_L^\alpha&=(P_L\se\psi)^\alpha
 =e^{\alpha\dot\alpha}\psi_{R\dot\alpha},
 \label{eq:one-form-matter-ansatz-left}\\
 \Psi_L^{(2)\alpha}&=
 [P_L(\se\wedge\se)\psi]^\alpha
 =(\se\wedge\se)_L{}^\alpha{}_\beta\psi_L^\beta.
 \label{eq:two-form-matter-ansatz-left}
\end{align}
Thus the two ansaetze involve opposite Weyl components of the same Dirac
spinor. In the conventions of Appendix~\ref{app:PlebanskiForms},
\begin{equation}
 (\se\wedge\se)_L{}^\alpha{}_\beta
 =2\mathcal B_L{}^\alpha{}_\beta
 =-2e^a\wedge e^b(\Sigma_{ab})^\alpha{}_\beta.
 \label{eq:sese-left-matter}
\end{equation}

The spinorial two-form can be paired with the odd curvature
$\rho_L=D_L\Psi_L$:
\begin{equation}
 S_{\mathrm{kin},L}= - \frac{1}{12} \int_M
 \left\langle
    \boldsymbol{\Psi}^{(2)}_L
    \wedge D_L\boldsymbol{\Psi}_L
    \right\rangle
 \label{eq:spinorial-BF-kinetic0}
\end{equation}

Here the brackets denote the standard invariant bilinear form on
$\mathfrak{osp}(1|2)_L$. Writing
\begin{equation}
    \boldsymbol{\Psi}^{(2)}_L
    =
    \Psi_L^{(2)\alpha}Q_\alpha,
    \qquad
    D_L\boldsymbol{\Psi}_L
    =
    D_L\Psi_L^\beta Q_\beta,
\end{equation}
one finds
\begin{equation}
    \left\langle
    \boldsymbol{\Psi}^{(2)}_L
    \wedge D_L\boldsymbol{\Psi}_L
    \right\rangle
    =
    -2 \,\epsilon_{\alpha\beta}
    \Psi_L^{(2)\alpha}\wedge D_L\Psi_L^\beta.
\end{equation}
Thus the standard left-chiral supertrace leads to
\begin{equation}
 S_{\mathrm{kin},L}= \frac{1}{6}\int_M
 \epsilon_{\alpha\beta}\Psi_L^{(2)\alpha}
 \wedge D_L\Psi_L^\beta.
 \label{eq:spinorial-BF-kinetic}
\end{equation}

Before restricting to the composite matter ansatz, one may instead regard $\Psi_L^{(2)}$ and $\Psi_L$ as independent spinor-valued forms. The resulting auxiliary spinorial BF theory gives, upon independent variation of the two-form and one-form, respectively,
\begin{equation}
 D_L\Psi_L^\alpha=0,
 \qquad
 D_L\Psi_L^{(2)\alpha}=0,
 \label{eq:unreduced-spinorial-BF-eom}
\end{equation}
up to the equivalent placement of $\epsilon_{\alpha\beta}$.

After imposing both ansaetze and using \eqref{Talphaalphadot}, we get
\begin{equation}
 D_L(e^{\alpha\dot\alpha}\psi_{R\dot\alpha})
 =T^{\alpha\dot\alpha}\psi_{R\dot\alpha}
 -e^{\alpha\dot\alpha}\wedge D_R\psi_{R\dot\alpha}, \label{Dsepsi}
\end{equation}
therefore the reduced term is
\begin{align}
 S_{\mathrm{kin},L}= \frac{1}{6} \int_M\big[&
 -\psi_L^\gamma H_\gamma{}^{\dot\beta}
 \wedge D_R\psi_{R\dot\beta}
 \nonumber\\
 &+\epsilon_{\alpha\beta}(\se\wedge\se)_L{}^\alpha{}_\gamma
 \wedge T^{\beta\dot\beta}\psi_L^\gamma\psi_{R\dot\beta}
 \big].
 \label{eq:reduced-spinor-kinetic}
\end{align}
where we defined the three-form
\begin{equation}
 H_\gamma{}^{\dot\beta}:=
 \epsilon_{\alpha\beta}(\se\wedge\se)_L{}^\alpha{}_\gamma
 \wedge e^{\beta\dot\beta}.
 \label{eq:H-threeform-definition}
\end{equation}
The Clifford identity for three coframes has the form (see appendix \ref{app:threecoframecliffordidentity})
\begin{equation}
 H_\alpha{}^{\dot\beta}= 6i (\ast e^a)
 (\sigma_a)_\alpha{}^{\dot\beta}.
 \label{eq:H-Dirac-identity}
\end{equation}
Hence, the first term in Eq.~\eqref{eq:reduced-spinor-kinetic} is the ordinary Weyl kinetic term in form notation \eqref{weylkineticterm}, whereas the second is its torsion coupling. Note that, although Eq.~\eqref{eq:reduced-spinor-kinetic} contains both chiral connections through $D_R\psi_R$ and the torsion $T^{\alpha\dot\alpha}$, this does not introduce an independent $\omega_R$ dependence. The two appearances originate from the covariant decomposition of $D_L(e^{\alpha\dot\alpha}\psi_{R\dot\alpha})$, and the $\omega_R$ contributions cancel identically between the kinetic and torsion terms. Thus the reduced expression retains the connection dependence of its parent left-chiral invariant, which involves only $\omega_L$.

Variation with respect to $\psi_L$ gives
\begin{equation}
 H_\gamma{}^{\dot\beta}\wedge D_R\psi_{R\dot\beta}
 -\epsilon_{\alpha\beta}(\se\wedge\se)_L{}^\alpha{}_\gamma
 \wedge T^{\beta\dot\beta}\psi_{R\dot\beta}=0.
 \label{eq:psiL-reduced-eom}
\end{equation}
For $T^a=0$ this is the Weyl equation
\begin{equation}
 (\sigma^a)_\gamma{}^{\dot\beta}D_a\psi_{R\dot\beta}=0.
\end{equation}
Variation with respect to $\psi_R$, followed by covariant integration by
parts, gives the compact equation
\begin{equation}
 e^{\alpha\dot\beta}\wedge D_L\!\left[
 (\se\wedge\se)_{L\alpha\gamma}\psi_L^\gamma\right]=0.
 \label{eq:psiR-reduced-eom}
\end{equation}
To derive this equation is convenient to use \eqref{eq:reduced-spinor-kinetic} without using the expansion \eqref{Dsepsi}.

In the purely left-chiral complex theory, $\psi_L$ and $\psi_R$ are treated as independent fields. A Lorentzian real action is obtained by adjoining the Hermitian-conjugate right-chiral term and then imposing $\psi_L=(\psi_R)^\dagger$; in that completion the two Weyl equations are complex conjugates.

\subsection{Original action restricted to the matter ansatz}
\label{sec:original-action-matter-ansatz}

We can now combine the geometric sector with the spinorial kinetic construction above in order to obtain a theory with nontrivial Weyl-spinor dynamics. It is useful, however, to separate the two contributions before forming the complete action.

We first restrict the original geometric action to the matter ansatz $\Psi_L=\se\psi_R$ and determine the equations that it induces on this constrained field space. As shown below, this restriction alone does not generate propagation: the original $\langle\se^2F\rangle$ contribution still produces an algebraic equation for the spinor zero-form.
The extended matter action is then obtained by adding the spinorial BF invariant constructed in the preceding subsection. Its first-order derivative contribution combines with the algebraic term inherited from the geometric action, so that the latter becomes a mass-like left--right mixing rather than a constraint forcing the spinor to vanish. The resulting reduced theory therefore couples the original gravitational geometry to a genuinely dynamical Weyl sector.

In this section we work in the complex left-chiral truncation, set $A_R=0$, and impose only
\begin{equation}
 \Psi_L^\alpha=e^{\alpha\dot\alpha}\psi_{R\dot\alpha}
 \label{eq:original-action-simple-matter-ansatz}
\end{equation}
in $S_L$. The reduced original action is
\begin{align}
 S_L^{\mathrm{MA}}=\int_M\big[&
 -8\lambda_4B^i\wedge B_i
 -4i\lambda_2B^i\wedge\mathcal F_i(\psi_R)
 \nonumber\\
 &+2\lambda_0\mathcal F^i(\psi_R)\wedge\mathcal F_i(\psi_R)
 -2\lambda_0\epsilon_{\alpha\beta}
 \rho_L^\alpha(\psi_R)\wedge\rho_L^\beta(\psi_R)\big],
 \label{eq:original-action-matter-ansatz}
\end{align}
where the reduced curvatures are those in Eq.~\eqref{eq:reduced-supercurvatures} below. The two $\lambda_0$ terms are the bosonic and fermionic components of $\langle F_L\wedge F_L\rangle$ and therefore remain a boundary term. They must not be varied separately as bulk terms.

In this truncation $D_R=d$ on right-handed spinors, while $D_L$ denotes the covariant derivative acting on the undotted index,
\begin{equation}
D_Le^{\alpha\dot\alpha} = d e^{\alpha\dot\alpha} + \omega_L{}^\alpha{}_\beta \wedge e^{\beta\dot\alpha}
=:T_L^{\alpha\dot\alpha}.
\end{equation}
This is a complex chiral description, not yet a Lorentzian real completion as described in the previous section.

Let $\mathcal E_{L\alpha}$ denote the unrestricted fermionic Euler--Lagrange three-form,
\begin{equation}
 \mathcal E_{L\alpha}:=\lambda_2^L e^a\wedge e^b
 (\Sigma_{ab})_{\alpha\beta}\Psi_L^\beta.
 \label{eq:EL-algebraic-definition}
\end{equation}
The variation induced by $\delta\psi_{R\dot\alpha}$ is
$\delta\Psi_L^\alpha=e^{\alpha\dot\alpha}
\delta\psi_{R\dot\alpha}$. Hence the matter equation is the pullback of the unrestricted vector--spinor equation (see appendix \ref{app:matteransatzpullback}),
\begin{equation}
 e^{\alpha\dot\alpha}\wedge
 \mathcal E_{L\alpha}
 \big|_{\Psi_L=\se\psi_R}=0.
 \label{eq:original-action-psiR-pullback}
\end{equation}

This equation is also algebraic. For an invertible coframe the three-coframe Clifford identity reduces it, up to a nonzero numerical factor, to
\begin{equation}
 \lambda_2^L\,\mathrm{vol}_e\,\psi_R^{\dot\alpha}=0.
 \label{eq:original-action-psiR-algebraic}
\end{equation}
Thus, for $\lambda_2^L\neq0$, imposing the matter ansatz in the original action weakens the unrestricted equation to its tangent projection, but still forces $\psi_R=0$; it does not by itself generate spinor propagation.

For completeness, the remaining reduced equations are most cleanly stated by the chain rule. Since the ansatz is independent of $\omega_L$, the connection equation is the pullback of the original connection equation,
\begin{equation}
 \left.\frac{\delta S_L}{\delta\omega_L^{ab}}\right|_{\Psi_L=\se\psi_R}=0.
 \label{eq:original-action-reduced-connection}
\end{equation}
Equivalently, the bulk equation is $D_LB^i=0$; for a nondegenerate coframe it fixes the active chiral connection to the corresponding Levi--Civita projection. The coframe equation acquires the tangent contribution caused by the explicit $e$ in the ansatz:
\begin{equation}
 \left.\frac{\delta S_L}{\delta e^a}\right|_{\Psi_L=\se\psi_R}
 +(\sigma_a)^{\alpha\dot\alpha}\psi_{R\dot\alpha}
 \wedge\left.\mathcal E_{L\alpha}\right|_{\Psi_L=\se\psi_R}=0.
 \label{eq:original-action-reduced-coframe}
\end{equation}
On the generic solution of
Eq.~\eqref{eq:original-action-psiR-algebraic}, the second term and all fermionic bilinears vanish, leaving the ordinary chiral gravitational equations. This establishes the baseline against which the additional spinorial BF term should be compared.

\subsection{Full action in the matter-ansatz sector}

We now add Eq.~\eqref{eq:spinorial-BF-kinetic} to Eq.~\eqref{eq:original-action-matter-ansatz} and impose both ansaetze at the level of the action:
\begin{equation}
 S_{\mathrm{red}}[e,\omega_L,\psi_L,\psi_R]
 :=S_L^{\mathrm{MA}}[e,\omega_L,\psi_R]+S_{\mathrm{kin},L}.
 \label{eq:full-reduced-matter-action}
\end{equation}
The distinction between unrestricted and tangent variations was made explicit in the preceding subsection. The new ingredient is that the algebraic equation \eqref{eq:original-action-psiR-pullback} is now coupled to the first-order equations supplied by $S_{\mathrm{kin},L}$.

Under $\Psi_L^\alpha=e^{\alpha\dot\alpha}\psi_{R\dot\alpha}$ define
\begin{align}
 \Theta_L^{ab}(\psi_R)&=(\Sigma^{ab})_{\alpha\beta}
 e^{\alpha\dot\alpha}\wedge e^{\beta\dot\beta}
 \psi_{R\dot\alpha}\psi_{R\dot\beta},\\
 \mathcal F_L^{ab}(\psi_R)&=R_L^{ab}-\Theta_L^{ab}(\psi_R),\\
 \rho_L^\alpha(\psi_R)&=T^{\alpha\dot\alpha}\psi_{R\dot\alpha}
 -e^{\alpha\dot\alpha}\wedge D_R\psi_{R\dot\alpha}.
 \label{eq:reduced-supercurvatures}
\end{align}
The full reduced action is therefore
\begin{align}
S_{\mathrm{red}}=\int_M\big[&
-8\lambda_4B^i\wedge B_i
-4i\lambda_2B^i\wedge\mathcal F_i(\psi_R)
\nonumber\\
&+2\lambda_0\mathcal F^i(\psi_R)\wedge\mathcal F_i(\psi_R)
-2\lambda_0\epsilon_{\alpha\beta}
 \rho_L^\alpha(\psi_R)\wedge\rho_L^\beta(\psi_R)
\nonumber\\
&+\frac{1}{6}\epsilon_{\alpha\beta}
 (\se\wedge\se)_L{}^\alpha{}_\gamma\psi_L^\gamma
 \wedge\rho_L^\beta(\psi_R)\big].
 \label{eq:full-reduced-action-expanded}
\end{align}
As in the preceding subsection, the two $\lambda_0$ terms are retained together and give no local bulk equation. The two spinor equations of the complete reduced model are
\begin{align}
&H_\gamma{}^{\dot\beta}\wedge D_R\psi_{R\dot\beta}
-\epsilon_{\alpha\beta}(\se\wedge\se)_L{}^\alpha{}_\gamma
 \wedge T^{\beta\dot\beta}\psi_{R\dot\beta}=0,
\label{eq:full-reduced-psiL}\\
&e^{\alpha\dot\alpha}\wedge
 \mathcal E_{L\alpha}\big|_{\Psi_L=\se\psi_R}
+\frac{1}{6}e^{\alpha\dot\alpha}\wedge D_L\!\left[
 (\se\wedge\se)_{L\alpha\gamma}\psi_L^\gamma\right]=0.
\label{eq:full-reduced-psiR}
\end{align}
Thus the gamma trace of the old algebraic equation is balanced by a first-order derivative of $\psi_L$. Together with Eq.~\eqref{eq:full-reduced-psiL}, it forms a nontrivial Dirac-type system;
the matter field is no longer forced to vanish. The algebraic contribution from $\langle\se^2F\rangle$ should be identified as a mass-like left--right mixing term. Its physical mass is fixed only after canonical normalization and depends on the dimensions and normalization of $\lambda_2^L$.

Finally, because $\Psi_L=\se\psi_R$ depends explicitly on the coframe and $D_L\Psi_L$ contains torsion, the new term supplies the usual spin current to the connection equation and energy--momentum to the coframe equation. In terms of the variational currents
\begin{equation}
 s_{ab}:=-\frac{\delta S_{\mathrm{kin},L}}{\delta\omega_L^{ab}},
 \qquad
 \tau_a:=-\frac{\delta S_{\mathrm{kin},L}}{\delta e^a},
 \label{eq:spinor-variational-currents}
\end{equation}
the connection and coframe equations are the corresponding gravitational equations of Sec.~\ref{sec:gravitational-field-equations-fermionic}, restricted to the matter ansatz, with sources $s_{ab}$ and $\tau_a$ and with the additional tangent contribution associated with the explicit coframe dependence of $\Psi_L=\se\psi_R$. The connection equation is algebraic in torsion, as in Einstein--Cartan theory. Eliminating the algebraic torsion is therefore expected to generate a local four-fermion interaction, as in Einstein--Cartan theory. The generalized two-form and one-form ansaetze therefore turn the spinorial BF invariant into genuine spin-$\tfrac12$ propagation, whereas the original three invariants alone provide only algebraic fermionic dynamics.

The construction above illustrates the utility of the matter ansatz within a chiral Lorentz superalgebra. The soldering form converts the fermionic one-form of the superconnection into a spin-$\tfrac12$ field and, together with the generalized spinorial two-form, allows a first-order kinetic invariant to be written without introducing an independent gravitino. At the same time, the original geometric action retains a direct dynamical imprint on the reduced fermionic sector: although by itself it provides no propagating spinor equation, its $\langle\se^2F\rangle$ contribution becomes an algebraic left--right mixing term and therefore plays the role of a geometrically generated mass-like interaction once the kinetic term is included. The Chern--Weil sector remains locally topological, while the new kinetic invariant makes the torsion equation sensitive to the spin current and hence allows the usual Einstein--Cartan-type induced four-fermion interaction after eliminating torsion. Thus the resulting spinor dynamics is not appended independently to the geometric theory: its kinetic, mass-like, and torsional structures arise from different but closely related pieces of the same soldered chiral construction. Since the matter ansatz nevertheless restricts the original vector-spinor field space, one must still determine which gauge transformations preserve this restricted sector and under what conditions it defines a consistent truncation; these questions are addressed next.

\section{Equivariance of the matter ansatz and residual supersymmetry}
\label{sec:matter-ansatz-equivariance}

This section examines the consistency of the matter ansatz at three logically distinct levels. Kinematically, the ansatz restricts the vector--spinor to its gamma-trace, spin-$\tfrac12$ component. At the level of symmetries, one must determine which transformations are tangent to this restricted field space: Lorentz equivariance is automatic, whereas residual supersymmetry imposes a twistor-spinor condition on its parameter, and invariance of the fermionic curvature gives a further background condition. Dynamically, the Euler--Lagrange equations obtained after imposing the ansatz are only the tangent projections of the unrestricted equations, so a genuine consistent truncation additionally requires the discarded normal components to vanish. These kinematical, symmetry, and dynamical requirements are independent and are considered in turn.

\subsection{Kinematics of the matter ansatz}

Suppressing chiral labels, the unconventional-supersymmetry matter ansatz identifies the fermionic one-form in the superconnection with a soldered spin-$\tfrac12$ field,
\begin{equation}
\Psi=\slashed e\,\psi,\qquad \Psi_\mu=\gamma_\mu\psi.
\label{eq:matter-ansatz-equivariance}
\end{equation}
where $\gamma_\mu$ is a soldered gamma matrix of the form \eqref{solderedgammaa}. This construction was introduced in Refs.~\cite{Alvarez:2011gd,Alvarez:2013tga} and has recently been reinterpreted through the dressing-field method in Ref.~\cite{Francois:2024xqi}. In four dimensions a vector-spinor decomposes into its gamma-trace and gamma-traceless parts, with projectors
\begin{equation}
(P_{1/2})_\mu{}^\nu=\frac{1}{4}\gamma_\mu\gamma^\nu,\qquad (P_{3/2})_\mu{}^\nu=\delta_\mu{}^\nu-\frac{1}{4}\gamma_\mu\gamma^\nu.
\label{eq:vector-spinor-projectors}
\end{equation}
The ansatz is therefore equivalent to $P_{3/2}\Psi=0$: it removes the independent spin-$\tfrac32$ component of the fermionic one-form.

Before imposing the matter ansatz, the odd curvature $\rho=D\Psi$ transforms covariantly because it is a component of the full supercurvature. This fact does not by itself imply that a gauge transformation preserves the restricted subspace defined by Eq.~\eqref{eq:matter-ansatz-equivariance}. For Lorentz transformations preservation is algebraic and follows from the intertwining identity $[J_{ab},\gamma_c]=(J^{(1)}_{ab})^d{}_c\gamma_d$. Supersymmetry is different because a single odd transformation does not preserve the vector Clifford span, as discussed towards the end of Sec.~\ref{sec:cliiford} and Appendix~\ref{app:osp-six-dimensional-realization}.

For a supersymmetry transformation of the form \(\delta_\epsilon\Psi=D\epsilon\), the sector defined by \eqref{eq:matter-ansatz-equivariance} admits induced variations \(\delta_\epsilon e^a\) and \(\delta_\epsilon\psi\) satisfying
\begin{equation}
D\epsilon=\delta_\epsilon\slashed e\,\psi+\slashed e\,\delta_\epsilon\psi.
\label{eq:full-tangency}
\end{equation}
Therefore, preservation of the matter ansatz requires the supersymmetry variation to be tangent to the constrained field space,
\begin{equation}
P_{3/2}\bigl(D\epsilon-\delta_\epsilon\slashed e\,\psi\bigr)=0.
\label{eq:matter-ansatz-tangency}
\end{equation}
On a bosonic background, for which \(\psi=0\), and with the coframe held fixed under the residual rigid supersymmetry transformation \cite{Alvarez:2013tga}, this condition reduces to
\begin{equation}
P_{3/2}D\epsilon=0.
\end{equation}
Equivalently,
\begin{equation}
\nabla_\mu\epsilon=\frac{1}{4}\gamma_\mu\slashed \nabla\epsilon=\gamma_\mu\eta,\qquad \eta:=\frac{1}{4}\slashed \nabla\epsilon,
\label{eq:twistor-from-matter-ansatz}
\end{equation}
where, for a spinor zero-form
\begin{equation}
 D\epsilon = dx^\mu \nabla_\mu \epsilon\,,
\end{equation}
where
\begin{equation}
 \nabla_\mu \epsilon = \partial_\mu \epsilon + \frac{1}{2} \omega^{ab}{}_\mu \tilde{\Sigma}_{ab}\epsilon,
\end{equation}
where it is understood that $\tilde{\Sigma}_{ab} = \tfrac{1}{2} \gamma_{ab}$ goes in the appropriate chiral representation for $\epsilon$. Thus, according to Eq. \eqref{eq:twistor-from-matter-ansatz}, the supersymmetry parameters compatible with the matter ansatz are twistor spinors, also known as conformal Killing spinors \cite{Baum:1998ra,deMedeiros:2012sb,Festuccia:2011ws,Lauria:2020rhc}. This identifies the equivariance condition associated with the no-gravitino constraint with the geometric characterization of rigid supersymmetry on curved backgrounds.  Indeed, for a twistor spinor we can write
\begin{equation}
 \delta \Psi = D \epsilon = \se \delta\psi, \qquad \delta \psi = \frac{1}{4} \slashed{\nabla} \epsilon\,,\label{deltaPsitwistorspinor}
\end{equation}
provided $\epsilon$ satistfies \eqref{eq:twistor-from-matter-ansatz}. Eq. \eqref{deltaPsitwistorspinor} implies that rigid supersymmetry preserves the Lorentz-equivariant matter-ansatz sector.

Killing spinors constitute the special subclass for which \(\eta\) is proportional to \(\epsilon\). Indeed, a Killing spinor satisfies
\begin{equation}
\nabla_\mu \epsilon
=
\lambda\,\gamma_\mu\epsilon,
\qquad
\lambda=\mathrm{const}.
\end{equation}
Therefore,
\begin{equation}
\eta=\lambda\epsilon,
\qquad
\slashed \nabla\epsilon=4\lambda\epsilon,
\label{eq:killing-spinor-subclass}
\end{equation}
and consequently
\begin{equation}
 \delta_\mathrm{Ks} \Psi = \se (\lambda \epsilon) = \se \delta\psi, \qquad \delta \psi_\mathrm{Ks} = \lambda \epsilon\,,\label{deltaPsikillingspinor}
\end{equation}

Thus, for a Killing-spinor parameter the residual supersymmetry acts on the spin-$\tfrac12$ field as the inhomogeneous fermionic shift $\delta_{\mathrm{Ks}}\psi=\lambda\epsilon$. This should not be interpreted as a spacetime translation or as a local fermionic gauge redundancy. After imposing the matter ansatz, the supersymmetry parameter is no longer an arbitrary local spinor, but is constrained by the twistor-spinor equation and, in the present subclass, by the Killing-spinor equation. Consequently, an arbitrary configuration $\psi(x)$ cannot in general be gauged away \cite{Valenzuela:2023aoa}, rather, the transformation relates configurations within the Lorentz-equivariant matter-ansatz sector. For $\lambda\neq0$, the configuration $\psi=0$ is not invariant under the residual supersymmetry, even though the transformation remains tangent to the matter-ansatz sector. The inhomogeneous form of the transformation is reminiscent of fermionic shift symmetries and nonlinear realizations of supersymmetry; however, identifying $\psi$ as a Goldstino would additionally require establishing spontaneous supersymmetry breaking and the closure of the residual supersymmetry algebra.

It is now clear that the two-form matter ansatz
\begin{equation}
\Psi^{(2)}=\se\wedge\se \psi, 
\end{equation}
used in Eq.~\eqref{eq:two-form-matter-ansatz-left}, imposes no additional condition on the residual supersymmetry. Indeed, since $\Psi^{(2)}=\se\wedge\Psi^{(1)}$, its equivariance follows directly from that of the one-form matter ansatz $\Psi^{(1)}=\se\psi$, provided $\delta_\epsilon\se=0$. Therefore, once the supersymmetry parameter satisfies the twistor-spinor condition required for $\Psi^{(1)}$, the transformation of $\Psi^{(2)}$ automatically remains within the corresponding two-form matter-ansatz sector. The two-form ansatz thus introduces no independent supersymmetry constraint.

\subsection{Susy invariance of the background within the matter ansatz}

Preservation of the ansatz must be distinguished from invariance of the fermionic curvature. On a bosonic background,
\begin{equation}
\delta_\epsilon\rho=D(\delta_\epsilon\Psi)=D^2\epsilon=\mathcal R\epsilon,
\label{eq:odd-curvature-susy-variation}
\end{equation}
where $\mathcal R$ denotes the complete bosonic curvature acting on the spinor in the appropriate representation; for a Lorentz and internal connection, $\mathcal R=\frac{1}{4}R^{ab}\gamma_{ab}+F^I t_I$ \footnote{In the present case we do not have internal sector and therefore $F^I=0$}. Consequently, $P_{3/2}D\epsilon=0$ does not generally imply $\delta_\epsilon\rho=0$.

The relation between the two conditions follows from the integrability of the twistor equation. Allowing torsion $T^a=D e^a$, one has $D\slashed e=T^a\gamma_a$ and Eq.~\eqref{eq:twistor-from-matter-ansatz} may be written as $D\epsilon=\slashed e\,\eta$. Applying $D$ once more gives
\begin{equation}
\mathcal R\epsilon=T^a\gamma_a\eta-\slashed e\wedge D\eta.
\label{eq:twistor-integrability-torsion}
\end{equation}
Therefore the odd component of the curvature is invariant precisely when $T^a\gamma_a\eta=\slashed e\wedge D\eta$. In particular, a flat Lorentz and internal connection gives $D^2\epsilon=0$ locally, while nonzero torsion remains constrained by Eq.~\eqref{eq:twistor-integrability-torsion}. If $D$ is instead an AdS or supercovariant derivative containing a term proportional to $\slashed e$, the relevant condition is the vanishing of the full supercurvature acting on $\epsilon$, not merely $R^{ab}(\omega)\epsilon=0$.

\subsection{Dynamical consistency of the matter ansatz}

From the dynamical point of view, the matter ansatz should also be regarded as a truncation of the original fermionic one-form. The question is then if the model is a consistent truncation of the parent theory or if the model should be regarded as a reduced theory on the constrained field space. Let
\begin{equation}
    \hat{\mathcal{E}}_{\Psi}^{\mu}=0\label{vectorspinorequation}
\end{equation}
denote the equation obtained by varying the unrestricted field \(\Psi_{\mu}\). Here we denoted $\hat{\mathcal{E}}_{\Psi}^{\nu}$ the Euler-Lagrange equation coming from a generic theory with a vector-spinor one form. After imposing
\begin{equation}
    \Psi_{\mu}=\gamma_{\mu}\psi
\end{equation}
and keeping the coframe fixed, the allowed variations are restricted to
\begin{equation}
    \delta\Psi_{\mu}=\gamma_{\mu}\delta\psi.
\end{equation}
Consequently, the reduced variational principle yields only the gamma-trace equation
\begin{equation}
    \gamma_{\mu}\hat{\mathcal{E}}_{\Psi}^{\mu}=0, \label{gamma-traceless-eq}
\end{equation}
rather than the complete vector-spinor equation \eqref{vectorspinorequation}.

Equation \eqref{gamma-traceless-eq} is equivalent to the Euler--Lagrange three-form equation obtained by varying the action with respect to the one-form $\Psi_L^\alpha$; see Eq.~\eqref{eq:original-action-psiR-pullback}. Indeed, for a left-chiral spinor,
\begin{equation}
e^{\alpha\dot\alpha}\wedge \hat{\mathcal{E}}_{L\alpha}=0
\end{equation}
is the two-component-spinor form of \eqref{gamma-traceless-eq}. Indeed, writing
\begin{equation}
\hat{\mathcal{E}}_{L\alpha}
=
\hat{\mathcal{E}}_{L\alpha}{}^{\mu}\,\ast e_\mu,
\qquad
e^{\alpha\dot\alpha}
=
e^\nu(\sigma_\nu)^{\alpha\dot\alpha},
\end{equation}
and using
\begin{equation}
e^\nu\wedge\ast e_\mu
=
\delta^\nu{}_\mu\,\mathrm{vol}_e,
\end{equation}
one obtains
\begin{equation}
e^{\alpha\dot\alpha}\wedge \hat{\mathcal{E}}_{L\alpha}
=
(\sigma_\mu)^{\alpha\dot\alpha}
\hat{\mathcal{E}}_{L\alpha}{}^\mu\,
\mathrm{vol}_e.
\end{equation}
Therefore,
\begin{equation}
e^{\alpha\dot\alpha}\wedge \hat{\mathcal{E}}_{L\alpha}=0
\qquad\Longleftrightarrow\qquad
(\sigma_\mu)^{\alpha\dot\alpha}
 \hat{\mathcal{E}}_{L\alpha}{}^\mu=0.
\end{equation}
In four-component notation this is the corresponding chiral block of $\gamma_\mu \hat{\mathcal{E}}_\Psi^\mu$ so that, up to the convention used for the chiral blocks,
\begin{equation}
e^{\alpha\dot\alpha}\wedge \hat{\mathcal{E}}_{L\alpha}=0
\qquad\Longleftrightarrow\qquad
\left(\gamma_\mu \hat{\mathcal{E}}_\Psi^\mu\right)_R=0.
\end{equation}
The full equation $\gamma_\mu \hat{\mathcal{E}}_\Psi^\mu=0$ additionally contains the conjugate right-handed vector-spinor equation.

For a solution of the reduced theory to lift to a solution of the original superconnection theory, one must therefore also require the vanishing of the discarded gamma-traceless component,
\begin{equation}
    \left(
        \delta^{\mu}{}_{\nu}
        -\frac{1}{d}\gamma^{\mu}\gamma_{\nu}
    \right)
    \hat{\mathcal{E}}_{\Psi}^{\nu}
    \bigg|_{\Psi=\slashed{e}\psi}
    =0. \label{truncationdynamics}
\end{equation}
Seen as a truncation of a parent theory, the matter ansatz sector is dynamically consistent if Eq.~\eqref{truncationdynamics} is satisfied.
To be clear, if Eq.~\eqref{truncationdynamics} is not satisfied, a reduced theory can still be perfectly well defined, however, in the constrained field space.

This dynamical requirement is distinct from the twistor-spinor condition
\begin{equation}
    \left(
        \delta_{\mu}{}^{\nu}
        -\frac{1}{d}\gamma_{\mu}\gamma^{\nu}
    \right)
    D_{\nu}\epsilon=0,
\end{equation}
which instead ensures that the rigid supersymmetry transformation is tangent to the matter-ansatz sector.

Let us focus now on the complete reduced system \eqref{eq:full-reduced-psiL}--\eqref{eq:full-reduced-psiR}. Before restricting the one-form equation to the matter-ansatz subspace, the corresponding three-form in the parent theory is
\begin{equation}
 \widehat{\mathcal E}_{L\alpha}
 :=
 \mathcal E_{L\alpha}
 +\frac{1}{6}D_L\!\left[
 (\se\wedge\se)_{L\alpha\gamma}\psi_L^\gamma
 \right].
 \label{eq:full-parent-oneform-eom}
\end{equation}
where $\mathcal E_{L\alpha}$ is defined in \eqref{eq:EL-algebraic-definition}. Equation~\eqref{eq:full-reduced-psiR} is precisely its tangent projection,
\begin{equation}
 e^{\alpha\dot\alpha}\wedge
 \widehat{\mathcal E}_{L\alpha}=0,
\end{equation}
and therefore fixes only the gamma trace. The algebraic contribution does not obstruct the normal equation. Indeed, using the three-coframe identity of Appendix~\ref{app:threecoframecliffordidentity},
\begin{equation}
 \left.
 \mathcal E_{L\alpha}
 \right|_{\Psi_L=\se\psi_R}
 =
 3i\lambda_2^L
 (\ast e^a)
 (\sigma_a)_\alpha{}^{\dot\beta}
 \psi_{R\dot\beta},
 \label{eq:algebraic-eom-pure-gamma-trace}
\end{equation}
so its vector--spinor coefficient is proportional to
$\gamma_\mu\psi_R$ and its gamma-traceless projection vanishes identically.

The derivative contribution is different. It is already sufficient to consider the torsion-free sector, where
$D_L(\se\wedge\se)_L=0$. Writing
\begin{equation}
 \frac{1}{6}D_L\!\left[
 (\se\wedge\se)_{L\alpha\gamma}\psi_L^\gamma
 \right]
 =
 K_{L\alpha}{}^\mu\,\ast e_\mu,
\end{equation}
the Clifford algebra gives, up to a nonzero convention-dependent overall factor,
\begin{equation}
 K_\mu
 \propto
 \gamma_{\mu\nu}D^\nu\psi_L
 =
 \gamma_\mu\slashed D\psi_L-D_\mu\psi_L.
\end{equation}
Consequently,
\begin{equation}
 (P_{3/2}K)_\mu
 \propto
 -\left(
 D_\mu-\frac{1}{4}\gamma_\mu\slashed D
 \right)\psi_L.
 \label{eq:normal-eom-twistor-psiL}
\end{equation}
Thus the discarded component of the one-form equation is a twistor-type equation for the dynamical field $\psi_L$.  Eq.~\eqref{eq:full-reduced-psiL} gives, for vanishing torsion, the Weyl equation for $\psi_R$ and does not imply \eqref{eq:normal-eom-twistor-psiL}. In the torsionful case additional algebraic torsion terms enter the normal projection, so the condition is not weakened.

There is an analogous loss of equations in the two-form ansatz. Before restriction, variation of $\Psi_L^{(2)}$ gives $D_L\Psi_L=0$, whereas Eq.~\eqref{eq:full-reduced-psiL} is only its projection along $\delta\Psi_L^{(2)}=(\se\wedge\se)_L\delta\psi_L$. Hence the two reduced spinor equations do not reconstruct the full unrestricted one-form and two-form equations. Generic solutions of the reduced Dirac-type system therefore do not lift to solutions of the unrestricted auxiliary spinorial BF theory. The matter ansaetze should in this sense be regarded as defining a reduced theory on the constrained field space, rather than a consistent truncation, unless the additional normal equations are imposed or follow from further dynamical conditions. A complementary strategy was developed by Torres-Gomez and Krasnov \cite{Torres-Gomez:2012tka}, who constructed a Lagrangian for an unrestricted spinor-valued one-form whose Hamiltonian constraints render the spin-$\tfrac32$ components nonpropagating, leaving precisely the degrees of freedom of a massive spin-$\tfrac12$ fermion.

\section{Conclusions and outlook}
\label{sec:conclusions-outlook}

In this work, we developed a systematic method for constructing Lorentz-covariant differential forms from the exterior algebra generated by the Clifford-valued coframe
\begin{equation}
\slashed e := e^{a}\gamma_{a}.
\end{equation}
The central observation is that the wedge product automatically removes the symmetric metric contractions appearing in the Clifford product. Consequently, the exterior powers of $\slashed e$ select the antisymmetrized Clifford generators,
\begin{equation}
\slashed e^{\wedge p}
=
e^{a_{1}}\wedge\cdots\wedge e^{a_{p}} \, \gamma_{a_{1}\cdots a_{p}},
\end{equation}
and provide a direct correspondence between differential forms and the graded components of the complexified Clifford algebra. Taking invariant traces of products involving $\slashed e$, the curvature, and their covariant derivatives then yields Lorentz-invariant forms without introducing their tensor contractions separately. In four dimensions, the chiral decomposition of this exterior algebra naturally isolates the self-dual and anti-self-dual Plebanski two-forms. This construction provides the organizing principle underlying the gravitational and spinorial invariants considered throughout the paper.

Applying this method to the left- and right-handed orthosymplectic superconnections, we constructed the three natural invariant four-forms
\begin{equation}
\left\langle \slashed e^{\,4}\right\rangle,
\qquad
\left\langle \slashed e^{\,2}\wedge F\right\rangle,
\qquad
\left\langle F\wedge F \right\rangle.
\end{equation}
Their chiral combinations reproduce the characteristic terms of Einstein--Cartan--Holst gravity with a cosmological contribution. Opposite relative weights between the left- and right-handed sectors generate the Einstein--Hilbert and cosmological terms, whereas equal weights produce the Holst contraction. The formalism therefore makes transparent how the parity-even and parity-odd gravitational invariants arise from the relative weighting of the two chiral sectors.

The fermionic analysis also reveals an important limitation of the original supercurvature construction. For unrestricted fermionic one-forms, the equations following from the three original invariants are algebraic and do not describe propagating spin-$\tfrac12$ degrees of freedom. After imposing the matter ansatz, which restricts the fermionic one-forms to frame-soldered Weyl spinors, the corresponding reduced equations remain algebraic and generically eliminate the spinors. Propagating fermion dynamics therefore requires an additional invariant. We introduced a spinorial BF-type term in which a spinorial two-form constructed from the coframe and a Weyl spinor is paired with the fermionic curvature. Under the matter ansatz, this term reduces to the standard first-order Weyl kinetic action, including its coupling to torsion. In the resulting dynamical theory, the algebraic contribution inherited from the original superconnection invariants is more appropriately interpreted as a mass-like mixing between the left- and right-handed spinors.

The matter ansatz is automatically equivariant under local Lorentz transformations, but its status under supersymmetry is subtler. Its preservation by a residual supersymmetry transformation imposes a twistor-spinor equation on the supersymmetry parameter, together with an additional integrability condition when invariance of the fermionic curvature is required. Moreover, imposing the ansatz before variation defines a reduced variational problem: the resulting spinor equations are the tangent, or gamma-trace, projections of the unrestricted vector-spinor equations. The reduced theory should therefore not be regarded as an automatically consistent truncation unless the complementary normal equations are also satisfied.

The connection equation remains algebraic in the torsion. Eliminating the algebraic torsion is therefore expected to generate a local four-fermion interaction, as in Einstein--Cartan theory. An explicit derivation of the resulting effective interaction, including its dependence on the chiral weights and the Immirzi parameter, would clarify the physical content of the reduced fermionic theory and constitutes a natural continuation of the present analysis.

A broader extension of the construction is suggested by replacing the Clifford-valued coframe with a super-coframe. The coframe used here takes values only in the vector sector of the relevant Clifford representation and intertwines Lorentz transformations in the expected way. It does not, by itself, form a representation space closed under the full adjoint action of the orthosymplectic superalgebra. A super-coframe containing both bosonic and fermionic components could provide the enlarged representation required for a genuinely supersymmetric soldering structure. Such a formulation may also furnish a geometric origin for the spinorial BF kinetic term, rather than introducing it as an additional Lorentz-invariant contribution.

A second direction is the construction of nontopological models by incorporating the spacetime Hodge dual. Curvature polynomials such as
\begin{equation}
\left\langle F \wedge F\right\rangle
\end{equation}
are of Chern--Weil type and are therefore topological when their invariant pairing is fixed. By contrast, invariants containing the Hodge dual, schematically
\begin{equation}
\left\langle F\wedge \ast F\right\rangle,
\end{equation}
depend on the metric determined by the coframe and can generate local propagating dynamics. Combining the exterior Clifford algebra with the Hodge dual may thus provide a systematic route to Yang--Mills-type, higher-curvature, and mixed bosonic--fermionic models while preserving manifest Lorentz covariance.

It would also be valuable to classify the solutions of the twistor-spinor equation selected by the present action. Their existence and integrability depend on the curvature and torsion of the background, and hence on the gravitational field equations obtained here. Determining the admissible twistor spinors on the classical backgrounds of the theory would identify the residual supersymmetries preserved by particular solutions and clarify the relation between the matter ansatz, background geometry, and fermionic curvature.

Finally, the supersymmetry equivariance problem should be revisited within the super-coframe formalism. The obstruction found for the ordinary coframe reflects the fact that the graded commutator of a supercharge with a Clifford vector generally leaves the vector subspace. Enlarging the soldering form to a supermultiplet may turn this obstruction into an exact intertwining relation for the full superalgebra. Establishing such an equivariance property would provide a natural setting in which the matter ansatz, its kinetic term, and its residual supersymmetry could be understood as different aspects of a single supergeometric construction.

\section*{Acknowledgements}
P.A. acknowledges support from ANID-FONDECYT Regular grant No. 1230112.
A.S. acknowledges support from ANID-FONDECYT postdoctoral grant 3240060.
J.H. was supported by Army Research Office grant W911NF-24-1-0058.

\begin{appendices}

\section{Preliminaries}\label{app:preliminaries}

In this appendix, we summarize the conventions and essential details of the representations used throughout the paper, working entirely in component notation and without invoking the superspace formalism.

\subsection{Notation}
We work in Minkowski metric $\eta_{ab} = \mathrm{diag} (-1,+1,+1,+1)$ and the orientation
\(\epsilon_{0123}=+1\).

$a,b=0,1,2,3$ flat spacetime indices

$i,j=1,2,3$ flat space indices

$\mu,\nu=0,1,2,3$ curved spacetime indices

$\alpha,\beta =1,2;\; \dot\alpha,\dot\beta=\dot 1,\dot 2$ two-component spinor indices

$(\sigma^a)_{\alpha\dot\alpha} = (\boldsymbol{1},\vec\sigma),\; (\bar\sigma^a)^{\dot\alpha\alpha}=(\boldsymbol{1},-\vec\sigma)$ where $\sigma^i$ are the Pauli matrices

$e^a = e^a{}_\mu dx^\mu$ is the vielbein, or coframe.

Withe mostly plus metric convention we have
\begin{equation}
 \sigma_a\bar\sigma_b+\sigma_b\bar\sigma_a
 =-2\eta_{ab}\mathbf 1_2,
 \qquad
 \bar\sigma_a\sigma_b+\bar\sigma_b\sigma_a
 =-2\eta_{ab}\mathbf 1_2.
 \label{eq:soldering-matrix-convention}
\end{equation}

Spinor conventions: $\epsilon_{\alpha\beta}$ and $\epsilon_{\dot\alpha\dot\beta}$ are both anti-symmetric with $\epsilon_{12}=\epsilon_{\dot 1\dot 2}=+1$ while $\epsilon^{12}=\epsilon^{\dot 1\dot 2}=-1$:
\begin{equation}
\epsilon_{\alpha\beta}=\begin{pmatrix}0&1\\-1&0\end{pmatrix}=-\epsilon^{\alpha\beta},\qquad
\epsilon_{\dot\alpha\dot\beta}=\begin{pmatrix}0&1\\-1&0\end{pmatrix}=-\epsilon^{\dot\alpha\dot\beta},
\end{equation}
Then
\begin{equation}
	\epsilon^{\alpha\beta}\epsilon_{\beta\gamma}=\tensor{\delta}{^\alpha_\gamma}, \quad
	\epsilon_{\alpha\beta}\epsilon^{\beta\gamma}=\tensor{\delta}{_\alpha^\gamma},\qquad
	\epsilon^{\dot\alpha\dot\beta}\epsilon_{\dot\beta\dot\gamma}=\tensor{\delta}{^{\dot\alpha}_{\dot\gamma}}, \quad
	\epsilon_{\dot\alpha\dot\beta}\epsilon^{\dot\beta\dot\gamma}=\tensor{\delta}{_{\dot\alpha}^{\dot\gamma}}.
\end{equation}
Raising and lowering can therefore be consistently defined through
\begin{equation}
\psi_\alpha = \epsilon_{\alpha\beta}\psi^\beta,\quad \psi^\alpha=\epsilon^{\alpha\beta}\psi_\beta,\qquad
\chi_{\dot\alpha} = \epsilon_{\dot\alpha\dot\beta}\chi^{\dot\beta},\quad \chi^{\dot\alpha}=\epsilon^{\dot\alpha\dot\beta}\chi_{\dot\beta}.
\end{equation}
Under these conventions,
$\tensor{X}{^\alpha_\alpha}=\epsilon^{\alpha\beta}\epsilon_{\alpha\gamma}\tensor{X}{_\beta^\gamma}=-\tensor{\delta}{^\beta_\gamma}\tensor{X}{_\beta^\gamma}=-\tensor{X}{_\alpha^\alpha}$ and likewise for dotted indices.
For example, $\tensor{(\bar\sigma^a)}{_{\dot\alpha}^\alpha}\tensor{(\sigma^b)}{_\alpha^{\dot\beta}}=-(\bar\sigma^a)_{\dot\alpha\alpha}(\sigma^b)^{\alpha\dot\beta}$.

Following standard two-component spinor conventions to raise and lower indiuces, \cite{VanProeyen:1999ni}  the Lorentz generators may be chosen as
\begin{align}
 (\Sigma_{ab})^\alpha{}_\beta
 &=
 \frac14  \bigl(\sigma_a\bar\sigma_b-\sigma_b\bar\sigma_a\bigr)^\alpha{}_\beta, \\
 (\bar\Sigma_{ab})_{\dot\alpha}{}^{\dot\beta}
 &=
 \frac14  \bigl(\bar\sigma_a\sigma_b-\bar\sigma_b\sigma_a\bigr)_{\dot\alpha}  {}^{\dot\beta}.  \label{eq:sigma-generator-definitions}
\end{align}
We lower or raise the remaining spinor index according to
\begin{equation}
 (\Sigma_{ab})_{\alpha\beta}
 :=\epsilon_{\alpha\gamma}(\Sigma_{ab})^\gamma{}_\beta, \qquad
 (\bar\Sigma_{ab})^{\dot\alpha\dot\beta}
 :=\epsilon^{\dot\alpha\dot\gamma}    (\bar\Sigma_{ab})_{\dot\gamma}{}^{\dot\beta}; \label{eq:symmetric-bivector-spinors}
\end{equation}
both objects are symmetric in their spinor indices. Our Lorentz-algebra convention is
\begin{equation}
 [\Sigma_{ab},\Sigma_{cd}]
 =
 -\eta_{ac}\Sigma_{bd}
 +\eta_{ad}\Sigma_{bc}
 +\eta_{bc}\Sigma_{ad}
 -\eta_{bd}\Sigma_{ac},
 \label{eq:lorentz-algebra-convention}
\end{equation}
with the identical relation for \(\bar\Sigma_{ab}\).

The (anti-)self-duality properties read
\begin{equation}
	\frac{1}{2}\tensor{\epsilon}{_a_b^c^d}(\Sigma_{cd})_{\alpha\beta}=-i(\Sigma_{ab})_{\alpha\beta},\qquad \frac{1}{2}\tensor{\epsilon}{_a_b^c^d}(\bar\Sigma_{cd})_{\dot\alpha\dot\beta}=+i(\bar\Sigma_{ab})_{\dot\alpha\dot\beta}.
\end{equation}

We distinguish throughout between the spacetime Hodge dual, denoted by $\ast$, and the internal Lorentz Hodge dual, denoted by $\star$. Let $X^{ab}$ be a Lorentz-bivector-valued $p$-form,
\begin{equation}
X^{ab}
=
\frac{1}{p!}
X^{ab}{}_{\mu_1\cdots\mu_p}
dx^{\mu_1}\wedge\cdots\wedge dx^{\mu_p}.
\end{equation}

The spacetime Hodge dual acts on the differential-form indices only:
\begin{equation}
(\ast X^{ab})_{\mu_1\cdots\mu_{4-p}}
=
\frac{1}{p!}
\epsilon_{\mu_1\cdots\mu_{4-p}}{}^{\nu_1\cdots\nu_p}
X^{ab}{}_{\nu_1\cdots\nu_p},
\end{equation}
or equivalently,
\begin{equation}
\ast
\left(
dx^{\mu_1}\wedge\cdots\wedge dx^{\mu_p}
\right)
=
\frac{1}{(4-p)!}
\epsilon^{\mu_1\cdots\mu_p}{}_{\nu_1\cdots\nu_{4-p}}
dx^{\nu_1}\wedge\cdots\wedge dx^{\nu_{4-p}}.
\end{equation}

By contrast, the internal Hodge dual acts only on the antisymmetric Lorentz indices:
\begin{equation}
(\star X)^{ab}
:=
\frac{1}{2}
\epsilon^{ab}{}_{cd}\,X^{cd}.
\end{equation}

In four-dimensional Lorentzian signature, both duality operators square to minus the identity on two-forms,
\begin{equation}
\ast^2=-1,
\qquad
\star^2=-1,
\end{equation}
when acting, respectively, on spacetime two-forms and internal Lorentz bivectors. Since they act on independent sets of indices, they commute:
\begin{equation}
[\ast,\star]X^{ab}=0.
\end{equation}

Accordingly, the internal self-dual and anti-self-dual projections are
\begin{equation}
X^{(\pm)ab}
=
P^{(\pm)ab}{}_{cd}X^{cd}
=
\frac{1}{2}
\left(
X^{ab}
\mp i\,(\star X)^{ab}
\right),
\end{equation}
and satisfy
\begin{equation}
\star X^{(\pm)}
=
\pm i\,X^{(\pm)}.
\end{equation}

These internal chiral projections should not be confused with the corresponding decomposition of a spacetime two-form into eigenspaces of the spacetime Hodge operator $\ast$. In the context of the chiral algebra, and with our conventions, we have denoted the anti-self dual part by Left and the self-dual part by Right, $X_L = X^{(-)}$ and $X_R = X^{(+)}$. The corresponding projectors are
\begin{equation}
\Pi_L=\frac{1}{2}(1+i\star),
\qquad
\Pi_R=\frac{1}{2}(1-i\star).
\label{eq:chiral-projectors-Immirzi}
\end{equation}

If the opposite convention for left and right duality is adopted, the labels \(\Pi_L\) and \(\Pi_R\) must be interchanged.

\section{Plebanski self-dual two-forms}
\label{app:PlebanskiForms}

The Plebanski two-forms arise directly from the chiral blocks of the matrix-valued exterior product \(\se\wedge\se\). Using the explicit block representation of the Lorentz generators, one
finds
\begin{equation}
\se\wedge\se
=
-2
\begin{pmatrix}
e^{a}\wedge e^{b}
\left(\Sigma_{ab}\right)^{\alpha}{}_{\beta}
&
0
\\[2mm]
0
&
e^{a}\wedge e^{b}
\left(\bar{\Sigma}_{ab}\right)_{\dot{\alpha}}{}^{\dot{\beta}}
\end{pmatrix}.
\end{equation}
This naturally defines the left- and right-handed Plebanski two-forms
by
\begin{align}
\mathcal{B}_{L}^{\alpha}{}_{\beta}
&:=
-e^{a}\wedge e^{b}
\left(\Sigma_{ab}\right)^{\alpha}{}_{\beta},
\\
\mathcal{B}_{R\dot{\alpha}}{}^{\dot{\beta}}
&:=
-e^{a}\wedge e^{b}
\left(\bar{\Sigma}_{ab}\right)_{\dot{\alpha}}{}^{\dot{\beta}}.
\end{align}
Consequently,
\begin{equation}
\se\wedge\se
=
2\begin{pmatrix}
\mathcal{B}_{L}^{\alpha}{}_{\beta} & 0
\\[2mm]
0 & \mathcal{B}_{R\dot{\alpha}}{}^{\dot{\beta}}
\end{pmatrix}.
\end{equation}

Equivalently, the Plebanski two-forms may be written as symmetric
rank-two spinors,
\begin{align}
\mathcal{B}_{L}^{\alpha\beta} &:= \epsilon^{\beta\gamma} \mathcal{B}_{L}^{\alpha}{}_{\gamma}
= -e^{a}\wedge e^{b} \left(\Sigma_{ab}\right)^{\alpha\beta},\\
\mathcal{B}_{R}^{\dot{\alpha}\dot{\beta}} &:= \epsilon^{\dot{\alpha}\dot{\gamma}} \mathcal{B}_{R\dot{\gamma}}{}^{\dot{\beta}} 
= -e^{a}\wedge e^{b} \left(\bar{\Sigma}_{ab}\right)^{\dot{\alpha}\dot{\beta}}.
\end{align}

The chirality of these spinor-valued two-forms is encoded in the duality properties of the Lorentz generators entering their definition. Indeed,
\begin{equation}
\frac{1}{2}\epsilon_{ab}{}^{cd} (\Sigma_{cd})_{\alpha\beta} =  -i(\Sigma_{ab})_{\alpha\beta},
\qquad
\frac{1}{2}\epsilon_{ab}{}^{cd} (\bar\Sigma_{cd})^{\dot\alpha\dot\beta} = +i(\bar\Sigma_{ab})^{\dot\alpha\dot\beta}.
\end{equation}
Thus, $\mathcal B_L^{\alpha\beta}$ and $\mathcal B_R^{\dot\alpha\dot\beta}$ are the spinorial representatives of the $(1,0)$ and $(0,1)$ components, respectively, of $e^a\wedge e^b$.

It is custom to define
\begin{equation}
E^{i}:=e^{0}\wedge e^{i},
\qquad
S^{i}:=\frac{1}{2}\epsilon^{i}{}_{jk}\,
e^{j}\wedge e^{k}.
\end{equation}
The spacetime Hodge operator acts as
\begin{equation}
\ast E^{i}=-S^{i},
\qquad
\ast S^{i}=E^{i},
\end{equation}
so that \(\ast^{2}=-1\) on two-forms. The chiral Plebanski two-forms are therefore
\begin{align}
B^{i}
&=
\frac{1}{2}\epsilon^{i}{}_{jk}\,
e^{j}\wedge e^{k}
+i\,e^{0}\wedge e^{i},
\\
\bar{B}^{i}
&=
\frac{1}{2}\epsilon^{i}{}_{jk}\,
e^{j}\wedge e^{k}
-i\,e^{0}\wedge e^{i}.
\end{align}
Their duality properties are
\begin{align}
\ast B^{i}
&=
-iB^{i},
\\
\ast\bar{B}^{i}
&=
+i\bar{B}^{i}.
\end{align}
Thus, in the conventions adopted here, \(B^{i}\) is anti-self-dual, whereas \(\bar{B}^{i}\) is self-dual. Moreover, for a real coframe,
\begin{equation}
\bar{B}^{i}=\left(B^{i}\right)^{*}.
\end{equation}

The relation between the symmetric-spinor two-forms and the usual Plebański basis follows directly by comparing the two representations of $\se\wedge\se$. One obtains
\begin{equation}
\mathcal B^L_{\alpha\beta}
=
2i\,B^i(\tau_i )_{\alpha\beta},
\qquad
\mathcal B_R^{\dot\alpha\dot\beta}
=
-2i\,\bar B^i
(\bar\tau_i )^{\dot\alpha\dot\beta}.
\end{equation}
where
\begin{equation}
 (\tau_i)_{\alpha \beta} = (\Sigma_{0i})_{\alpha \beta}, \qquad (\bar{\tau}_i)^{\dot{\alpha} \dot{\beta}} = (\bar\Sigma_{0i})^{\dot\alpha\dot\beta}\,.
\end{equation}
Equivalently, at the level of the endomorphism-valued two-forms,
\begin{equation}
\mathcal B_L
=
2i B^i J^L_{0i},
\qquad
\mathcal B_R
=
-2i\bar B^i J^R_{0i}.
\end{equation}
Consequently,
\begin{equation}
\se\wedge\se
=
4iB^iJ^L_{0i}
-
4i\bar B^iJ^R_{0i}.
\end{equation}

\section{The \(6\times 6\) graded realization of \(\mathfrak{osp}(1|2,\mathbb C)_L\oplus
\mathfrak{osp}(1|2,\mathbb C)_R\)}
\label{app:osp-six-dimensional-realization}

This section constructs an explicit \(6\times6\) graded matrix realization of \(\mathfrak{osp}(1|2,\mathbb C)_L\oplus \mathfrak{osp}(1|2,\mathbb C)_R\), obtained by combining the fundamental \(2|1\) representations of its left and right chiral factors. Let us remark that, when lifted to the $6 \times 6$ Clifford realization the exterior product $\se \wedge \se$ also belong to the chiral diagonal blocks,
\begin{align}
\se\wedge\se
&=
e^{a}\wedge e^{b}
\left(
\Gamma_{a}^{LR}\Gamma_{b}^{LR}
+
\Gamma_{a}^{LR}\Gamma_{b}^{RL}
+
\Gamma_{a}^{RL}\Gamma_{b}^{LR}
+
\Gamma_{a}^{RL}\Gamma_{b}^{RL}
\right)
\nonumber\\
&=
e^{a}\wedge e^{b}
\left(
\Gamma_{a}^{LR}\Gamma_{b}^{RL}
+
\Gamma_{a}^{RL}\Gamma_{b}^{LR}
\right),
\end{align}
where we used
\begin{equation}
\Gamma_{a}^{LR}\Gamma_{b}^{LR}
=
\Gamma_{a}^{RL}\Gamma_{b}^{RL}
=
0.
\end{equation}
The remaining products satisfy
\begin{align}
\Gamma_{a}^{LR}\Gamma_{b}^{RL}
&=
-\eta_{ab}P_{L}-2J_{ab}^{L},
\\
\Gamma_{a}^{RL}\Gamma_{b}^{LR}
&=
-\eta_{ab}P_{R}-2J_{ab}^{R},
\end{align}
where \(P_L\) and \(P_R\) denote the projectors onto the two chiral
blocks. Since
\begin{equation}
\eta_{ab}\,e^{a}\wedge e^{b}=0,
\end{equation}
the terms proportional to the metric vanish, and therefore
\begin{equation}
\se\wedge\se
=
-2e^{a}\wedge e^{b}
\left(
J_{ab}^{L}+J_{ab}^{R}
\right).
\end{equation}

Now we will define the resulting \(4|2\)-dimensional carrier superspace and embed the Lorentz generators, supercharges, and Dirac matrices into it. We then study the graded adjoint action of the supercharges on the lifted vector and chiral bivector sectors. This action does not preserve either bosonic sector separately: it maps the bosonic matrices into odd partner matrices. Consequently, the induced supercharges are naturally described by rectangular maps between the even and odd sectors, which combine into endomorphisms of the complete \(3|2\) adjoint superspaces. Finally, we show that the anticommutators of these induced odd endomorphisms reproduce the corresponding even Lorentz transformations, thereby realizing explicitly the closure of the two chiral \(\mathfrak{osp}(1|2,\mathbb C)\) algebras.

\subsection{The fundamental superspace}

Let each factor \(\mathfrak{osp}(1|2,\mathbb C)\) act in its fundamental \(2|1\)-dimensional representation \cite{Kac:1977em,Frappat:1996pb}.  The carrier space of the direct sum is
\begin{equation}
 V=V_L\oplus V_R,
 \qquad
 V_L=S_L\oplus\Pi\mathbb C_L,
 \qquad
 V_R=S_R\oplus\Pi\mathbb C_R, \label{eq:fundamental-superspace}
\end{equation}
where \(S_L\simeq\mathbb C^2\) and \(S_R\simeq\mathbb C^2\) carry, respectively, the Lorentz representations \((\tfrac12,0)\) and \((0,\tfrac12)\).  The symbol \(\Pi\) reverses parity; hence \(\Pi\mathbb C_L\) and \(\Pi\mathbb C_R\) are one-dimensional odd Lorentz-singlet spaces.  Thus
\begin{equation}
 V\simeq\mathbb C^{4|2},
 \qquad
 V=S_L\oplus\Pi\mathbb C_L\oplus
   S_R\oplus\Pi\mathbb C_R.  \label{eq:ordered-six-dimensional-space}
\end{equation}
We shall use the \(2+1+2+1\) ordering displayed in \eqref{eq:ordered-six-dimensional-space}.  The grading operator and supertrace are then
\begin{equation}
 \Gamma_{\mathrm{gr}}
 =\operatorname{diag}(\mathbf 1_2,-1,\mathbf 1_2,-1),
 \qquad
 \operatorname{STr}_V X
 =\operatorname{Tr}(\Gamma_{\mathrm{gr}}X).
 \label{eq:grading-and-supertrace}
\end{equation}

\subsection{Embedding of the two superalgebras}

In the ordered basis \eqref{eq:ordered-six-dimensional-space}, the bosonic generators are lifted as
\begin{align}
 \widehat J^{\,L}_{ab}
 &=
 \operatorname{diag}\bigl(\Sigma_{ab},0,0_2,0\bigr),
 \\
 \widehat J^{\,R}_{ab}
 &=
 \operatorname{diag}\bigl(0_2,0,\bar\Sigma_{ab},0\bigr).
 \label{eq:lifted-lorentz-generators}
\end{align}
Introduce \(2\times1\) columns \(u_\alpha,\bar u^{\dot\alpha}\) and
\(1\times2\) rows \(v_\alpha,\bar v^{\dot\alpha}\), normalized by
\begin{equation}
 (u_\alpha)^\gamma=\delta_\alpha{}^\gamma,
 \qquad
 (v_\alpha)_\gamma=\epsilon_{\alpha\gamma},
 \qquad
 (\bar u^{\dot\alpha})_{\dot\gamma}
   =\delta_{\dot\gamma}{}^{\dot\alpha},
 \qquad
 (\bar v^{\dot\alpha})^{\dot\gamma}
   =\epsilon^{\dot\alpha\dot\gamma}.
 \label{eq:odd-block-normalization}
\end{equation}
The odd generators \(Q_\alpha\) and \(Q^{\dot\alpha}\) are represented by
\begin{align}
 \widehat Q_\alpha
 &=
 \begin{pmatrix}
  0_2&u_\alpha&0&0\\
  v_\alpha&0&0&0\\
  0&0&0_2&0\\
  0&0&0&0
 \end{pmatrix},
 &
 \widehat Q^{\dot\alpha}
 &=
 \begin{pmatrix}
  0_2&0&0&0\\
  0&0&0&0\\
  0&0&0_2&\bar u^{\dot\alpha}\\
  0&0&\bar v^{\dot\alpha}&0
 \end{pmatrix}.
 \label{eq:lifted-supercharges}
\end{align}
They obey
\begin{align}
 [\widehat J^{\,L}_{ab},\widehat Q_\alpha]
 &=(\Sigma_{ab})^\beta{}_\alpha\widehat Q_\beta,
 \\
 [\widehat J^{\,R}_{ab},\widehat Q^{\dot\alpha}]
 &=(\bar\Sigma_{ab})_{\dot\beta}{}^{\dot\alpha}
   \widehat Q^{\dot\beta},
 \label{eq:lorentz-action-on-supercharges}
\end{align}
and every graded commutator between a left and a right generator vanishes.

For later use, define symmetric spinor generators by
\begin{align}
 \{\widehat Q_\alpha,\widehat Q_\beta\}
 &=\widehat J^{\,L}_{\alpha\beta},
 &
 (\widehat J^{\,L}_{\alpha\beta})^\gamma{}_\delta
 &=
 \delta_\alpha{}^\gamma\epsilon_{\beta\delta}
 +\delta_\beta{}^\gamma\epsilon_{\alpha\delta},
 \\
 \{\widehat Q^{\dot\alpha},\widehat Q^{\dot\beta}\}
 &=\widehat J_R^{\dot\alpha\dot\beta},
 &
 (\widehat J_R^{\dot\alpha\dot\beta})_{\dot\gamma}
 {}^{\dot\delta}
 &=
 \delta_{\dot\gamma}{}^{\dot\alpha}
   \epsilon^{\dot\beta\dot\delta}
 +\delta_{\dot\gamma}{}^{\dot\beta}
   \epsilon^{\dot\alpha\dot\delta}.
 \label{eq:fundamental-odd-odd-brackets}
\end{align}
Equivalently, in the bivector basis fixed by
\eqref{eq:sigma-generator-definitions},
\begin{align}
 \{\widehat Q_\alpha,\widehat Q_\beta\}
 &=
 -(\Sigma^{ab})_{\alpha\beta}\widehat J^{\,L}_{ab},
 \\
 \{\widehat Q^{\dot\alpha},\widehat Q^{\dot\beta}\}
 &=
 -(\bar\Sigma^{ab})^{\dot\alpha\dot\beta}
   \widehat J^{\,R}_{ab},
 \qquad
 \{\widehat Q_\alpha,\widehat Q^{\dot\beta}\}=0.
 \label{eq:bivector-form-odd-odd-brackets}
\end{align}

\subsection{Lifted Dirac matrices}

The ordinary Weyl-basis Dirac matrices,
\begin{equation}
 \Gamma_a=
 \begin{pmatrix}
  0&\sigma_a\\
  \bar\sigma_a&0
 \end{pmatrix},
 \qquad
 \{\Gamma_a,\Gamma_b\}=2\eta_{ab}\mathbf 1_4,
 \label{eq:four-dimensional-dirac-matrices}
\end{equation}
lift to \(V\) as
\begin{equation}
 \widehat\Gamma_a=
 \begin{pmatrix}
  0_2&0&\sigma_a&0\\
  0&0&0&0\\
  \bar\sigma_a&0&0_2&0\\
  0&0&0&0
 \end{pmatrix}.
 \label{eq:lifted-gamma-matrices}
\end{equation}
They vanish on the odd lines and therefore satisfy the projected Clifford relation
\begin{equation}
 \{\widehat\Gamma_a,\widehat\Gamma_b\}
 =2\eta_{ab}\widehat P,
 \qquad
 \widehat P
 =\operatorname{diag}(\mathbf 1_2,0,\mathbf 1_2,0).
 \label{eq:projected-clifford-relation}
\end{equation}
Similarly,
\begin{equation}
 \widehat\Gamma_5
 =\operatorname{diag}(-\mathbf 1_2,0,\mathbf 1_2,0),
 \qquad
 \widehat\Gamma_5^2=\widehat P.
 \label{eq:lifted-gamma-five}
\end{equation}
Consequently, \(\widehat\Gamma_a\) do not define a unital six-dimensional Clifford module: the Clifford unit is represented by \(\widehat P\), rather than by \(\mathbf 1_6\).  They instead give the ordinary Clifford representation on the even subspace \(S_L\oplus S_R\), extended by zero on \(\Pi\mathbb C_L\oplus\Pi\mathbb C_R\).

\subsection{Odd action on the vector sector}

Let an odd generator act on a matrix \(X\) through the graded adjoint action
\begin{equation}
 \mathcal Q_\alpha(X)
 :=[\widehat Q_\alpha,X]_{\mathrm{gr}},
 \qquad
 \mathcal Q^{\dot\alpha}(X)
 :=[\widehat Q^{\dot\alpha},X]_{\mathrm{gr}}.
 \label{eq:graded-adjoint-action}
\end{equation}
Because \(\widehat\Gamma_a\) is even,
\begin{equation}
 \mathcal Q_\alpha(\widehat\Gamma_a)
 =
 [\widehat Q_\alpha,\widehat\Gamma_a]
 =
 -[\widehat\Gamma_a,\widehat Q_\alpha].
 \label{eq:left-induced-vector-action}
\end{equation}
The resulting odd matrix is
\begin{equation}
 \Xi_{\alpha a}
 :=
 \mathcal Q_\alpha(\widehat\Gamma_a)
 =
 \begin{pmatrix}
  0&0&0&0\\
  0&0&v_\alpha\sigma_a&0\\
  0&-\bar\sigma_a u_\alpha&0&0\\
  0&0&0&0
 \end{pmatrix}.
 \label{eq:left-vector-superpartner}
\end{equation}
Likewise, in the right sector,
\begin{equation}
 \Xi^{\dot\alpha}{}_a
 :=
 \mathcal Q^{\dot\alpha}(\widehat\Gamma_a)
 =
 \begin{pmatrix}
  0&0&0&-\sigma_a\bar u^{\dot\alpha}\\
  0&0&0&0\\
  0&0&0&0\\
  \bar v^{\dot\alpha}\bar\sigma_a&0&0&0
 \end{pmatrix}.
 \label{eq:right-vector-superpartner}
\end{equation}

Equations \eqref{eq:left-vector-superpartner} and \eqref{eq:right-vector-superpartner} also exhibit an important representation-theoretic point.  Every linear combination \(C^b{}_a\widehat\Gamma_b\), with scalar coefficients \(C^b{}_a\), has support only in the \(S_L\leftrightarrow S_R\) blocks.  By contrast, \(\Xi_{\alpha a}\) and \(\Xi^{\dot\alpha}{}_a\) connect a Weyl space to an odd line.  Therefore no ordinary tensor \((Q_\alpha^{(\mathrm{vector})})^b{}_a\) can satisfy
\begin{equation}
 (Q_\alpha^{(\mathrm{vector})})^b{}_a\widehat\Gamma_b
 =
 -[\widehat\Gamma_a,\widehat Q_\alpha].
 \label{eq:impossible-vector-endomorphism}
\end{equation}
The correct object is the odd linear map
\begin{equation}
 \mathcal Q_\alpha^{(\mathrm{vector})}:
 \widehat\Gamma_a\longmapsto\Xi_{\alpha a},
 \label{eq:correct-vector-odd-map}
\end{equation}
whose codomain is the odd partner space spanned by \(\Xi_{\alpha a}\), rather than the original bosonic vector space. The same statement applies to \(\mathcal Q^{\dot\alpha}\).

\subsection{Odd action on the self-dual bivector sector}

Consider the left anti-self-dual Lorentz sector \(\mathfrak g_{\bar 0}^{L}\simeq(1,0)\), spanned either by \(\widehat J^{\,L}_{ab}\), with the anti-self-duality condition understood, or by the symmetric generators
\(\widehat J^{\,L}_{\alpha\beta}\).  From
\eqref{eq:lorentz-action-on-supercharges},
\begin{equation}
 \mathcal Q_\alpha(\widehat J^{\,L}_{ab})
 =
 [\widehat Q_\alpha,\widehat J^{\,L}_{ab}]
 =
 -(\Sigma_{ab})^\beta{}_\alpha\widehat Q_\beta.
 \label{eq:odd-action-on-left-bivector}
\end{equation}
The right-hand side is odd and hence cannot be expanded in the even basis \(\widehat J^{\,L}_{cd}\).  Thus there is no scalar tensor \((Q_\alpha^{(1,0)})^{cd}{}_{ab}\) satisfying
\begin{equation}
 (Q_\alpha^{(1,0)})^{cd}{}_{ab}\widehat J^{\,L}_{cd}
 =
 -[\widehat J^{\,L}_{ab},\widehat Q_\alpha].
 \label{eq:impossible-bivector-endomorphism}
\end{equation}
Instead one obtains the rectangular map from the even bivector sector to the odd spinor sector,
\begin{equation}
 (M_\alpha)^\beta{}_{ab}
 =-(\Sigma_{ab})^\beta{}_\alpha,
 \qquad
 M_\alpha:\mathfrak g_{\bar0}^{L}\longrightarrow
 \mathfrak g_{\bar1}^{L}.
 \label{eq:left-rectangular-R-map}
\end{equation}
The reverse rectangular map follows from the odd--odd bracket:
\begin{equation}
 (N_\alpha)^{cd}{}_\beta
 =-(\Sigma^{cd})_{\alpha\beta},
 \qquad
 N_\alpha:\mathfrak g_{\bar1}^{L}\longrightarrow
 \mathfrak g_{\bar0}^{L}.
 \label{eq:left-rectangular-L-map}
\end{equation}
The genuine odd endomorphism acts on the full adjoint superspace \(\mathfrak g_{\bar0}^{L}\oplus\mathfrak g_{\bar1}^{L}\simeq(3|2)\):
\begin{equation}
 \operatorname{ad}_{\widehat Q_\alpha}
 \equiv \mathbb Q_\alpha
 =
 \begin{pmatrix}
  0_{3 \times 3} & N_\alpha\\
  M_\alpha & 0_{2 \times 2}
 \end{pmatrix}.  \label{eq:left-full-adjoint-supercharge}
\end{equation}

The same construction in the self-dual right sector gives
\begin{align}
 (\bar M^{\dot\alpha})^{\dot\beta}{}_{ab}  &= -(\bar\Sigma_{ab})_{\dot\beta}{}^{\dot\alpha},  \\
 (\bar N^{\dot\alpha})^{cd}{}_{\dot\beta}  &=  -(\bar\Sigma^{cd})^{\dot\alpha\dot\beta},      \\
 \operatorname{ad}_{\widehat Q^{\dot\alpha}}
 \equiv \mathbb Q^{\dot\alpha}
 &=
 \begin{pmatrix}
  0_{3 \times 3} & \bar N^{\dot\alpha}\\
  \bar M^{\dot\alpha} & 0_{2 \times 2}
 \end{pmatrix},
 \label{eq:right-full-adjoint-supercharge}
\end{align}
acting on
\(\mathfrak g_{\bar0}^{R}\oplus\mathfrak g_{\bar1}^{R}
\simeq(3|2)\).

\subsection{Closure of the induced supercharges}

The fact that the left and right Lorentz generators satisfy the same Lorentz commutation relations does not by itself distinguish the representations \((1,0)\) and \((0,1)\). Their distinction is determined by the chiral bivector subspaces on which they act.

Let
\begin{equation}
 \mathcal B:=\Lambda^2_{\mathbb C}
 =\mathcal B_L\oplus\mathcal B_R
 \simeq(1,0)\oplus(0,1)
\end{equation}
be the complexified bivector space.

First define the Lorentz generator on the full bivector space by
\begin{equation}
 [J_{ef},J_{ab}]
 =
 (\mathcal J_{ef}^{(2)})^{cd}{}_{ab}J_{cd},
 \label{eq:full-bivector-generator-definition}
\end{equation}
where, with \(\delta^{cd}_{ab}:=\delta^c_{[a}\delta^d_{b]}\),
\begin{equation}
 (\mathcal J_{ef}^{(2)})^{cd}{}_{ab}
 =
 -\eta_{ea}\delta^{cd}_{fb}
 +\eta_{eb}\delta^{cd}_{fa}
 +\eta_{fa}\delta^{cd}_{eb}
 -\eta_{fb}\delta^{cd}_{ea}.
 \label{eq:full-bivector-generator}
\end{equation}
Since the Lorentz action commutes with the Hodge operator,
\begin{equation}
 [\mathcal J_{ef}^{(2)},\ast]=0,
\end{equation}
the two chiral bivector subspaces are invariant. The generators on these subspaces are therefore the restrictions
\begin{align}
 J_{ef}^{(1,0)}
 &=
 P_L\mathcal J_{ef}^{(2)}P_L,
 &
 J_{ef}^{(0,1)}
 &=
 P_R\mathcal J_{ef}^{(2)}P_R.
 \label{eq:chiral-bivector-generators}
\end{align}
In components,
\begin{align}
 (J_{ef}^{(1,0)})^{cd}{}_{ab}
 &=
 (P_L)^{cd}{}_{gh}
 (\mathcal J_{ef}^{(2)})^{gh}{}_{ij}
 (P_L)^{ij}{}_{ab},
 \\
 (J_{ef}^{(0,1)})^{cd}{}_{ab}
 &=
 (P_R)^{cd}{}_{gh}
 (\mathcal J_{ef}^{(2)})^{gh}{}_{ij}
 (P_R)^{ij}{}_{ab}.
 \label{eq:chiral-bivector-generators-components}
\end{align}
where
\begin{align}
\left(P_L\right)^{ab}{}_{cd}
&=
\frac{1}{4}
\left(
\eta^{a}{}_{c}\eta^{b}{}_{d}
-
\eta^{b}{}_{c}\eta^{a}{}_{d}
\right)
+
\frac{i}{4}\epsilon^{ab}{}_{cd},
\\[4pt]
\left(P_R\right)^{ab}{}_{cd}
&=
\frac{1}{4}
\left(
\eta^{a}{}_{c}\eta^{b}{}_{d}
-
\eta^{b}{}_{c}\eta^{a}{}_{d}
\right)
-
\frac{i}{4}\epsilon^{ab}{}_{cd}.
\end{align}

Consequently,
\begin{align}
 [\widehat J^{\,L}_{ef},\widehat J^{\,L}_{ab}]
 &=
 (J_{ef}^{(1,0)})^{cd}{}_{ab}
 \widehat J^{\,L}_{cd},
 \\
 [\widehat J^{\,R}_{ef},\widehat J^{\,R}_{ab}]
 &=
 (J_{ef}^{(0,1)})^{cd}{}_{ab}
 \widehat J^{\,R}_{cd}.
 \label{eq:chiral-bivector-commutators}
\end{align}
Here the bivector indices in the first and second equations are understood as projected with \(P_L = P_{(1,0)}\) and \(P_R = P_{(0,1)}\), respectively. Although these equations have the same Lorentz-algebraic form, they act on inequivalent three-dimensional chiral carrier spaces. Since \(\mathcal J_{ef}^{(2)}\), \(\eta_{ab}\), and \(\epsilon_{abcd}\) are real, the two chiral generators are related by complex conjugation:
\begin{equation}
\left(J_{ef}^{(0,1)}\right)^{cd}{}_{ab}
=
\overline{
\left(J_{ef}^{(1,0)}\right)^{cd}{}_{ab}
}.
\end{equation}

The even element \(\widehat J^L_{ab}\) acts on both graded
components of the adjoint superspace. On the even component it acts
in the self-dual spin-one representation, whereas on the odd component
it acts in the left-handed spinor representation:
\begin{align}
 [\widehat J^L_{ab},\widehat J^L_{cd}]
 &=
 \left(J_{ab}^{(1,0)}\right)^{ef}{}_{cd}
 \widehat J^L_{ef},
 \\
 [\widehat J^L_{ab},\widehat Q_\alpha]
 &=
 (\Sigma_{ab})^\beta{}_\alpha
 \widehat Q_\beta.
\end{align}
Therefore, its adjoint representation on
\(\mathfrak g_{\bar0}^L\oplus\mathfrak g_{\bar1}^L\) is
\begin{equation}
 \operatorname{ad}_{\widehat J^L_{ab}}
 \equiv
 \mathbb J^L_{ab}
 =
 \begin{pmatrix}
  J_{ab}^{(1,0)}&0\\
  0&\Sigma_{ab}
 \end{pmatrix}.
\end{equation}
Note that the two diagonal blocks are not duplicate generators: they are the
representations of the same Lorentz generator on the even and odd
components, respectively.

Similarly, on the right \(3|2\) adjoint superspace,
\begin{equation}
 \mathbb J^{\,R}_{ab}
 =
 \begin{pmatrix}
  J_{ab}^{(0,1)}&0\\
  0&\bar\Sigma_{ab}
 \end{pmatrix},
 \label{eq:right-even-adjoint-generator}
\end{equation}
where
\begin{equation}
 [\widehat J^{\,R}_{ab},\widehat Q^{\dot\alpha}]
 =
 (\bar\Sigma_{ab})_{\dot\beta}{}^{\dot\alpha}
 \widehat Q^{\dot\beta}.
\end{equation}

The rectangular maps do not possess anticommutators separately: \(M_\alpha M_\beta\) and \(N_\alpha N_\beta\) are not composable. The full odd matrices \(\mathbb Q_\alpha\), however, obey
\begin{align}
 \{\mathbb Q_\alpha,\mathbb Q_\beta\}
 &=
 \begin{pmatrix}
  N_\alpha M_\beta+N_\beta M_\alpha&0\\
  0&M_\alpha N_\beta+M_\beta N_\alpha
 \end{pmatrix}
 \nonumber\\
 &=
 -(\Sigma^{ab})_{\alpha\beta}
 \begin{pmatrix}
  J_{ab}^{(1,0)}&0\\
  0&\Sigma_{ab}
 \end{pmatrix}
 =
 -(\Sigma^{ab})_{\alpha\beta}\mathbb J^{\,L}_{ab}.
 \label{eq:left-induced-supercharge-closure}
\end{align}
This is simply the adjoint representation of the original \(\mathfrak{osp}(1|2,\mathbb C)_L\) odd--odd bracket.

Analogously, in the right sector,
\begin{equation}
 \{\mathbb Q^{\dot\alpha},\mathbb Q^{\dot\beta}\}
 =
 -(\bar\Sigma^{ab})^{\dot\alpha\dot\beta}
 \begin{pmatrix}
  J_{ab}^{(0,1)}&0\\
  0&\bar\Sigma_{ab}
 \end{pmatrix}
 =
 -(\bar\Sigma^{ab})^{\dot\alpha\dot\beta}
 \mathbb J^{\,R}_{ab}.
 \label{eq:right-induced-supercharge-closure}
\end{equation}
Because the left and right superalgebras form a direct sum,
\begin{equation}
 \{\mathbb Q_\alpha,\mathbb Q^{\dot\beta}\}=0.
\end{equation}

More generally, define
\begin{equation}
 \mathcal Q_\alpha X
 :=
 [\widehat Q_\alpha,X]_{\mathrm{gr}}.
\end{equation}
The graded Jacobi identity implies that the induced action closes on every matrix \(X\) in the module generated from \(\widehat\Gamma_a\):
\begin{equation}
 \bigl\{\mathcal Q_\alpha,\mathcal Q_\beta\bigr\}X
 =
 \bigl[
   \{\widehat Q_\alpha,\widehat Q_\beta\},X
 \bigr]_{\mathrm{gr}}.
 \label{eq:induced-adjoint-closure}
\end{equation}

Thus the induced supersymmetry transformations close on the enlarged
graded module containing both the bosonic matrices and their odd
partners. This does not contradict the closure of
\(\mathfrak{osp}(1|2,\mathbb C)_L\): the complete graded space spanned
by \(\widehat J^L_{ab}\) and \(\widehat Q_\alpha\) is closed under the
graded bracket. However, the bosonic subspaces separately are not
invariant under an individual supersymmetry transformation. In
particular,
\begin{equation}
 [\widehat Q_\alpha,\widehat J^L_{ab}]
 \propto
 \widehat Q_\beta,
 \qquad
 [\widehat Q_\alpha,\widehat\Gamma_a]_{\mathrm{gr}}
 =
 \Xi_{\alpha a}.
\end{equation}

Thus, neither the self-dual bivector span nor the span of \(\widehat\Gamma_a\) is preserved by a single supersymmetry transformation. Nevertheless, the graded composition of two supersymmetry transformations closes onto an even Lorentz transformation.

\section{Three-coframe Clifford identity}
\label{app:threecoframecliffordidentity}

We derive here the three-coframe identity used in Eq.~\eqref{eq:H-Dirac-identity}. Starting from
Eqs.~\eqref{eq:H-threeform-definition} and \eqref{eq:sese-left-matter},
\begin{align}
H_\gamma{}^{\dot\beta}
&=
-2e^a\wedge e^b\wedge e^c\,
\epsilon_{\alpha\beta}
(\Sigma_{ab})^\alpha{}_\gamma
(\sigma_c)^{\beta\dot\beta}
\nonumber\\
&=
2e^a\wedge e^b\wedge e^c\,
(\Sigma_{ab})_{\beta\gamma}
(\sigma_c)^{\beta\dot\beta}
\nonumber\\
&=
2e^a\wedge e^b\wedge e^c\,
(\Sigma_{ab})_{\gamma\beta}
(\sigma_c)^{\beta\dot\beta},
\end{align}
where in the last step we used the symmetry of
$(\Sigma_{ab})_{\alpha\beta}$.

From the definition of the Lorentz generators in Eq.~\eqref{eq:sigma-generator-definitions}, together with the Clifford relations, one obtains
\begin{equation}
(\Sigma_{ab})_{\gamma\beta}
(\sigma_c)^{\beta\dot\beta}
=
\frac{1}{2}
\left[
\eta_{bc}(\sigma_a)_\gamma{}^{\dot\beta}
-\eta_{ac}(\sigma_b)_\gamma{}^{\dot\beta}
-i\epsilon_{abcd}
(\sigma^d)_\gamma{}^{\dot\beta}
\right].
\end{equation}
The sign of the $\epsilon$ term is consistent with the left-handed duality convention $\star\Sigma_{ab}=-i\Sigma_{ab}$ in Eq.~\eqref{starSigmas}.

After contraction with $e^a\wedge e^b\wedge e^c$, the two terms proportional to the metric vanish, since they identify two coframe indices. Therefore,
\begin{equation}
e^a\wedge e^b\wedge e^c\,
(\Sigma_{ab})_{\gamma\beta}
(\sigma_c)^{\beta\dot\beta}
=
-\frac{i}{2}
\epsilon_{abcd}\,
e^a\wedge e^b\wedge e^c\,
(\sigma^d)_\gamma{}^{\dot\beta}.
\end{equation}
Substituting this result into the definition of $H$ gives
\begin{equation}
H_\gamma{}^{\dot\beta}
=
-i\epsilon_{abcd}\,
e^a\wedge e^b\wedge e^c\,
(\sigma^d)_\gamma{}^{\dot\beta}.
\end{equation}

Using the spacetime Hodge-dual convention adopted in
Appendix~\ref{app:preliminaries},
\begin{equation}
\epsilon_{abcd}\,
e^a\wedge e^b\wedge e^c
=
-3!\,\ast e_d,
\end{equation}
we finally obtain
\begin{equation}
H_\gamma{}^{\dot\beta}
=
6i\,(\ast e_d)
(\sigma^d)_\gamma{}^{\dot\beta}
=
6i\,(\ast e^a)
(\sigma_a)_\gamma{}^{\dot\beta},
\end{equation}
which proves Eq.~\eqref{eq:H-Dirac-identity}.

\section{Matter ansatz subspace}
\label{app:matteransatzpullback}

The terminology ``pullback'' used in \eqref{eq:original-action-psiR-pullback}, refers to the restriction of the variational problem to the matter-ansatz subspace of field space.  More precisely, for fixed coframe the ansatz defines the embedding
\begin{equation}
 \iota_e:\psi_R\longmapsto\Psi_L=\slashed e\,\psi_R,
 \qquad
 (\iota_e\psi_R)^\alpha
 =
 e^{\alpha\dot\alpha}\psi_{R\dot\alpha}.
\end{equation}
If the unrestricted vector--spinor variation is written as
\begin{equation}
 \delta S_L
 =
 \int_M
 \delta\Psi_L^\alpha\wedge\mathcal E_{L\alpha}
 +\cdots,
\end{equation}
then, on the matter-ansatz subspace and with $\delta e=0$,
\begin{equation}
 \delta\Psi_L^\alpha
 =
 e^{\alpha\dot\alpha}\delta\psi_{R\dot\alpha}.
\end{equation}
Consequently,
\begin{equation}
 \iota_e^*(\delta S_L)
 =
 \int_M
 \delta\psi_{R\dot\alpha}\,
 e^{\alpha\dot\alpha}\wedge
 \left.
 \mathcal E_{L\alpha}
 \right|_{\Psi_L=\slashed e\psi_R}.
\end{equation}
Thus the Euler--Lagrange equation of the reduced matter field is the projection, or equivalently the pullback in field space, of the unrestricted vector--spinor equation,
\begin{equation}
 e^{\alpha\dot\alpha}\wedge
 \left.
 \mathcal E_{L\alpha}
 \right|_{\Psi_L=\slashed e\psi_R}
 =0.
\end{equation}

\end{appendices}

\bibliographystyle{ieeetr}
\bibliography{ch-gr_paper_v10.bib}

@article{Tung:1999ic,
    author = "Tung, Roh Suan",
    title = "{Gravitation as a supersymmetric gauge theory}",
    eprint = "gr-qc/9904008",
    archivePrefix = "arXiv",
    doi = "10.1016/S0375-9601(99)00845-2",
    journal = "Phys. Lett. A",
    volume = "264",
    pages = "341--345",
    year = "2000"
}

@article{Chou:2005ht,
    author = "Chou, Chung-Hsien and Tung, Roh-Suan and Yu, Hoi-Lai",
    title = "{Origin of the Immirzi parameter}",
    eprint = "gr-qc/0509028",
    archivePrefix = "arXiv",
    doi = "10.1103/PhysRevD.72.064016",
    journal = "Phys. Rev. D",
    volume = "72",
    pages = "064016",
    year = "2005"
}

@article{Lisi:2010td,
    author = "Lisi, A. Garrett and Smolin, Lee and Speziale, Simone",
    title = "{Unification of gravity, gauge fields, and Higgs bosons}",
    eprint = "1004.4866",
    archivePrefix = "arXiv",
    primaryClass = "gr-qc",
    doi = "10.1088/1751-8113/43/44/445401",
    journal = "J. Phys. A",
    volume = "43",
    pages = "445401",
    year = "2010"
}

@article{Torres-Gomez:2012tka,
    author = "Torres-Gomez, Alexander and Krasnov, Kirill",
    title = "{Fermions via spinor-valued one-forms}",
    eprint = "1212.3452",
    archivePrefix = "arXiv",
    primaryClass = "hep-th",
    doi = "10.1142/S0217751X13501133",
    journal = "Int. J. Mod. Phys. A",
    volume = "28",
    number = "24",
    pages = "1350113",
    year = "2013"
}

@article{Valenzuela:2023aoa,
    author = "Valenzuela, Mauricio and Zanelli, Jorge",
    title = "{Massless Rarita-Schwinger equations: Half and three halves spin solution}",
    eprint = "2305.00106",
    archivePrefix = "arXiv",
    primaryClass = "hep-th",
    doi = "10.21468/SciPostPhys.16.3.065",
    journal = "SciPost Phys.",
    volume = "16",
    number = "3",
    pages = "065",
    year = "2024"
}

@article{Immirzi:1996di,
    author = "Immirzi, Giorgio",
    title = "{Real and complex connections for canonical gravity}",
    eprint = "gr-qc/9612030",
    archivePrefix = "arXiv",
    reportNumber = "DFUP-118-96",
    doi = "10.1088/0264-9381/14/10/002",
    journal = "Class. Quant. Grav.",
    volume = "14",
    pages = "L177--L181",
    year = "1997"
}

@article{Holst:1995pc,
    author = "Holst, Soren",
    title = "{Barbero's Hamiltonian derived from a generalized Hilbert-Palatini action}",
    eprint = "gr-qc/9511026",
    archivePrefix = "arXiv",
    reportNumber = "USITP-95-10",
    doi = "10.1103/PhysRevD.53.5966",
    journal = "Phys. Rev. D",
    volume = "53",
    pages = "5966--5969",
    year = "1996"
}

@article{Perez:2005pm,
    author = "Perez, Alejandro and Rovelli, Carlo",
    title = "{Physical effects of the Immirzi parameter}",
    eprint = "gr-qc/0505081",
    archivePrefix = "arXiv",
    doi = "10.1103/PhysRevD.73.044013",
    journal = "Phys. Rev. D",
    volume = "73",
    pages = "044013",
    year = "2006"
}

@article{Eder:2021rgt,
    author = "Eder, Konstantin and Sahlmann, Hanno",
    title = "{Holst-MacDowell-Mansouri action for (extended) supergravity with boundaries and super Chern-Simons theory}",
    eprint = "2104.02011",
    archivePrefix = "arXiv",
    primaryClass = "gr-qc",
    doi = "10.1007/JHEP07(2021)071",
    journal = "JHEP",
    volume = "07",
    pages = "071",
    year = "2021"
}

@article{Chagoya:2016zhy,
    author = "Chagoya, Javier and Sabido, Miguel",
    title = "{Topological M-theory, self-dual gravity and the Immirzi parameter}",
    eprint = "1612.04002",
    archivePrefix = "arXiv",
    primaryClass = "gr-qc",
    doi = "10.1088/1361-6382/aacebf",
    journal = "Class. Quant. Grav.",
    volume = "35",
    number = "16",
    pages = "165002",
    year = "2018"
}

@article{Mercuri:2006um,
    author = "Mercuri, Simone",
    title = "{Fermions in Ashtekar-Barbero connections formalism for arbitrary values of the Immirzi parameter}",
    eprint = "gr-qc/0601013",
    archivePrefix = "arXiv",
    doi = "10.1103/PhysRevD.73.084016",
    journal = "Phys. Rev. D",
    volume = "73",
    pages = "084016",
    year = "2006"
}

@book{Lauria:2020rhc,
    author = "Lauria, Edoardo and Van Proeyen, Antoine",
    title = "{${\cal N}=2$ Supergravity in $D=4,5,6$ Dimensions}",
    eprint = "2004.11433",
    archivePrefix = "arXiv",
    primaryClass = "hep-th",
    doi = "10.1007/978-3-030-33757-5",
    isbn = "978-3-030-33755-1, 978-3-030-33757-5",
    volume = "966",
    month = "3",
    year = "2020"
}

@article{Baum:1998ra,
    author = "Baum, Helga",
    title = "{Twistor spinors on Lorentzian symmetric spaces}",
    eprint = "math/9803089",
    archivePrefix = "arXiv",
    reportNumber = "SFB-288-313",
    doi = "10.1016/S0393-0440(99)00069-8",
    journal = "J. Geom. Phys.",
    volume = "34",
    pages = "270--286",
    year = "2000"
}

@article{Francois:2024xqi,
    author = "Fran{\c{c}}ois, J. and Ravera, L.",
    title = "{Unconventional supersymmetry via the dressing field method}",
    eprint = "2412.01898",
    archivePrefix = "arXiv",
    primaryClass = "hep-th",
    doi = "10.1103/76n5-4mg1",
    journal = "Phys. Rev. D",
    volume = "111",
    number = "12",
    pages = "125022",
    year = "2025"
}

@article{deMedeiros:2012sb,
    author = "de Medeiros, Paul",
    title = "{Rigid supersymmetry, conformal coupling and twistor spinors}",
    eprint = "1209.4043",
    archivePrefix = "arXiv",
    primaryClass = "hep-th",
    doi = "10.1007/JHEP09(2014)032",
    journal = "JHEP",
    volume = "09",
    pages = "032",
    year = "2014"
}

@article{Festuccia:2011ws,
    author = "Festuccia, Guido and Seiberg, Nathan",
    title = "{Rigid Supersymmetric Theories in Curved Superspace}",
    eprint = "1105.0689",
    archivePrefix = "arXiv",
    primaryClass = "hep-th",
    doi = "10.1007/JHEP06(2011)114",
    journal = "JHEP",
    volume = "06",
    pages = "114",
    year = "2011"
}

@article{Cartan:1923zea,
    author = "Cartan, E.",
    title = "{Sur les vari{\'e}t{\'e}s {\`a} connexion affine et la th{\'e}orie de la relativit{\'e} g{\'e}n{\'e}ralis{\'e}e. (premi{\`e}re partie)}",
    journal = "Annales Sci. Ecole Norm. Sup.",
    volume = "40",
    pages = "325--412",
    year = "1923"
}

@article{Sciama:1964jqa,
    author = "Sciama, D. W.",
    title = "{The Physical Structure of General Relativity}",
    doi = "10.1103/RevModPhys.36.463",
    journal = "Rev. Mod. Phys.",
    volume = "36",
    number = "1",
    pages = "463",
    year = "1964"
}

@article{Kibble:1961ba,
    author = "Kibble, T. W. B.",
    editor = "Hsu, Jong-Ping and Fine, D.",
    title = "{Lorentz invariance and the gravitational field}",
    doi = "10.1063/1.1703702",
    journal = "J. Math. Phys.",
    volume = "2",
    pages = "212--221",
    year = "1961"
}

@article{Hehl:1976kj,
    author = "Hehl, F. W. and Von Der Heyde, P. and Kerlick, G. D. and Nester, J. M.",
    title = "{General Relativity with Spin and Torsion: Foundations and Prospects}",
    doi = "10.1103/RevModPhys.48.393",
    journal = "Rev. Mod. Phys.",
    volume = "48",
    pages = "393--416",
    year = "1976"
}

@article{Weyl:1929fm,
    author = "Weyl, H.",
    title = "{Electron and Gravitation. 1. (In German)}",
    doi = "10.1007/BF01339504",
    journal = "Z. Phys.",
    volume = "56",
    pages = "330--352",
    year = "1929"
}

@article{MacDowell:1977jt,
    author = "MacDowell, S. W. and Mansouri, F.",
    title = "{Unified Geometric Theory of Gravity and Supergravity}",
    reportNumber = "COO-3075-164",
    doi = "10.1103/PhysRevLett.38.739",
    journal = "Phys. Rev. Lett.",
    volume = "38",
    pages = "739",
    year = "1977",
    note = "[Erratum: Phys.Rev.Lett. 38, 1376 (1977)]"
}

@article{Townsend:1977xw,
    author = "Townsend, Paul K.",
    title = "{Small Scale Structure of Space-Time as the Origin of the Gravitational Constant}",
    reportNumber = "ITP-SB-77-9",
    doi = "10.1103/PhysRevD.15.2795",
    journal = "Phys. Rev. D",
    volume = "15",
    pages = "2795",
    year = "1977"
}

@article{Troncoso:1997va,
    author = "Troncoso, Ricardo and Zanelli, Jorge",
    title = "{New gauge supergravity in seven-dimensions and eleven-dimensions}",
    eprint = "hep-th/9710180",
    archivePrefix = "arXiv",
    doi = "10.1103/PhysRevD.58.101703",
    journal = "Phys. Rev. D",
    volume = "58",
    pages = "101703",
    year = "1998"
}

@article{Troncoso:1998ng,
    author = "Troncoso, Ricardo and Zanelli, Jorge",
    editor = "Castagnino, M. A. and Nunez, Carmen A.",
    title = "{Gauge supergravities for all odd dimensions}",
    eprint = "hep-th/9807029",
    archivePrefix = "arXiv",
    doi = "10.1023/A:1026614631617",
    journal = "Int. J. Theor. Phys.",
    volume = "38",
    pages = "1181--1206",
    year = "1999"
}

@inproceedings{Zanelli:2005sa,
    author = "Zanelli, Jorge",
    title = "{Lecture notes on Chern-Simons (super-)gravities. Second edition (February 2008)}",
    booktitle = "{7th Mexican Workshop on Particles and Fields}",
    eprint = "hep-th/0502193",
    archivePrefix = "arXiv",
    month = "2",
    year = "2005"
}

@article{Alvarez:2011gd,
    author = "Alvarez, Pedro D. and Valenzuela, Mauricio and Zanelli, Jorge",
    title = "{Supersymmetry of a different kind}",
    eprint = "1109.3944",
    archivePrefix = "arXiv",
    primaryClass = "hep-th",
    doi = "10.1007/JHEP04(2012)058",
    journal = "JHEP",
    volume = "04",
    pages = "058",
    year = "2012"
}

@article{Alvarez:2013tga,
    author = "Alvarez, Pedro D. and Pais, Pablo and Zanelli, Jorge",
    title = "{Unconventional supersymmetry and its breaking}",
    eprint = "1306.1247",
    archivePrefix = "arXiv",
    primaryClass = "hep-th",
    reportNumber = "PAGES-314-321",
    doi = "10.1016/j.physletb.2014.06.031",
    journal = "Phys. Lett. B",
    volume = "735",
    pages = "314--321",
    year = "2014"
}

@article{Alvarez:2020qmy,
    author = "Alvarez, Pedro D. and Valenzuela, Mauricio and Zanelli, Jorge",
    title = "{Chiral gauge theory and gravity from unconventional supersymmetry}",
    eprint = "2005.04178",
    archivePrefix = "arXiv",
    primaryClass = "hep-th",
    doi = "10.1007/JHEP07(2020)205",
    journal = "JHEP",
    volume = "07",
    number = "07",
    pages = "205",
    year = "2020"
}

@article{Plebanski:1977zz,
    author = "Plebanski, Jerzy F.",
    title = "{On the separation of Einsteinian substructures}",
    doi = "10.1063/1.523215",
    journal = "J. Math. Phys.",
    volume = "18",
    pages = "2511--2520",
    year = "1977"
}

@article{Ashtekar:1986yd,
    author = "Ashtekar, A.",
    title = "{New Variables for Classical and Quantum Gravity}",
    doi = "10.1103/PhysRevLett.57.2244",
    journal = "Phys. Rev. Lett.",
    volume = "57",
    pages = "2244--2247",
    year = "1986"
}

@article{Jacobson:1988yy,
    author = "Jacobson, Ted and Smolin, Lee",
    title = "{Covariant Action for Ashtekar's Form of Canonical Gravity}",
    reportNumber = "BRX-TH-229",
    doi = "10.1088/0264-9381/5/4/006",
    journal = "Class. Quant. Grav.",
    volume = "5",
    pages = "583",
    year = "1988"
}

@article{Capovilla:1989ac,
    author = "Capovilla, Riccardo and Jacobson, Ted and Dell, John",
    title = "{General Relativity Without the Metric}",
    reportNumber = "UMDGR-90-044, TJHSST-89-1",
    doi = "10.1103/PhysRevLett.63.2325",
    journal = "Phys. Rev. Lett.",
    volume = "63",
    pages = "2325",
    year = "1989"
}

@article{Capovilla:1991qb,
    author = "Capovilla, R. and Jacobson, T. and Dell, J. and Mason, L. J.",
    title = "{Selfdual two forms and gravity}",
    doi = "10.1088/0264-9381/8/1/009",
    journal = "Class. Quant. Grav.",
    volume = "8",
    pages = "41--57",
    year = "1991"
}

@article{Krasnov:2011pp,
    author = "Krasnov, Kirill",
    title = "{Pure Connection Action Principle for General Relativity}",
    eprint = "1103.4498",
    archivePrefix = "arXiv",
    primaryClass = "gr-qc",
    doi = "10.1103/PhysRevLett.106.251103",
    journal = "Phys. Rev. Lett.",
    volume = "106",
    pages = "251103",
    year = "2011"
}

@article{Urbantke:1984eb,
    author = "Urbantke, H.",
    title = "{ON INTEGRABILITY PROPERTIES OF SU(2) YANG-MILLS FIELDS. I. INFINITESIMAL PART}",
    reportNumber = "UWTHPH-1984-02",
    doi = "10.1063/1.526402",
    journal = "J. Math. Phys.",
    volume = "25",
    number = "7",
    pages = "2321--2324",
    year = "1984"
}

@article{Fine:2013qta,
    author = "Fine, Joel and Krasnov, Kirill and Panov, Dmitri",
    title = "{A gauge theoretic approach to Einstein 4-manifolds}",
    eprint = "1312.2831",
    archivePrefix = "arXiv",
    primaryClass = "math.DG",
    journal = "New York J. Math.",
    volume = "20",
    pages = "293--323",
    year = "2014"
}

@article{Kac:1977em,
    author = "Kac, V. G.",
    title = "{Lie Superalgebras}",
    doi = "10.1016/0001-8708(77)90017-2",
    journal = "Adv. Math.",
    volume = "26",
    pages = "8--96",
    year = "1977"
}

@article{Frappat:1996pb,
    author = "Frappat, L. and Sorba, P. and Sciarrino, A.",
    title = "{Dictionary on Lie superalgebras}",
    eprint = "hep-th/9607161",
    archivePrefix = "arXiv",
    reportNumber = "ENSLAPP-AL-600-96, DSF-T-30-96",
    month = "7",
    year = "1996"
}

@article{VanProeyen:1999ni,
    author = "Van Proeyen, Antoine",
    title = "{Tools for supersymmetry}",
    eprint = "hep-th/9910030",
    archivePrefix = "arXiv",
    reportNumber = "KUL-TF-99-37",
    journal = "Ann. U. Craiova Phys.",
    volume = "9",
    number = "I",
    pages = "1--48",
    year = "1999"
}

\end{document}